\documentclass[iop]{emulateapj}

\shorttitle{FIR SEDs of mm detected quasars at $z>5$}
\shortauthors{Leipski et al.}

\begin{document}



\title{Complete infrared spectral energy distributions of mm detected quasars at $z>5$}


\author{C. Leipski\altaffilmark{1}}
\author{K. Meisenheimer\altaffilmark{1}}
\author{F. Walter\altaffilmark{1}}
\author{M.-A. Besel\altaffilmark{1}}
\author{H. Dannerbauer\altaffilmark{2}}
\author{X. Fan\altaffilmark{3}}
\author{M. Haas\altaffilmark{4}}
\author{U. Klaas\altaffilmark{1}}
\author{O. Krause\altaffilmark{1}}
\author{H.-W. Rix\altaffilmark{1}}


\altaffiltext{1}{Max-Planck Institut f\"ur Astronomie (MPIA),
     K\"onigstuhl 17, D-69117 Heidelberg, Germany; email: {\tt leipski@mpia-hd.mpg.de}}
\altaffiltext{2}{Universit\"at Wien, Institut f\"ur Astronomie, T\"urkenschanzstra{\ss}e 17, 1080 Wien, Austria}
\altaffiltext{3}{Steward Observatory, University of Arizona, Tucson, AZ 85721, USA}
\altaffiltext{4}{Astronomisches Institut Ruhr-Universit\"at Bochum, Universit\"atsstra{\ss}e
     150, D-44801 Bochum, Germany}


\begin{abstract}

We present {\it Herschel} far-infrared (FIR) photometry of eleven quasars 
at redshift $z>5$ that have previously been detected at 1.2\,mm. We perform 
full spectral energy distribution (SED) fits over the wavelength range
$\lambda_{\rm rest}$\,$\sim$\,0.1-400\,$\mu$m for those objects with good
{\it Herschel} detections. These fits reveal the need for an
additional far-infrared (FIR) component besides the emission from a
dusty AGN-powered torus. This additional FIR component has
temperatures of  T$_{\rm FIR}$\,$\sim$\,40-60\,K with luminosities of
L$_{8-1000\mu{\rm m}}$\,$\sim$\,10$^{13}$\,L$_{\odot}$ (accounting for 
25-60\% of the bolometric FIR luminosity). If the FIR dust emission 
is due to star formation it would suggest star formation rates in excess 
of 1000 solar masses per year. We show that at long wavelengths 
($\lambda_{\rm rest} \gtrsim 50\,\mu$m) the contribution of the AGN-powered 
torus emission is negligible. This explains how previous FIR studies of 
high-redshift quasars that relied on single component fits to (ground-based)
observations at $\lambda_{\rm obs} \gtrsim 350\,\mu$m reached
T$_{\rm FIR}$ and L$_{\rm FIR}$ values similar to our complete SED
fits. Stacking the {\it Herschel} data of four individually
undetected sources reveals a significant average signal in the PACS
bands but not in SPIRE. The average SED of sources with 
individual {\it Herschel} detections shows a striking surplus in near- and
mid-infrared emission when compared to common AGN templates. 
The comparison between two average SEDs (sources with and without 
individual {\it Herschel} detections) matched in the UV/optical indicates 
that for these objects the strength of the MIR emission may correlate 
with the strength of the FIR emission.

\end{abstract}





\section{Introduction}

The presence of dust seems to be a ubiquitous property of galaxies
throughout the observable universe. Even the most distant quasars at
$z\sim6$ show evidence for copious amounts of dust 
\citep[e.g.][]{ber03a,bee06,wan08a,lei10a}.
This indicates rapid metal enrichment of the interstellar medium
within the first billion years after the big bang. About 30\% of the
known luminous $z\sim6$ quasars are detected in the mm continuum with many 
of them also detected in CO \citep[e.g.][and references therein]{wan08a}. 
Such studies confirm the notion that most of the rest-frame far-infrared 
(FIR) emission comes from massive
star formation, possibly indicating the formation of early galactic
bulges. Thus, these objects signify an important stage in the
connection between the build-up of stellar mass and black hole growth.

Most high-redshift objects lack full FIR/sub-mm
spectral energy distributions (SEDs). L$_{\rm FIR}$ and M$_{\rm dust}$
are commonly determined using single photometric measurements, typically 
obtained at 1.2\,mm (=\,250\,GHz) and applying standard values for the dust
temperature as determined from lower redshift objects. It is unknown
whether this assumption is appropriate for high-redshift
objects. Ground-based 350\,$\mu$m observations of a few 
$z>5$ quasars tentatively support the assumed values for the 
dust temperatures when combined with further sub-mm and mm data
\citep{wan08b}.

For a more comprehensive picture of the dust emission at high
redshifts we have obtained PACS (100+160\,$\mu$m) and SPIRE
(250+350+500\,$\mu$m) photometry of 69 quasars at $z>5$ 
as part of our {\it Herschel}\footnote{{\it Herschel} is an ESA space observatory with science instruments provided by European-led Principal Investigator consortia and with important participation from NASA.} Space Observatory
\citep{pil10}  key project "The Dusty Young Universe".  
{\it Spitzer} Space Telescope \citep{wer04} observations complement
these data. This enables the study of the full optical  through infrared
SED of these objects in the rest frame wavelength range
$0.5-80\,\mu$m, which -- most importantly -- covers the FIR peak
of the SED. While the photometry for the complete key project sample 
will be presented in a forthcoming paper, we here report on the 
{\it Herschel} observations and SED analysis of the eleven objects in the sample 
which were previously detected at 1.2\,mm (Tab.\,\ref{sample}). 
Currently, this sub-sample is best suited to explore the relative 
importance of FIR and sub-mm/mm measurements for the interpretation of the 
total infrared SED and the contribution from the coolest dust components. Moreover, 
the increased wavelength coverage allows us to develop our fitting procedure with 
optimal constraints.


We outline the data reduction in Section 2. In Section 3 we 
describe how we extract physical properties from our measurements,
which are then discussed in Section 4. We summarize and conclude in
Section 5. Throughout the paper we use a $\Lambda$CDM cosmology with 
$H_0=71$~km~s$^{-1}$~Mpc$^{-1}$, $\Omega_{{\rm m}} = 0.27$, and
$\Omega_{\Lambda} = 0.73$.

\begin{table*}[t]
\begin{center}
\caption{The sample.\label{sample}}
\begin{tabular}{lc r@{.}l c r@{ , }l }
\tableline\tableline
name & redshift & \multicolumn{2}{c}{$m_{1450}$} & $f_{250\,{\rm GHz}}$ & \multicolumn{2}{c}{References} \\
SDSS &          & \multicolumn{2}{c}{mag}       & mJy                & \multicolumn{2}{c}{}\\
(1)  & (2)      & \multicolumn{2}{c}{(3)}       & (4)                & \multicolumn{2}{c}{(5)} \\ 
\tableline
J020332.35+001228.6 & 5.72 & 20&94                   & $1.85 \pm 0.46$ & \hspace*{0.4cm}1  & 2    \\
J033829.30+002156.2 & 5.03 & 20&01                   & $3.7 \pm 0.3$   &  4 &  5    \\
J075618.13+410408.5 & 5.11 & 20&15                   & $5.5 \pm 0.5$   &  6 &  7    \\
J081827.40+172251.8 & 6.00 & 19&34                   & $1.19 \pm 0.38$ &  8 &  3    \\
J084035.09+562419.9 & 5.84 & 20&04                   & $3.20 \pm 0.64$ &  8 &  9     \\
J092721.82+200123.7 & 5.77 & 19&87                   & $4.98 \pm 0.75$ &  8 &  3    \\
J104433.04$-$012502.2 & 5.78 & 19&21                 & $1.82 \pm 0.43$ & 10 &  3    \\
J104845.05+463718.3 & 6.23 & 19&25                   & $3.0 \pm 0.4$   & 11 & 12  \\
J114816.64+525150.2 & 6.42 & 19&03                   & $5.0 \pm 0.6$   & 11 & 12  \\
J133550.80+353315.8 & 5.90 & 19&89                   & $2.34 \pm 0.50$ &  8 &  9    \\
J205406.42$-$000514.8 & 6.04 & 20&60                 & $2.38 \pm 0.53$ &  1 &  3    \\
\tableline
\end{tabular}
\tablecomments{(1) SDSS name ordered by R.A.; (2) redshift confirmed by CO measurements or NIR 
spectroscopy where available (see Appendix); (3) Apparent AB magnitude at 1450\,\AA~in the rest frame of the
quasar, corrected for galactic extinction; 
(4) Observed 250\,GHz flux in mJy. Errors are 1$\sigma$;
(5) References for columns (3) and (4), respectively. 
}
\end{center}
\tablerefs{  
  (1) \citealt{jia08};
  (2) \citealt{wan11};
  (3) \citealt{wan08a};
  (4) \citealt{fan99}; 
  (5) \citealt{car01};
  (6) \citealt{wan08b};
  (7) \citealt{pet03};
  (8) \citealt{fan06};
  (9) \citealt{wan07};
  (10) \citealt{fan01};
  (11) \citealt{fan03};
  (12) \citealt{ber03a}}
\end{table*}

\section{Observations and data reduction}

\subsection{PACS}

All objects were observed with the Photodetector Array Camera and
Spectrometer  (PACS; \citealt{pog10}) at 100 and 160\,$\mu$m using the
mini-scan map observing template. For each source we obtained two
maps with different scan angles using observing parameters as
recommended in the mini-scan map Astronomical Observation Template (AOT) 
release note\footnote{\url{http://herschel.esac.esa.int/twiki/bin/view/Public/PacsAotReleaseNotes}}. For each scan direction, five repetitions were executed. 
This resulted in a total on-source integration time of $\sim$\,900\,s for 
each object.

Data reduction was performed within the Herschel Interactive
Processing  Environment (HIPE, \citealt{ott10}), version 8.0.1. We
followed standard procedures  for deep field data reduction, including
source masking and high-pass filtering. The  maps of the two scan
directions were processed individually and mosaicked at the end of the
work flow. A first version of the combined map was used to create a
source mask for high-pass
filtering. Masking was performed by hand through visual inspection of
the mosaicked maps. This proved to be more reliable than a strict
sigma cut, as it also allowed the masking of faint structures which
can potentially influence the fluxes of faint targets if
located close to the object of interest. A second processing was then
performed, including the mask. The resulting maps (individual and
mosaic) were inspected visually and if necessary the mask or width of
the high-pass filter was adjusted. Images around the target positions 
at 100 and 160\,$\mu$m are presented in Fig.\,\ref{images}.

Aperture photometry of the final mosaics was performed in IDL. We used
relatively small apertures of 6\arcsec - 10\arcsec~radius to maximize
the signal-to-noise ratio. Appropriate aperture corrections were determined
from tabulated values of the Encircled Energy Fraction of unresolved sources. 
Details on the properties of the PACS Point Spread Function (PSF) can be found 
on the PACS calibration web pages\footnote{\url{http://herschel.esac.esa.int/twiki/bin/view/Public/PacsCalibrationWeb}}.

The uncertainty in the photometry of our maps cannot be
  determined directly from the pixel-to-pixel variations because the
 final PACS maps suffer from correlated noise 
(where the level of pixel noise correlation depends on the details of the data 
reduction and final map projection). In order to estimate robust
photometric uncertainties we therefore implemented the following
procedure: For any given map a set of 500 
apertures of the same size as used
for the QSO photometry was placed on random positions on the sky 
\citep[see also][]{lut11,pop12}. The
only constraint on the placement of these background apertures was that 
the central pixel of the aperture has to have at least 75\,\% 
of the integration time as the position of the QSO (to exclude noisy 
areas at the edge of the map). 
The distribution of the measured fluxes in these 500
apertures was then fitted by a Gaussian. The sigma value of this
Gaussian was taken as the 1$\sigma$ photometric  uncertainty of this
map. The resulting flux values 
for the quasars are given in Table\,\ref{photometry}. 


\begin{table*}[t]
\begin{center}
\caption{Infrared photometry.\label{photometry}}
\begin{tabular}{c r@{$\pm$}l r@{$\pm$}l r@{$\pm$}l r@{$\pm$}l r@{$\pm$}l r@{$\pm$}l r@{$\pm$}l r@{$\pm$}l r@{$\pm$}l r@{$\pm$}l
    r@{$\pm$}l }
\tableline\tableline
name &
\multicolumn{2}{c}{$F_{3.6\,\mu{\rm m}}$} &
\multicolumn{2}{c}{$F_{4.5\,\mu{\rm m}}$} &
\multicolumn{2}{c}{$F_{5.8\,\mu{\rm m}}$} &
\multicolumn{2}{c}{$F_{8.0\,\mu{\rm m}}$} &
\multicolumn{2}{c}{$F_{12\,\mu{\rm m}}$} &
\multicolumn{2}{c}{$F_{24\,\mu{\rm m}}$} &
\multicolumn{2}{c}{$F_{100\,\mu{\rm m}}$} &
\multicolumn{2}{c}{$F_{160\,\mu{\rm m}}$} &
\multicolumn{2}{c}{$F_{250\,\mu{\rm m}}$} &
\multicolumn{2}{c}{$F_{350\,\mu{\rm m}}$} &
\multicolumn{2}{c}{$F_{500\,\mu{\rm m}}$} \\
 &
\multicolumn{2}{c}{$\mu$Jy} & 
\multicolumn{2}{c}{$\mu$Jy} & 
\multicolumn{2}{c}{$\mu$Jy} & 
\multicolumn{2}{c}{$\mu$Jy} & 
\multicolumn{2}{c}{$\mu$Jy} & 
\multicolumn{2}{c}{$\mu$Jy} & 
\multicolumn{2}{c}{mJy} & 
\multicolumn{2}{c}{mJy} &
\multicolumn{2}{c}{mJy} & 
\multicolumn{2}{c}{mJy} &
\multicolumn{2}{c}{mJy} \\ 
(1)  & 
\multicolumn{2}{c}{(2)} & 
\multicolumn{2}{c}{(3)} & 
\multicolumn{2}{c}{(4)} & 
\multicolumn{2}{c}{(5)} & 
\multicolumn{2}{c}{(6)} & 
\multicolumn{2}{c}{(7)} & 
\multicolumn{2}{c}{(8)} &
\multicolumn{2}{c}{(9)} & 
\multicolumn{2}{c}{(10)} &
\multicolumn{2}{c}{(11)} &
\multicolumn{2}{c}{(12)} \\ 
\tableline
\multicolumn{1}{l}{J0203+0012}                    &  80&1 &  88&1 & 106&6 & 106&7  & 353&111                     &    680&44 & \multicolumn{2}{c}{$<3.3$} & \multicolumn{2}{c}{$<5.4$} & \multicolumn{2}{c}{$<15.6$} & \multicolumn{2}{c}{$<13.5$} & \multicolumn{2}{c}{$<18.0$} \\ 
\multicolumn{1}{l}{J0338+0021}                    &  81&2 &  71&2 &  82&7 & 158&9  & \multicolumn{2}{c}{$<355$}  &   1187&52 & 10.7&1.0                   & 18.5&2.0                   & 19.6&5.9 & 18.5&6.2 & 12.6&6.5 \\
\multicolumn{1}{l}{J0756+4104}                    &  61&2 &  62&2 &  70&6 & 123&7  & \multicolumn{2}{c}{$<732$}  &    698&36 & 6.2&0.8                    & 9.0&1.0                    & 11.4&5.3 & 19.0&4.8 & 19.9&5.0 \\
\multicolumn{1}{l}{J0818+1722\tablenotemark{a}}   & 168&2 & 200&2 & 167&8 & 216&10 & 425&127                     &   1004&30 & \multicolumn{2}{c}{$<3.0$} & \multicolumn{2}{c}{$<5.1$} & \multicolumn{2}{c}{$<14.7$} & \multicolumn{2}{c}{$<13.8$} & \multicolumn{2}{c}{$<15.3$} \\
\multicolumn{1}{l}{J0840+5624\tablenotemark{b}}   &  58&1 &  80&1 &  61&7 &  62&6  & \multicolumn{2}{c}{\nodata} &    440&29 & \multicolumn{2}{c}{$<2.7$} & \multicolumn{2}{c}{$<4.2$} & \multicolumn{2}{c}{$<15.3$} & \multicolumn{2}{c}{$<13.5$} & \multicolumn{2}{c}{$<15.3$} \\
\multicolumn{1}{l}{J0927+2001}                    &  47&2 &  50&2 &  42&7 &  76&7  & \multicolumn{2}{c}{$<757$}  &    639&40 & \multicolumn{2}{c}{$<3.0$} & \multicolumn{2}{c}{$<3.9$} & 13.1&5.3 & 15.3&5.0 & 19.5&5.8 \\
\multicolumn{1}{l}{J1044$-$0125}                  & 106&2 & 131&2 & 108&8 & 186&9  & \multicolumn{2}{c}{$<398$}  &   1436&39 & 6.7&0.8                    & 8.5&1.0                    & \multicolumn{2}{c}{$<15.3$} & \multicolumn{2}{c}{$<12.6$} & \multicolumn{2}{c}{$<16.5$} \\
\multicolumn{1}{l}{J1048+4637}                    & 110&2 & 120&2 &  95&7 & 128&7  & \multicolumn{2}{c}{$<315$}  &    818&41 & \multicolumn{2}{c}{$<2.1$} & \multicolumn{2}{c}{$<3.0$} & \multicolumn{2}{c}{$<14.4$} & \multicolumn{2}{c}{$<14.1$} & \multicolumn{2}{c}{$<18.6$} \\
\multicolumn{1}{l}{J1148+5251\tablenotemark{c}}   & 137&3 & 146&2 & 143&8 & 214&8  & 304&100                     &   1349&39 & 3.9&0.6   & 7.4&1.7                    & 21.0&5.3 & 21.8&4.9 & 12.4&5.7 \\
\multicolumn{1}{l}{J1335+3533}                    &  66&1 &  69&1 &  55&4 &  57&6  & \multicolumn{2}{c}{$<311$}  &    483&32 & \multicolumn{2}{c}{$<2.7$} & \multicolumn{2}{c}{$<3.0$} & \multicolumn{2}{c}{$<13.5$} & \multicolumn{2}{c}{$<14.1$} & \multicolumn{2}{c}{$<18.6$} \\
\multicolumn{1}{l}{J2054$-$0005\tablenotemark{d}} & \multicolumn{2}{c}{$<$\,18} & \multicolumn{2}{c}{$<$\,48}    & \multicolumn{2}{c}{\nodata} &  \multicolumn{2}{c}{\nodata} & \multicolumn{2}{c}{$<$\,162} & \multicolumn{2}{c}{$<$\,1932} & \multicolumn{2}{c}{$<2.7$} & 9.8&1.3   & 15.2&5.4 & 12.0&4.9 & \multicolumn{2}{c}{$<19.5$} \\
\tableline
\end{tabular}
\tablecomments{Upper limits correspond to 3$\sigma$. Photometry in
  columns (2)-(5) and (7) are from {\it Spitzer} observations, except
  for J2054$-$0005. 
Column (6) is based on data from the WISE All-Sky Survey. Columns (8)-(12) refer to {\it Herschel} data. {\bf Notes:} (a) The 
{\it Spitzer} values of this source may include some contamination from a nearby galaxy (see text). The WISE observations 
do not separate the objects and the quoted catalog flux has to be taken with caution; (b) No 12\,$\mu$m photometry could 
be performed due to severe blending with bright nearby source; (c) This is the only source also observed at 70\,$\mu$m where 
we measure a flux of $2.9\pm0.6$\,mJy; (d) This object was not observed with {\it Spitzer}. Data in columns (2), (3), and (7) 
are based on aperture photometry on WISE All-Sky Survey atlas images at 3.4, 4.6, and 22\,$\mu$m, respectively.}
\end{center}
\end{table*}

 
\subsection{SPIRE}

All quasars in the sample were also observed with the Spectral and
Photometric Imaging Receiver (SPIRE, \citealt{gri10}) at  250, 350,
and 500\,$\mu$m in small scan map mode for five repetitions and a total
on-source integration time of $\sim$\,190\,s per source.  Data reduction
followed standard procedures in HIPE as recommended by the SPIRE instrument
team.  Source extraction was performed with the HIPE build-in task
'sourceExtractorSussextractor' \citep{sav07} using information  on the
PSF (e.g. FWHM) given in the SPIRE Observer's
Manual\footnote{\url{http://herschel.esac.esa.int/Docs/SPIRE/html/spire\_om.html}}.

Our observations are dominated by confusion noise which is on the
order of  $6-7$\,mJy\,beam$^{-1}$ in the SPIRE photometric bands \citep{ngu10}
as determined  from deep extragalactic fields. In order to estimate
the uncertainties  due to confusion noise specifically in our target fields, 
we implemented the following procedure 
\citep[see also][]{elb11,pas11}: First, the source
extractor was run over the full calibrated maps. Detections within 
less than half the FWHM from the nominal target position 
were tentatively considered to belong to the quasar, pending 
further confirmation from our check for confusion with nearby 
FIR bright sources. We then created an
artificial image which included all the sources found by the  source
extractor and subtracted this ``source image'' from the observed map.
On this ``residual map'' we determined the pixel-to-pixel rms in a box
with a size of 8 times the FWHM (FWHM size: 18.2\arcsec, 24.9\arcsec,
and  36.3\arcsec~for default map pixel sizes of 6, 10, and 14\arcsec~at 
250, 350, and 500\,$\mu$m, respectively), centered
on the  nominal position of the QSO. The size  of this box was chosen
large enough to allow an appropriate sampling of the surroundings of
the source, but small enough to avoid including the lower coverage
areas at the edges of the map even for the longest wavelengths. In
addition,  the number of pixels per FWHM is approximately constant for the three wavelengths in the
final maps  ($2.5-3.0$\,px/FWHM) which translates into a similar number of
pixels used for determining the rms in the background box. 
The resulting estimates for the noise (limited by confusion) are comparable to the average values given in 
\citet{ngu10}, but have a tendency to be slightly lower. The fluxes and
uncertainties we determine 
are given in Table\,\ref{photometry} and the 250\,$\mu$m maps are presented in 
Fig.\,\ref{images}.

We note that Table\,\ref{photometry} lists a number of SPIRE flux measurements 
which are nominally below the estimated 3$\sigma$ value of the noise. In these 
cases, the inspection of the images revealed a clear excess of flux at the 
position of the quasar. 
It has been shown that the use of positional priors can reduce the effect of 
confusion noise by 20-30\,\% \citep{ros10}. While our strategy is 
somewhat different from that work, we benefit not only from accurate 
(relative and absolute) positional information, but also from information on 
the SEDs of the quasar and potential confusing sources in the field via 
our multi-wavelength data.  
This leads us to 
include these flux measurements in this study, although they have to 
be treated with caution. Similarly, fluxes at 500\,$\mu$m should be 
considered tentative because at this wavelength the beam is large 
($\sim$\,36\,\arcsec~FWHM), the confusion noise is high, and the 
significance of the detections is often low.

\subsection{Spitzer}

For all {\it Herschel} targets (except J2054$-$0005) we also  have
available mid-infrared (MIR) imaging  from {\it Spitzer} at 3.6, 4.5,
5.8, and 8.0\,$\mu$m with the InfraRed Array Camera (IRAC,
\citealt{faz04}) as well as at 24\,$\mu$m with the Multiband Imaging
Photometer for Spitzer  (MIPS, \citealt{rie04}).  For the redshifts of
our sources these passbands cover the rest frame optical and
near-infared (NIR) wavelengths ($\sim$\,$0.5-4$\,$\mu$m). The {\it
Spitzer} data have been processed  using standard procedures within
the {\sc Mopex} software package  provided by the {\it Spitzer}
Science Center (SSC). Aperture photometry  was performed in IDL using
standard sets of aperture radii and  appropriate aperture corrections
as outlined in the respective  instrument handbooks (also available
from the SSC website). Errors were  estimated in a similar fashion as
for PACS by measuring the flux in randomly  placed apertures on empty
parts of the background and determining the variations between these
background flux measurements.  All objects in this paper with {\it
  Spitzer}  data are detected at high significance in the five bands. The resulting
photometry is summarized in Table \ref{photometry} and is usually
consistent  with measurements published previously, where available
\citep[e.g.][]{jia06,hin06,jia10}.  The 24\,$\mu$m images are
presented alongside the PACS and SPIRE 250\,$\mu$m images in Fig.\,\ref{images}.

Our multi-wavelength data set, and in particular the {\it Spitzer} 
24\,$\mu$m images, provide a tool for determining the
exact position of the quasar in the {\it Herschel} bands.  Since we
can often identify several sources per field that are visible both at 
{\it Spitzer} and at {\it Herschel} wavelengths, the exact location of
the quasar in the FIR maps can be determined from the relative positional 
information.  With this procedure 
we can robustly identify faint {\it Herschel} detections with the quasars as well as 
avoid mis-identifications due to nearby objects.
During this exercise we observe absolute spatial offsets between {\it Spitzer}
and  {\it Herschel} of typically $\lesssim$\,2\arcsec, in line with
expectations  from the absolute pointing accuracies.

\begin{figure*}[t!]
\centering
\includegraphics[angle=0,scale=.23]{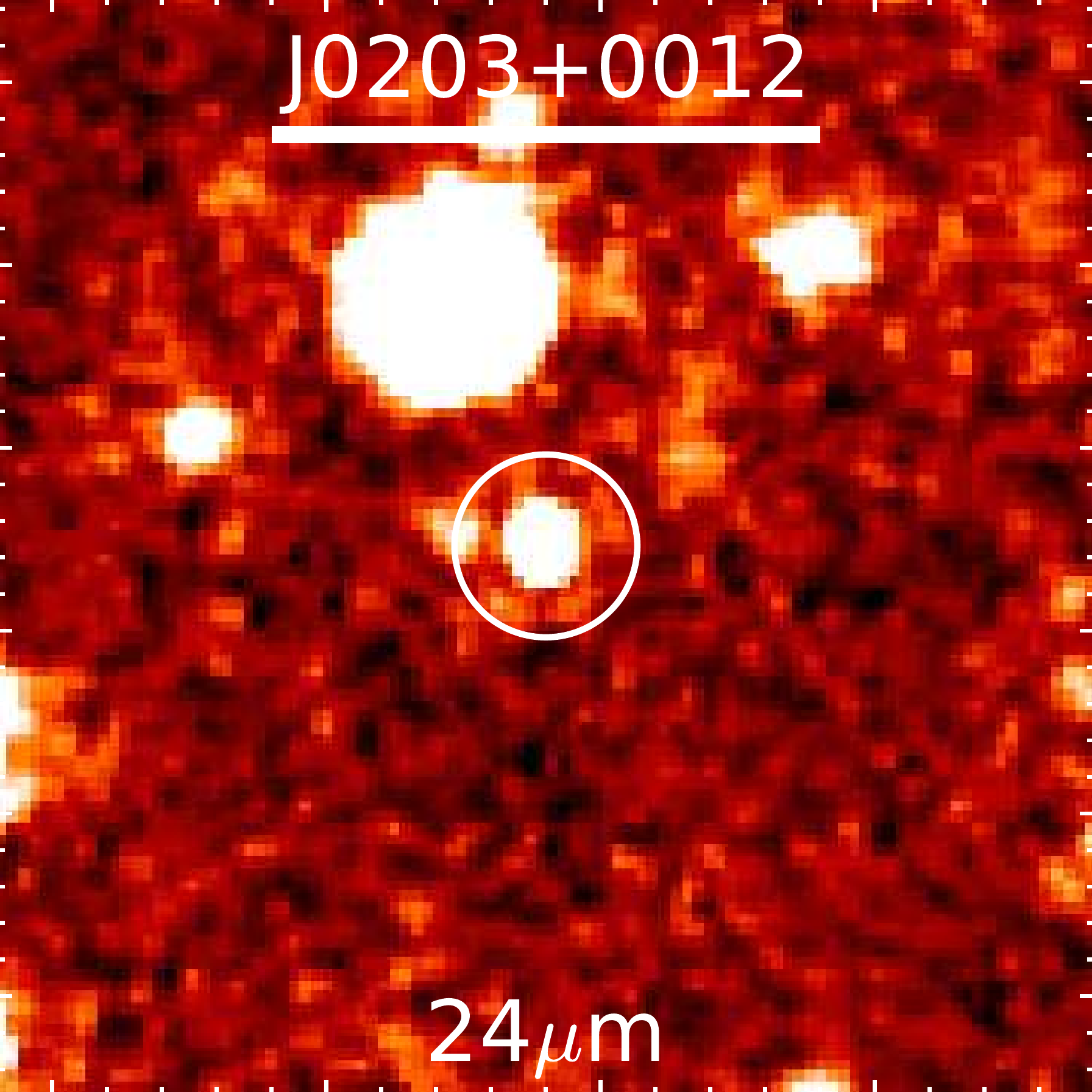}
\includegraphics[angle=0,scale=.23]{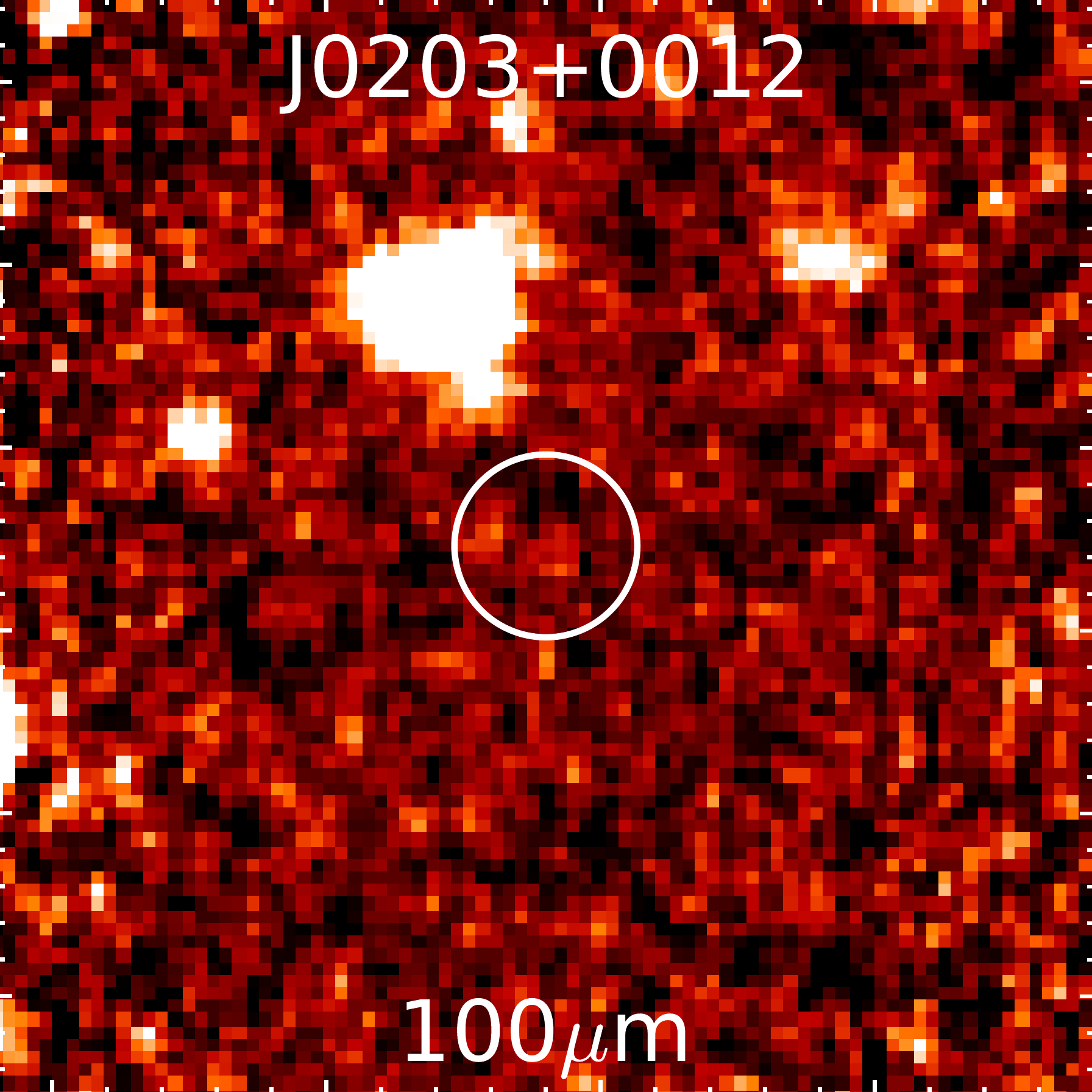}
\includegraphics[angle=0,scale=.23]{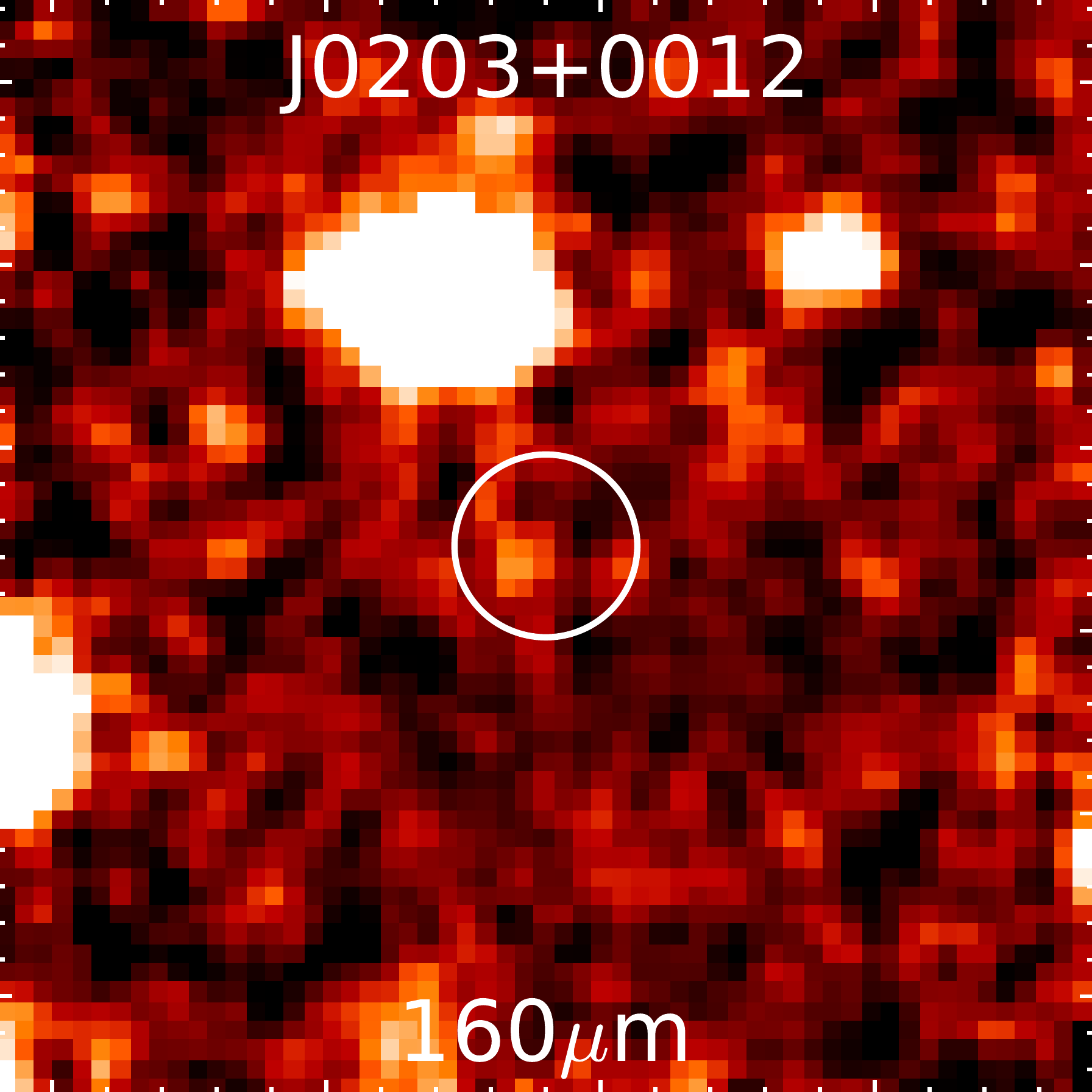}
\includegraphics[angle=0,scale=.23]{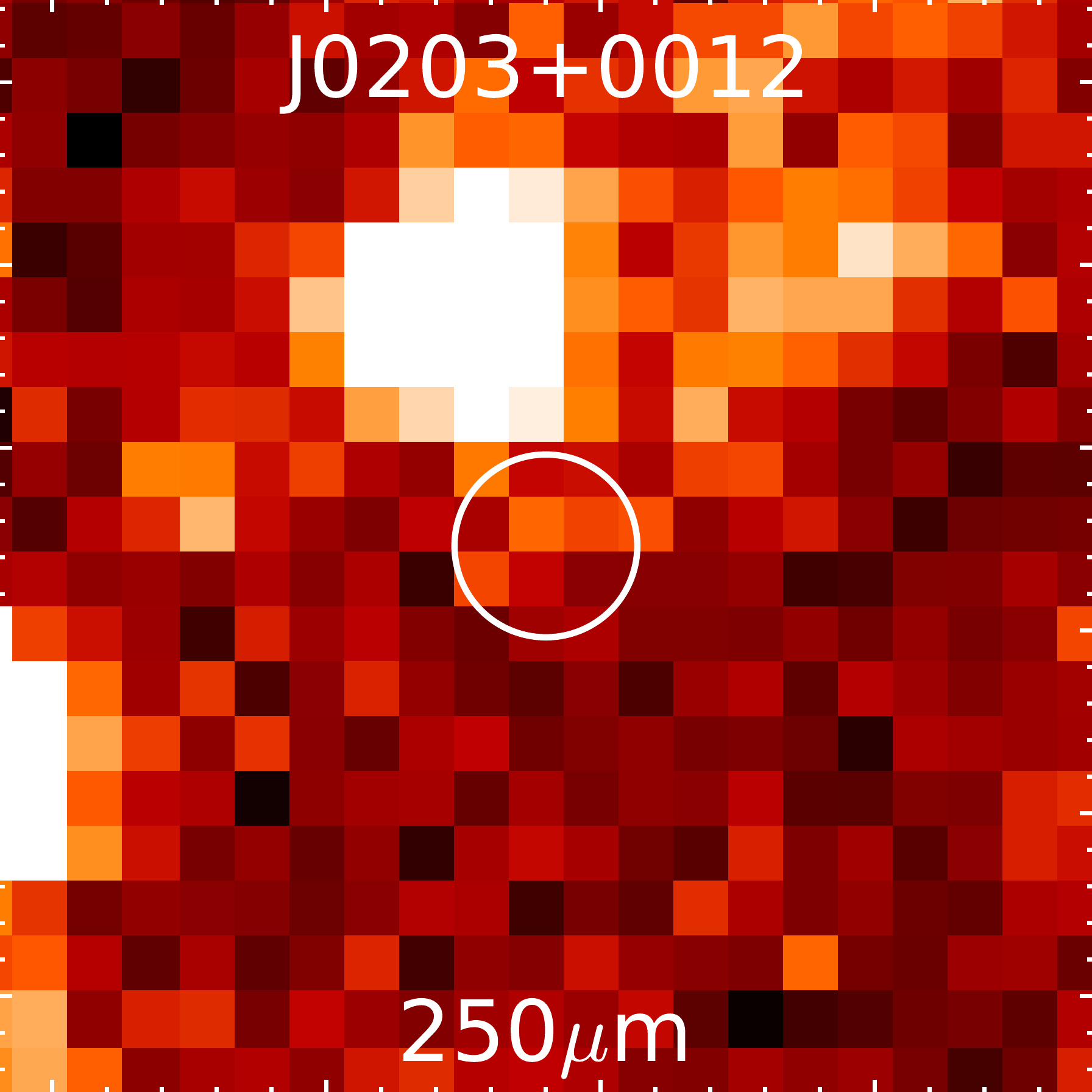}\\
\includegraphics[angle=0,scale=.23]{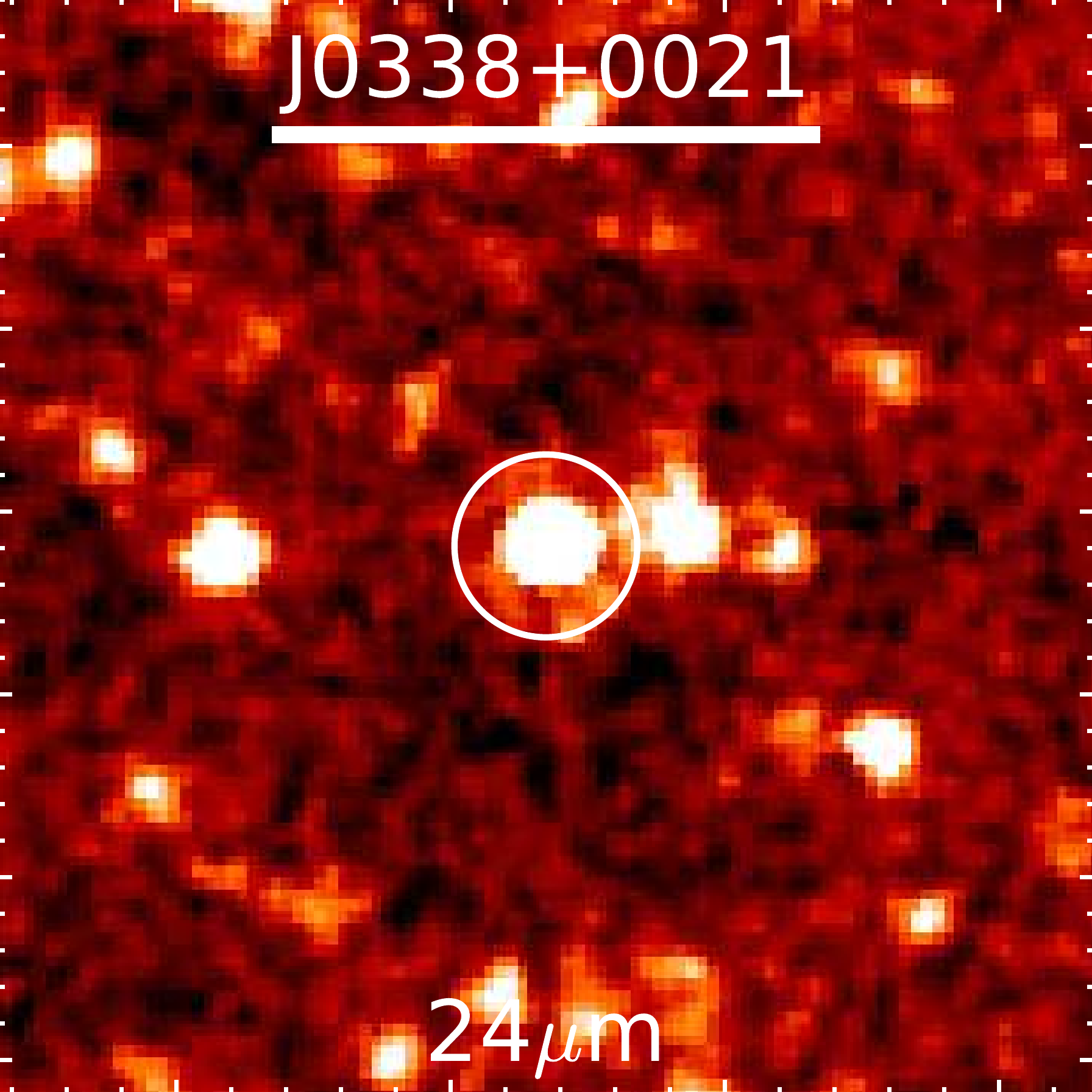}
\includegraphics[angle=0,scale=.23]{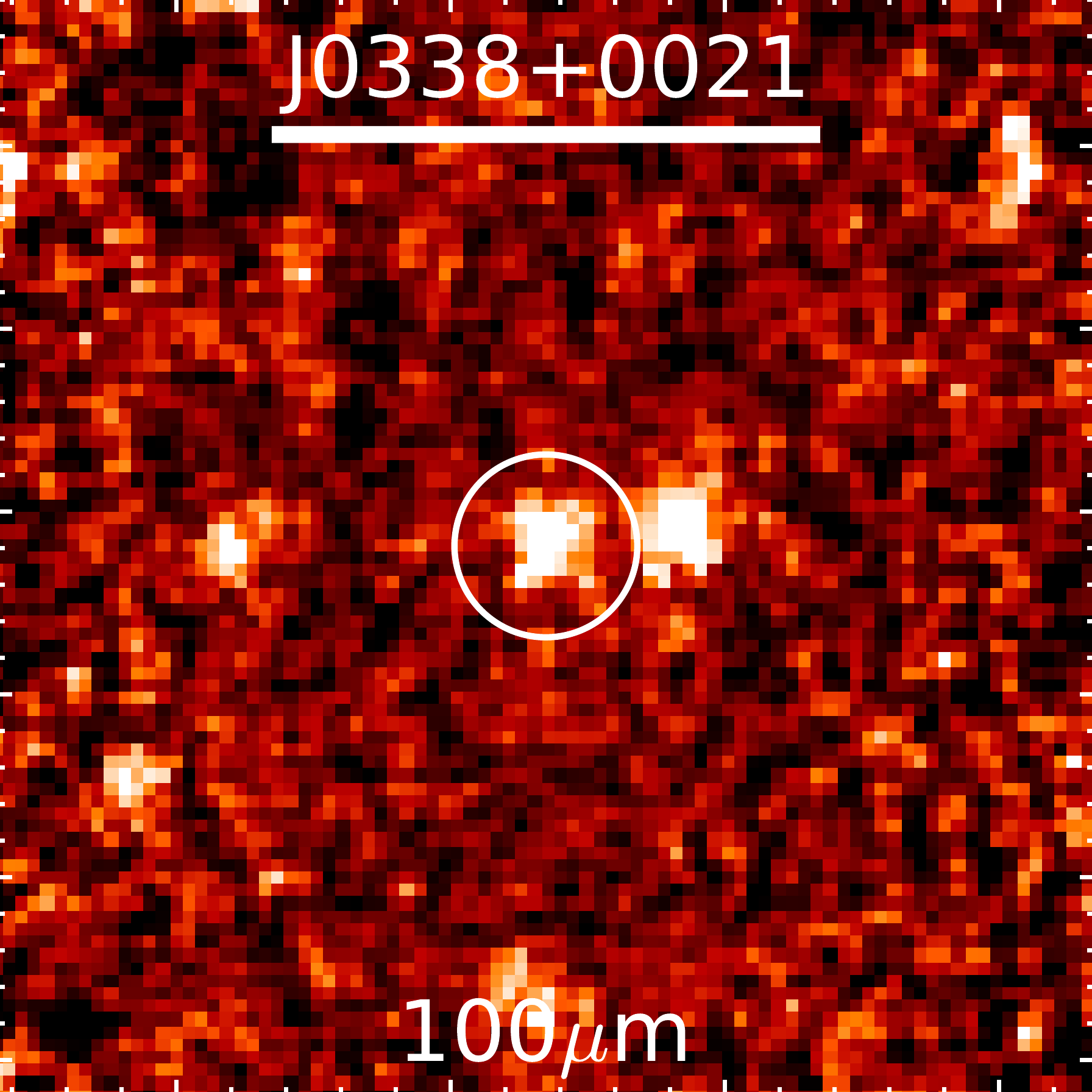}
\includegraphics[angle=0,scale=.23]{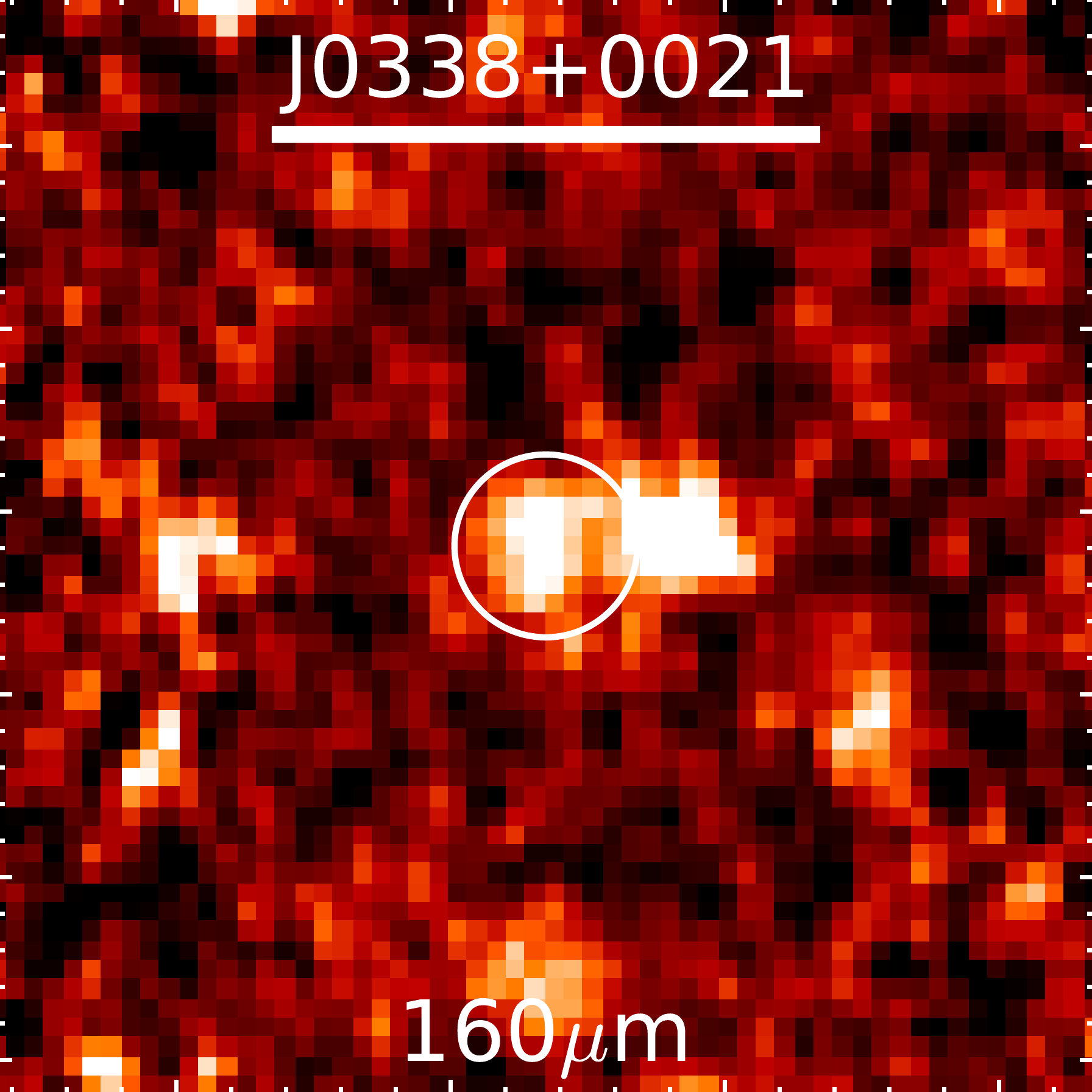}
\includegraphics[angle=0,scale=.23]{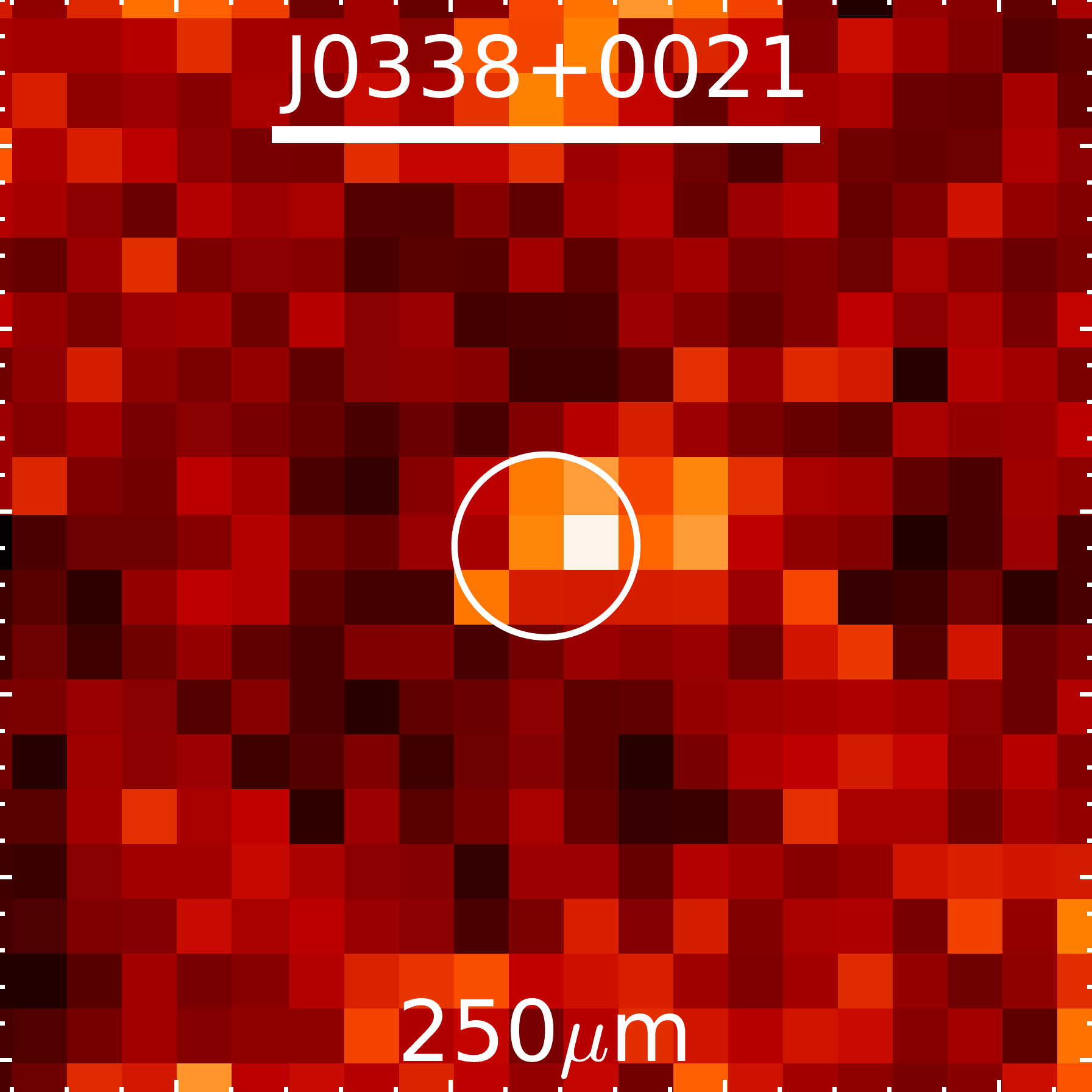}\\
\includegraphics[angle=0,scale=.23]{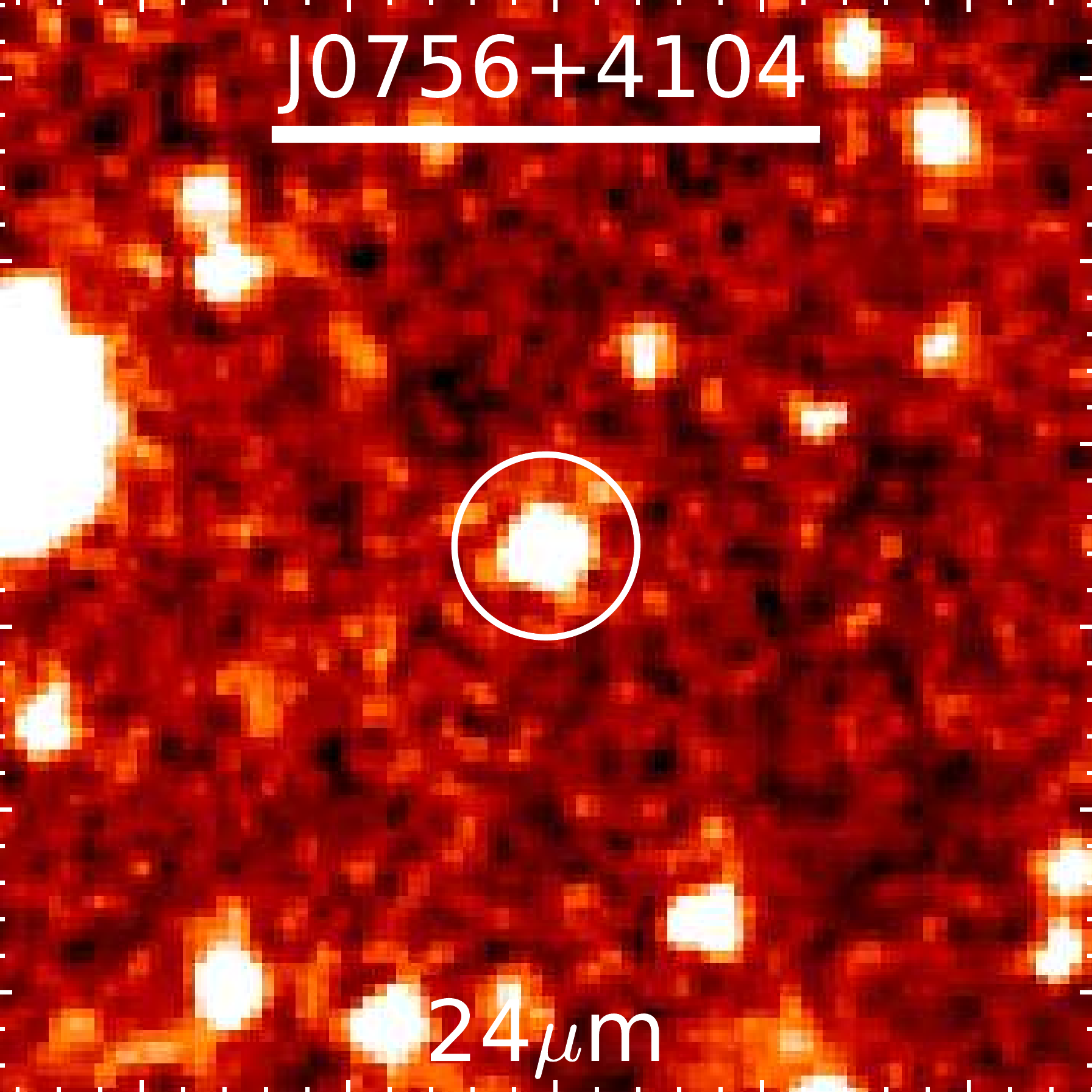}
\includegraphics[angle=0,scale=.23]{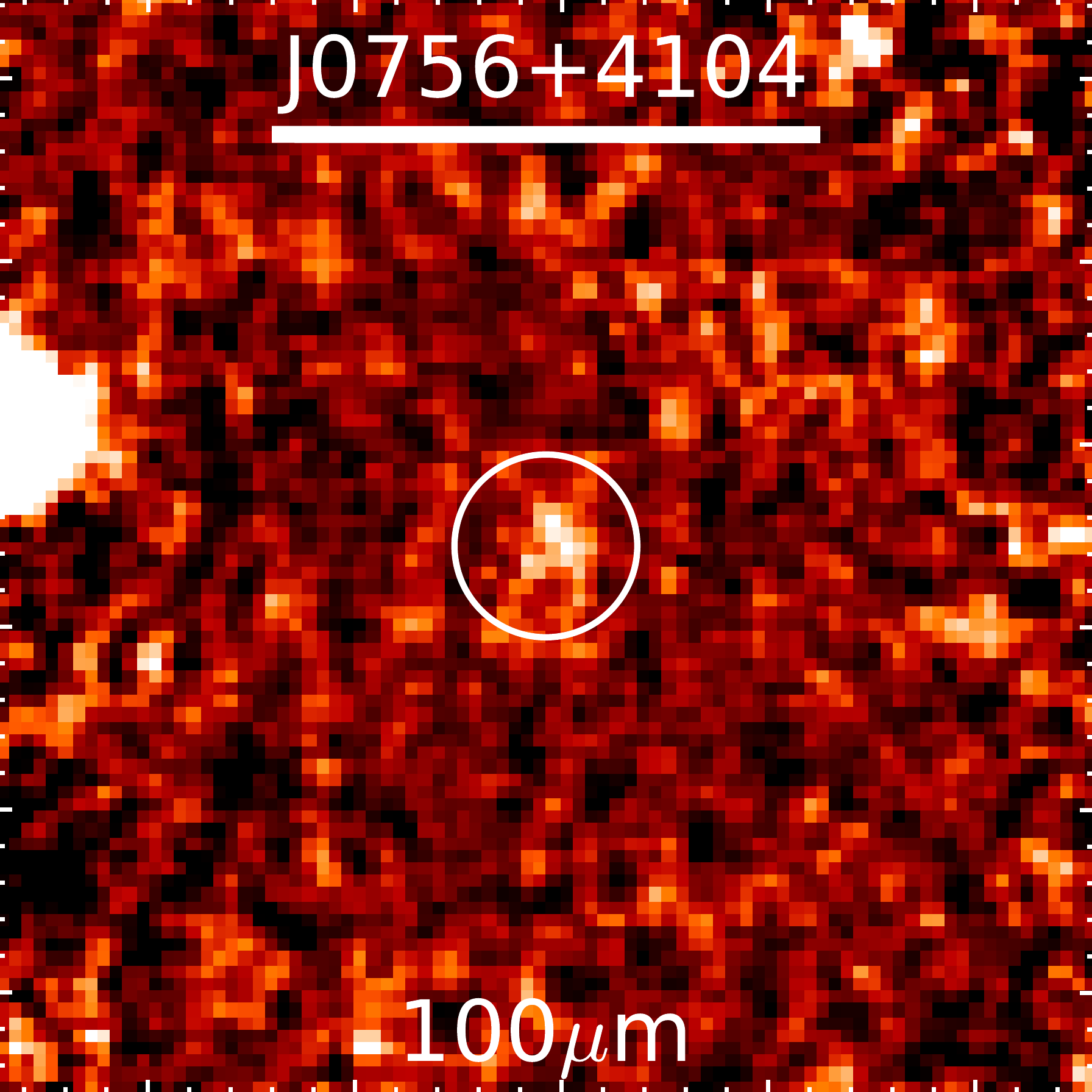}
\includegraphics[angle=0,scale=.23]{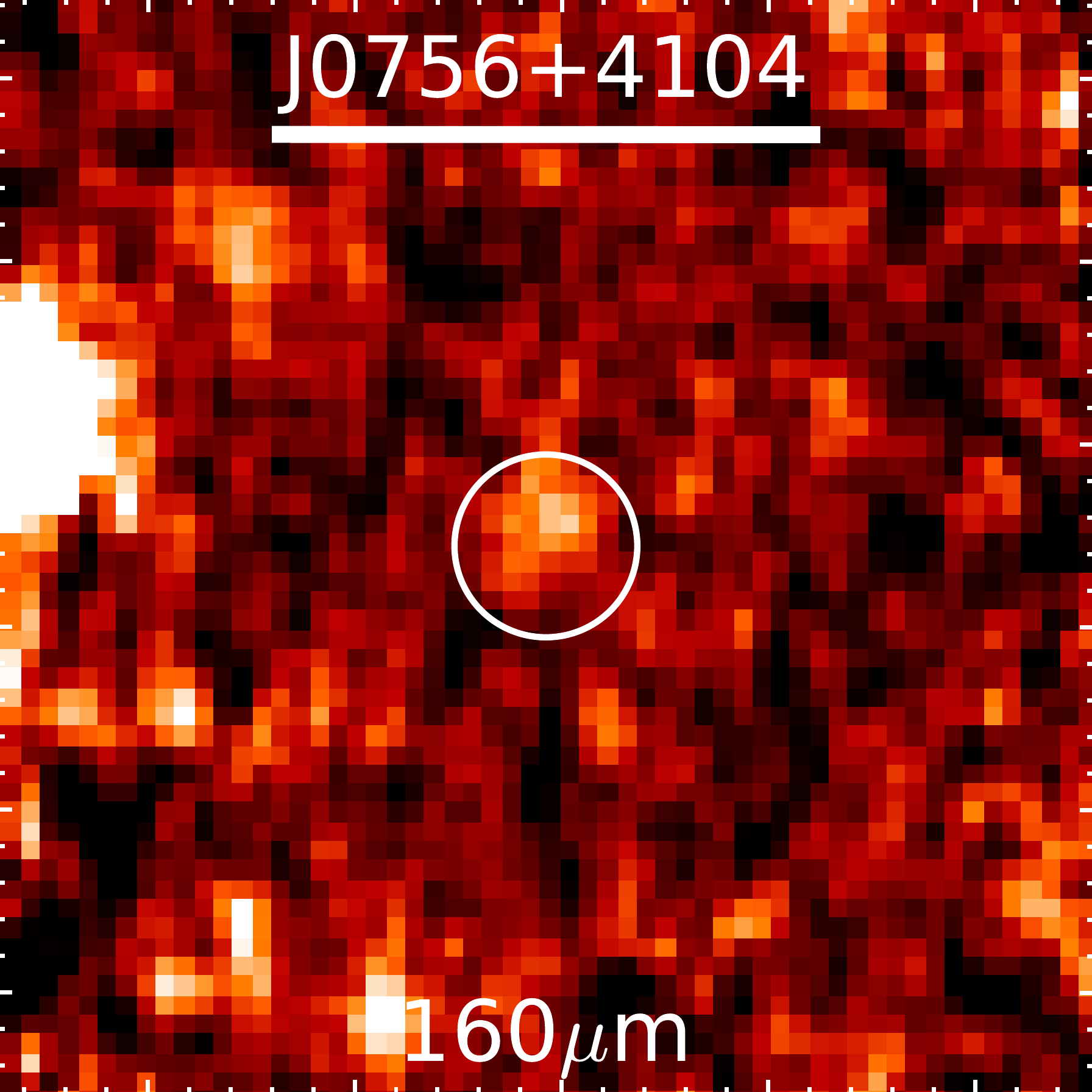}
\includegraphics[angle=0,scale=.23]{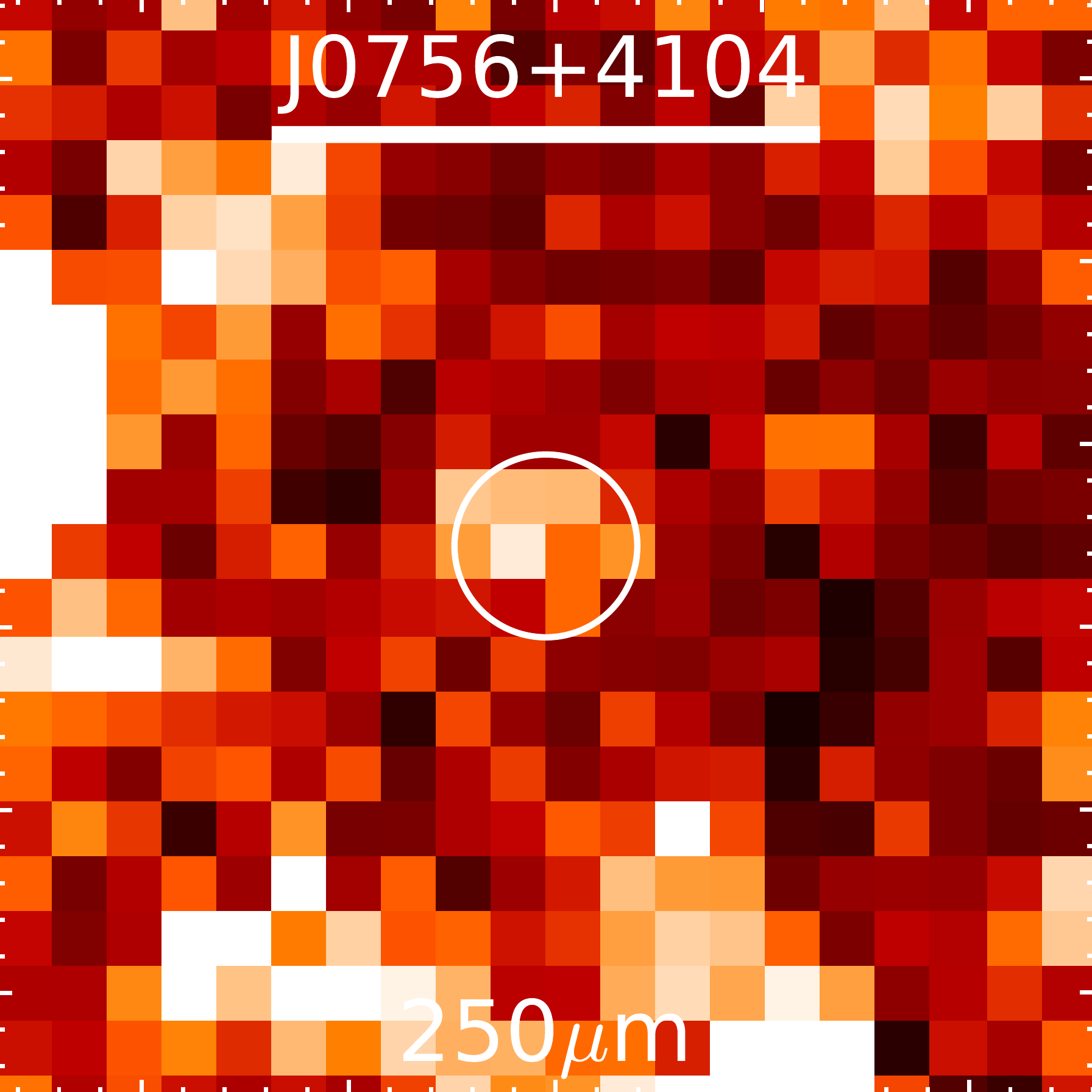}\\
\includegraphics[angle=0,scale=.23]{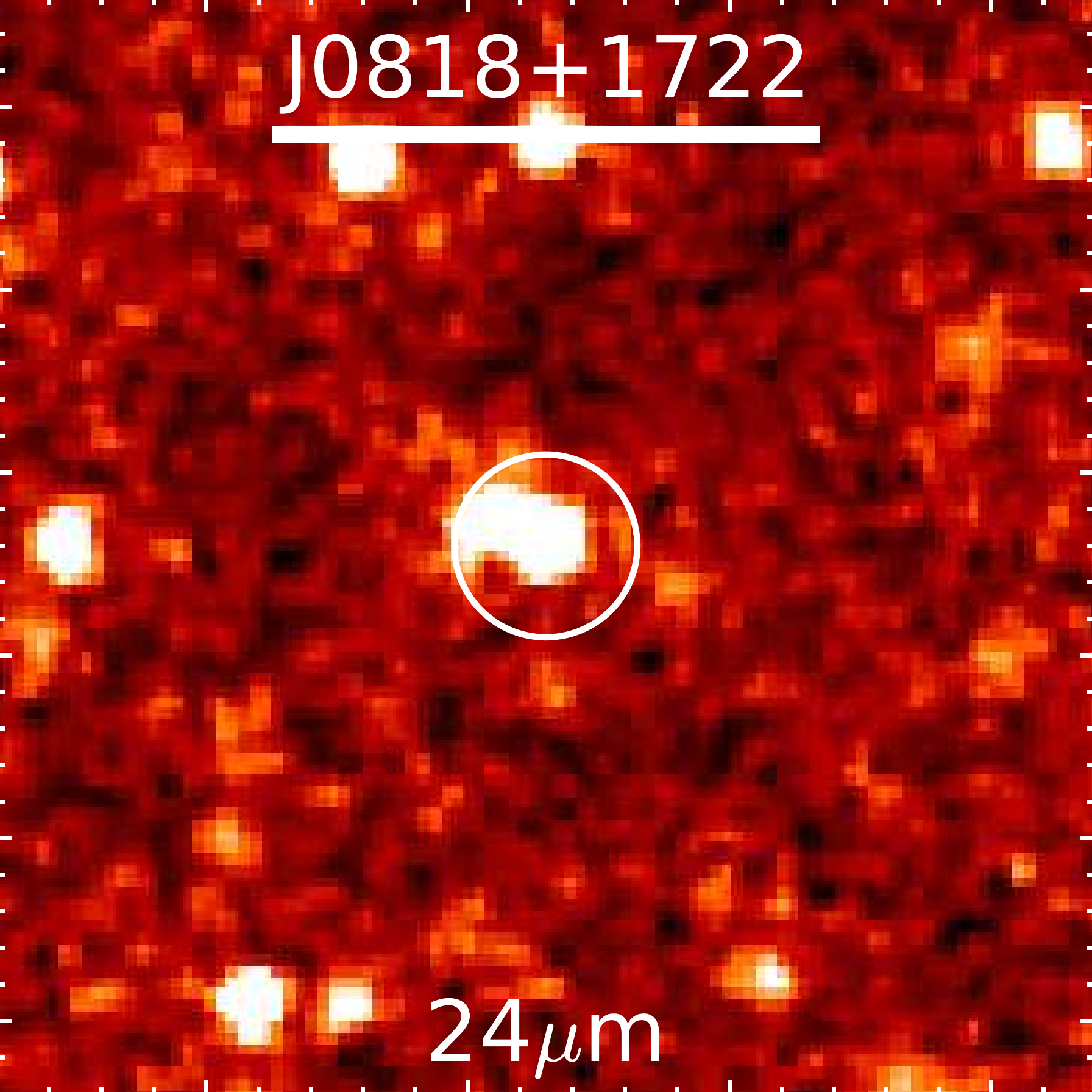}
\includegraphics[angle=0,scale=.23]{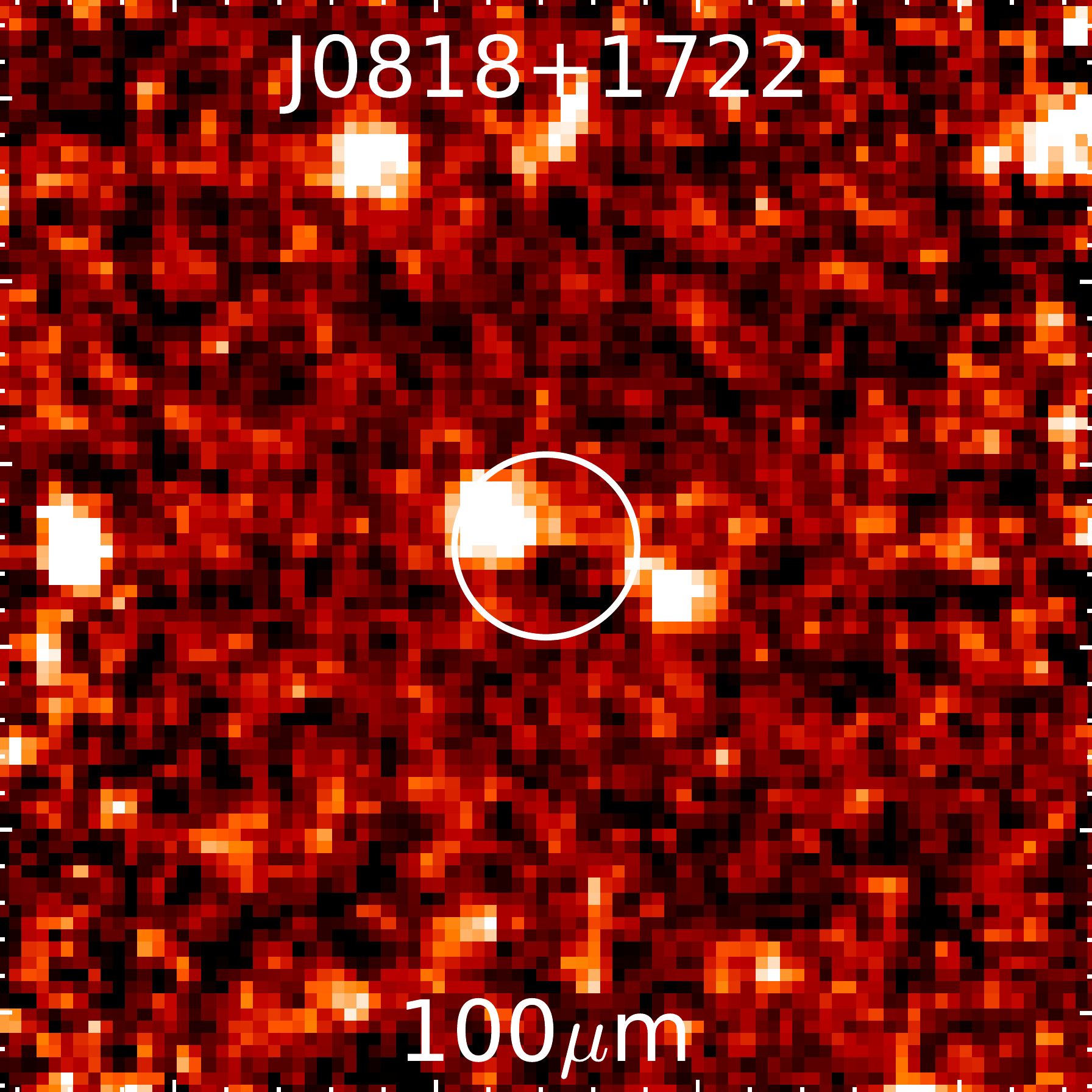}
\includegraphics[angle=0,scale=.23]{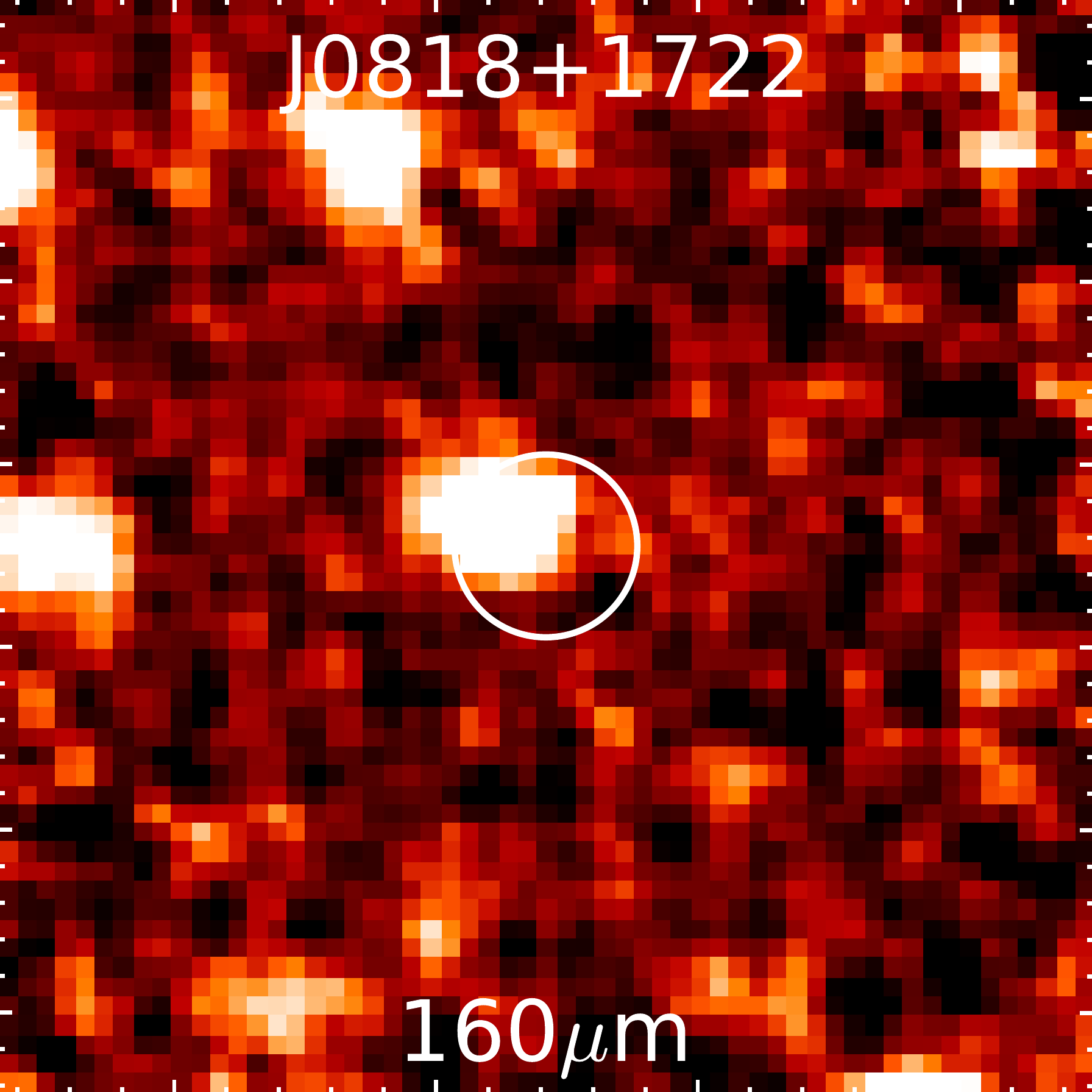}
\includegraphics[angle=0,scale=.23]{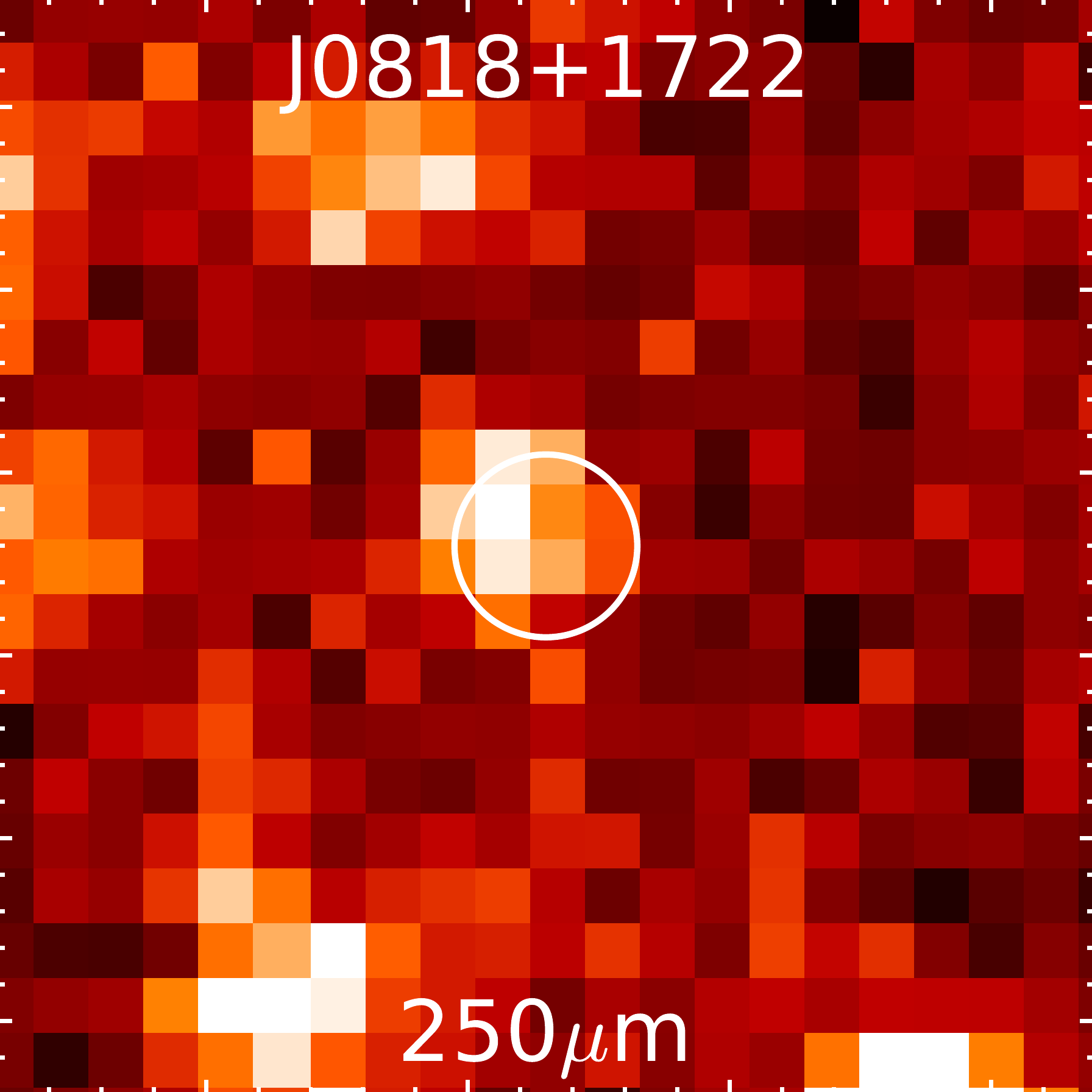}
\caption{The final maps of the quasars at 24, 100, 160 and 250\,$\mu$m (from 
left to right). All images are 2\arcmin~on a side and North is to the top 
with East to the left. The circle indicating the position of the quasar 
has a diameter of 20\arcsec. Sources detected in a particular band have 
their source name underlined in the corresponding image. In the case of 
J0818+1722 the bright source close to the QSO position at 100, 160, and 
250\,$\mu$m is identified with a foreground object. The QSO itself is 
undetected at these wavelengths (see the Appendix for details).
 \label{images} }
\end{figure*}
\addtocounter{figure}{-1}
\begin{figure*}[t!]
\centering
\includegraphics[angle=0,scale=.23]{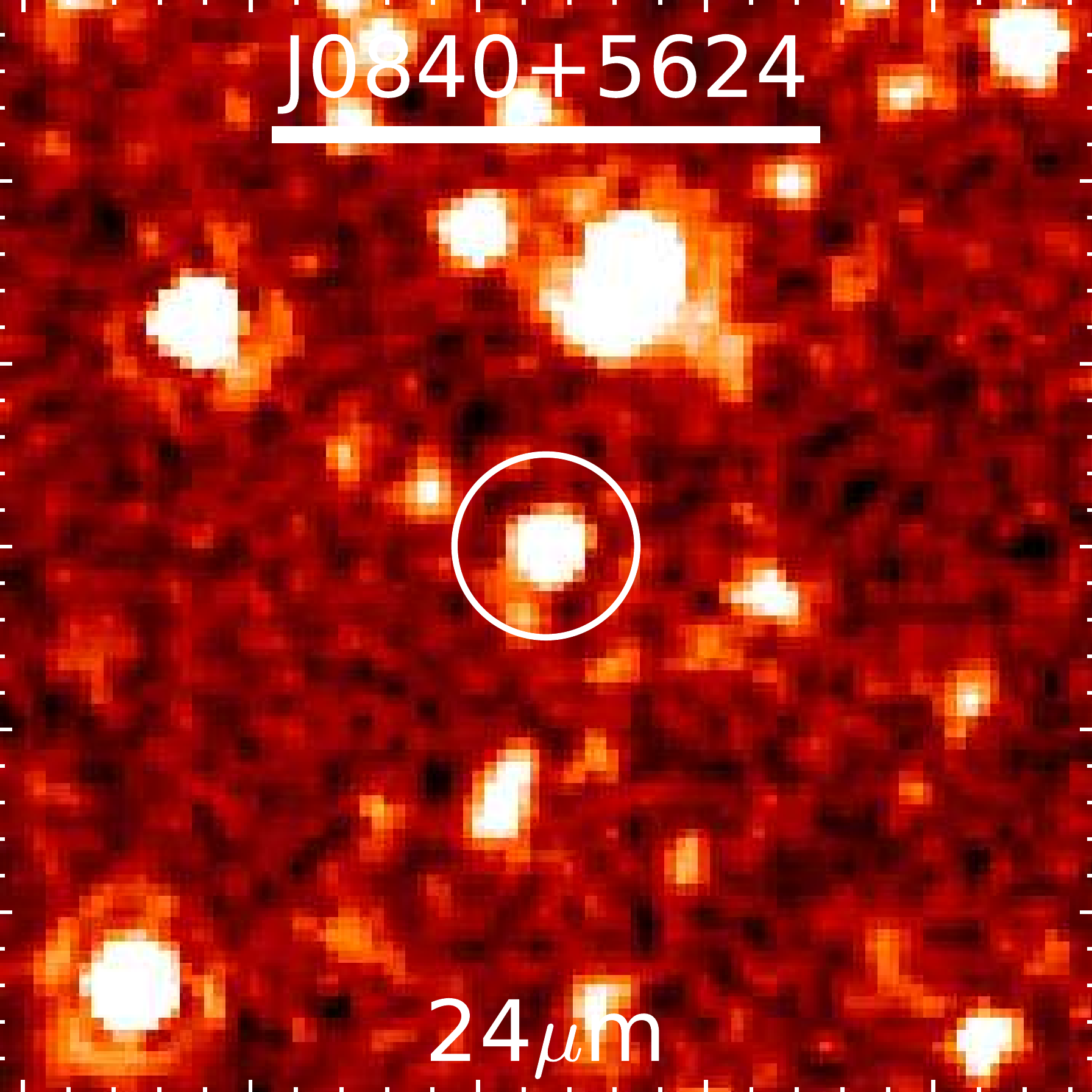}
\includegraphics[angle=0,scale=.23]{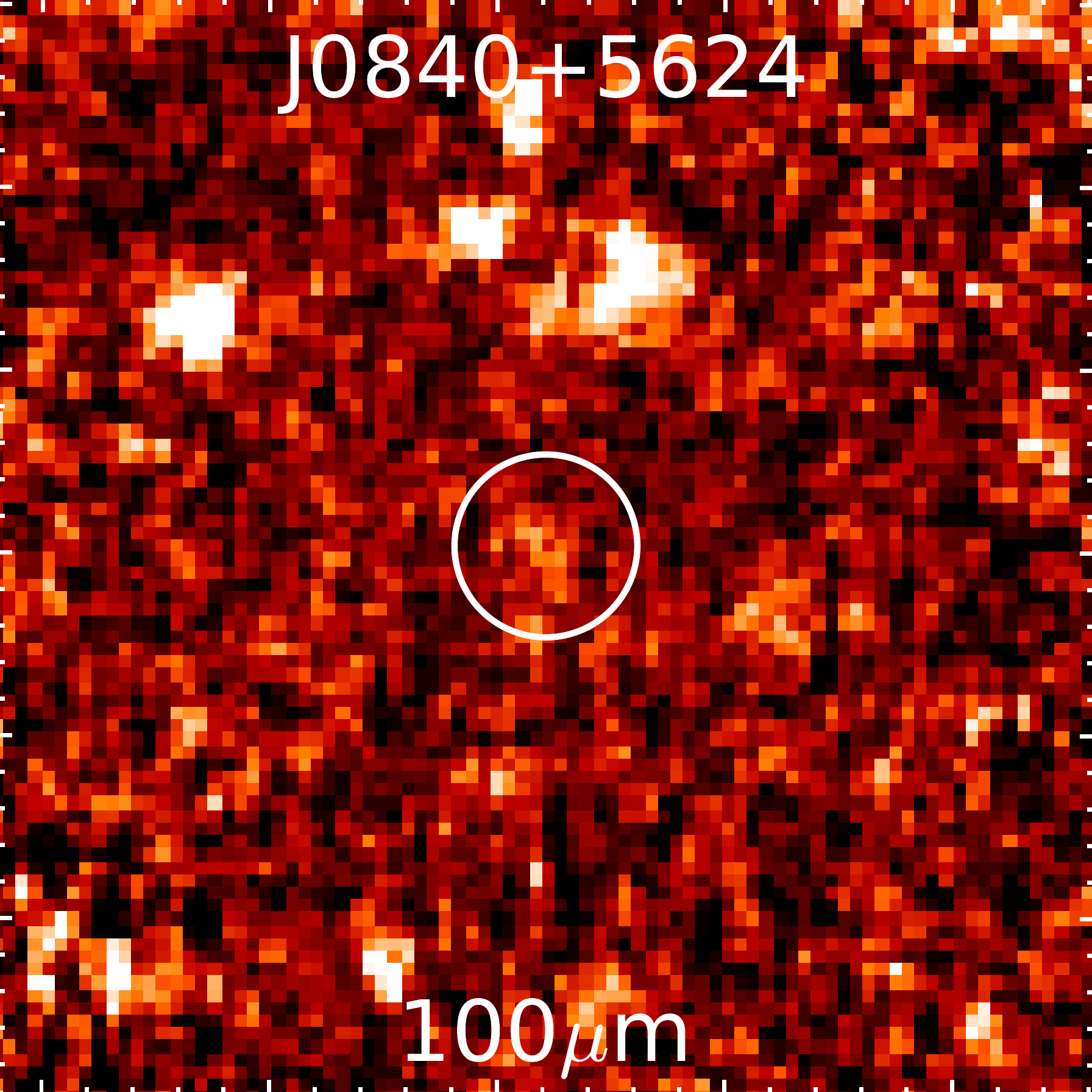}
\includegraphics[angle=0,scale=.23]{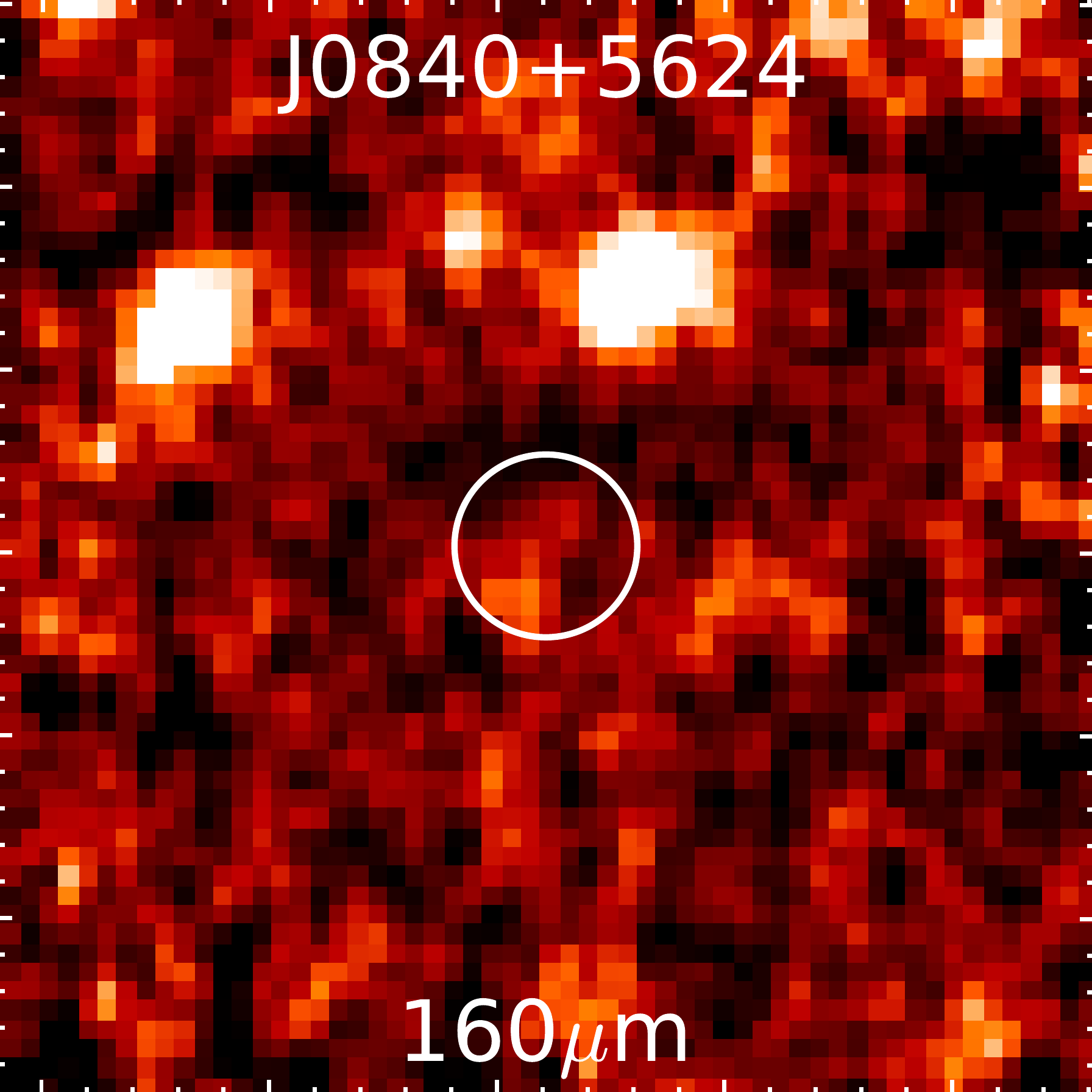}
\includegraphics[angle=0,scale=.23]{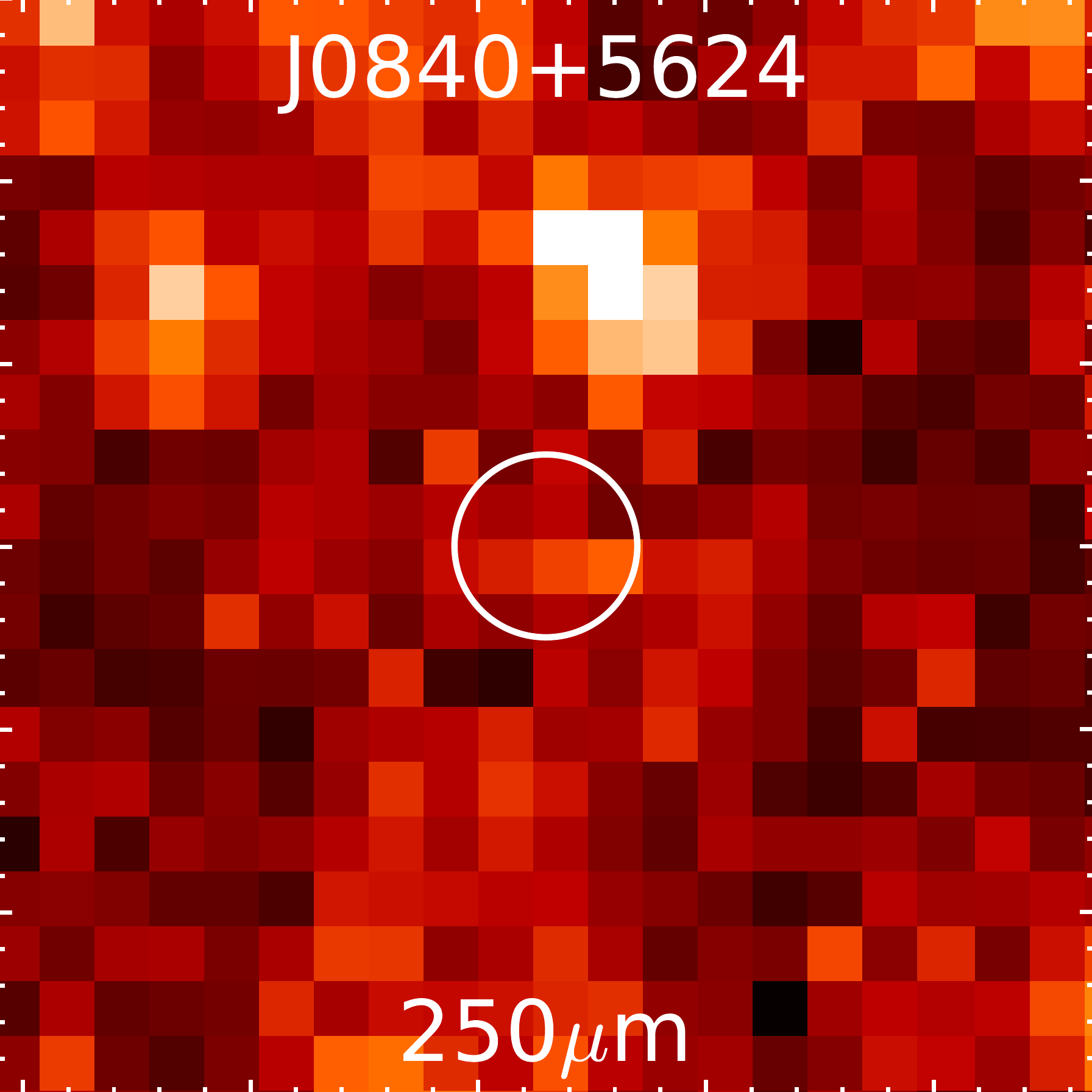}\\
\includegraphics[angle=0,scale=.23]{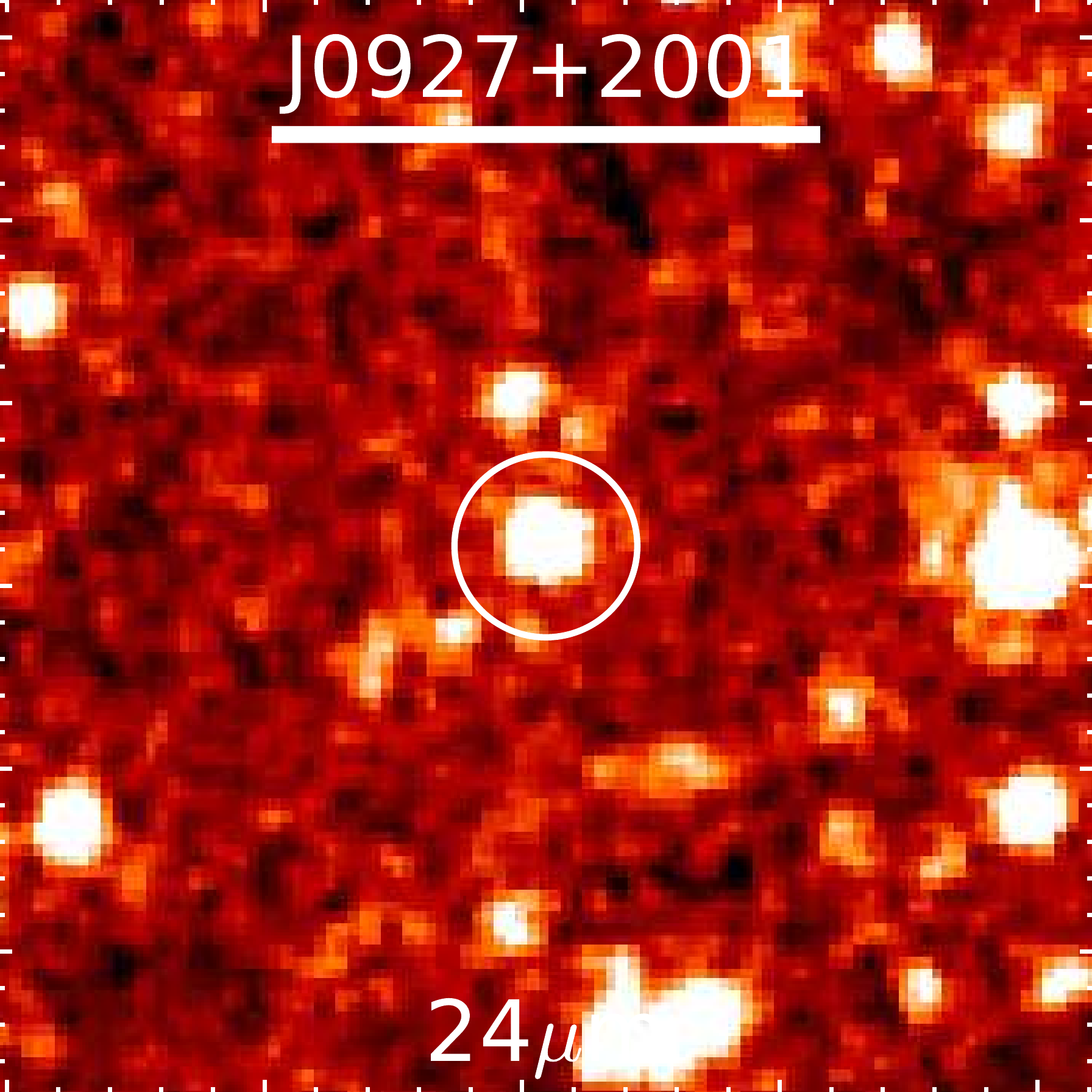}
\includegraphics[angle=0,scale=.23]{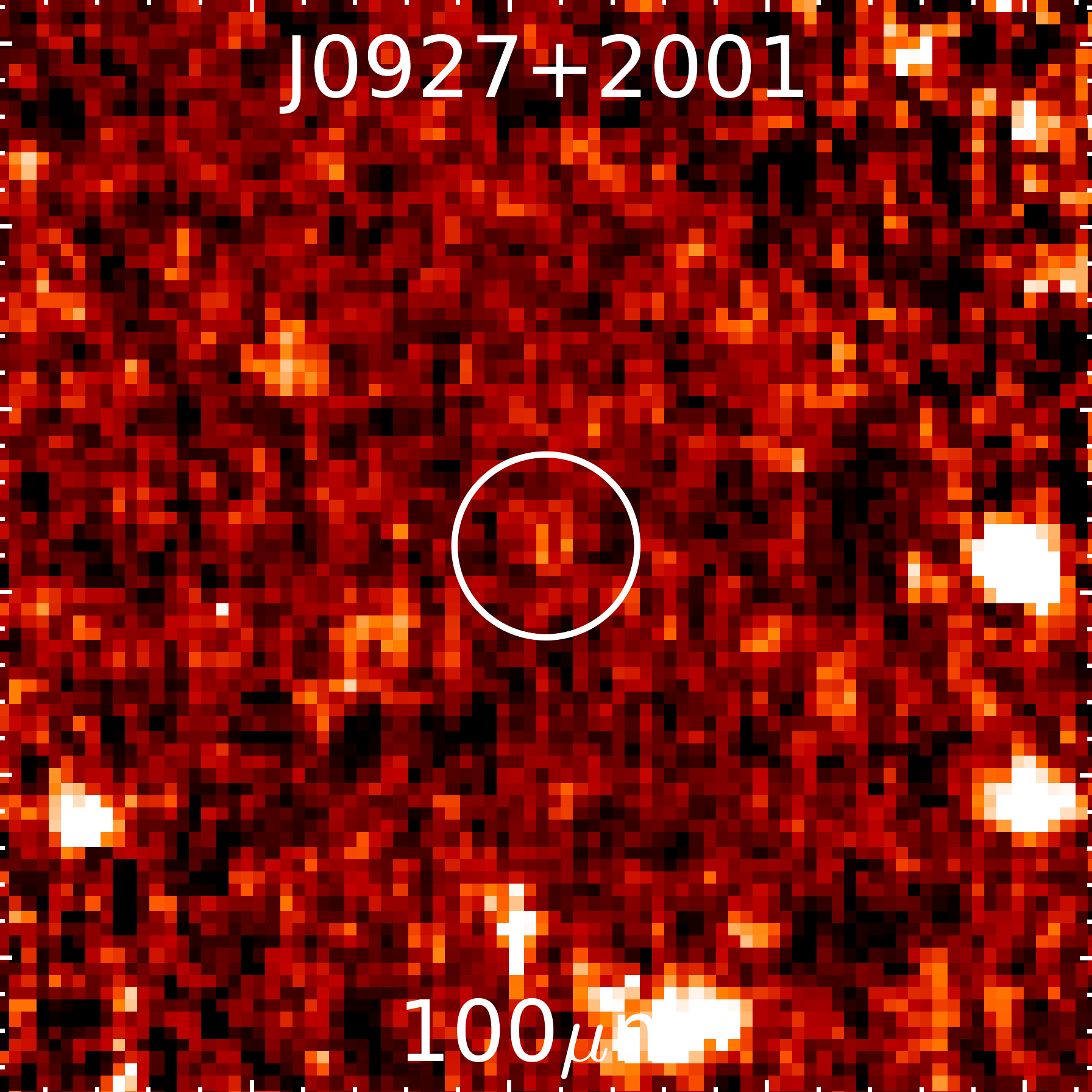}
\includegraphics[angle=0,scale=.23]{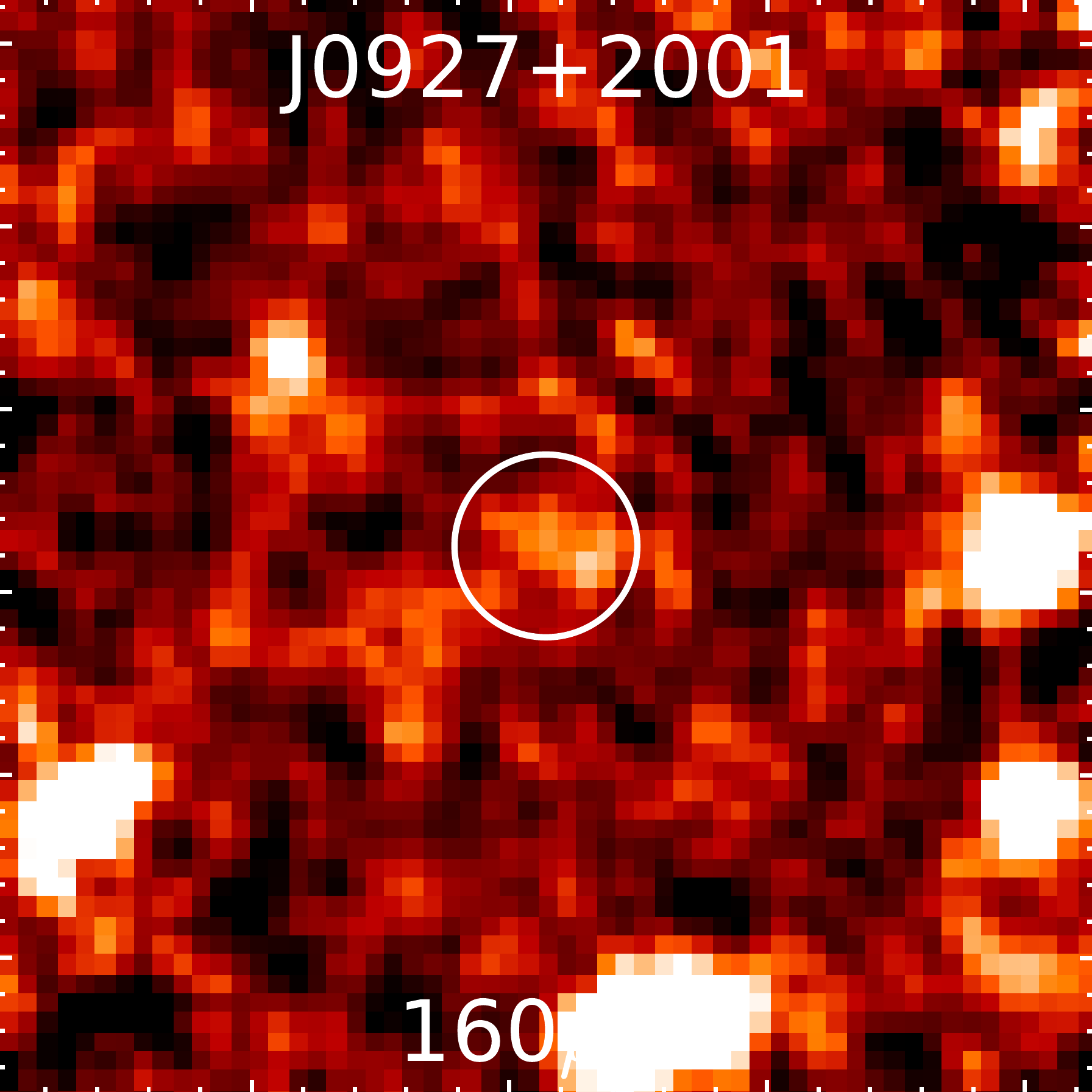}
\includegraphics[angle=0,scale=.23]{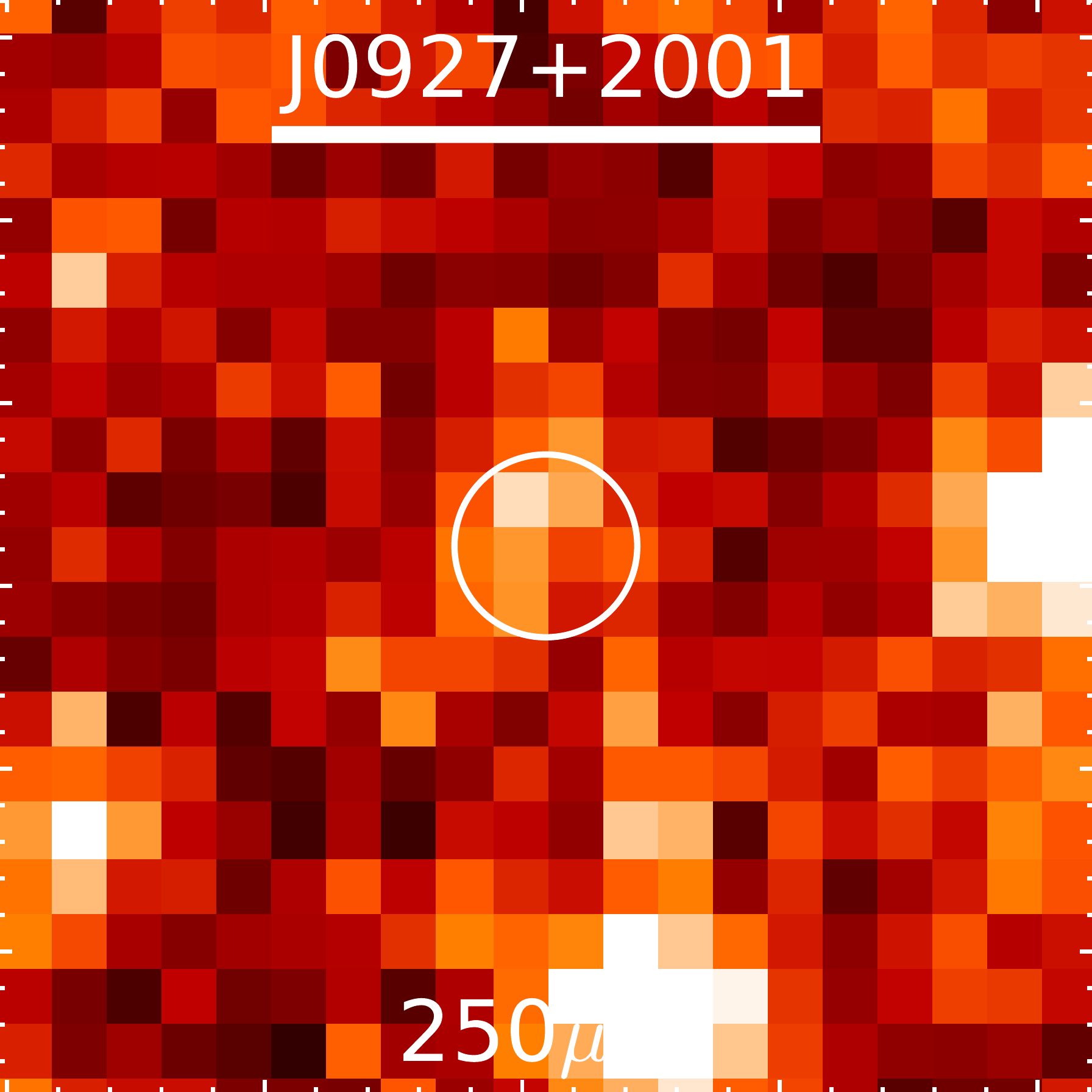}\\
\includegraphics[angle=0,scale=.23]{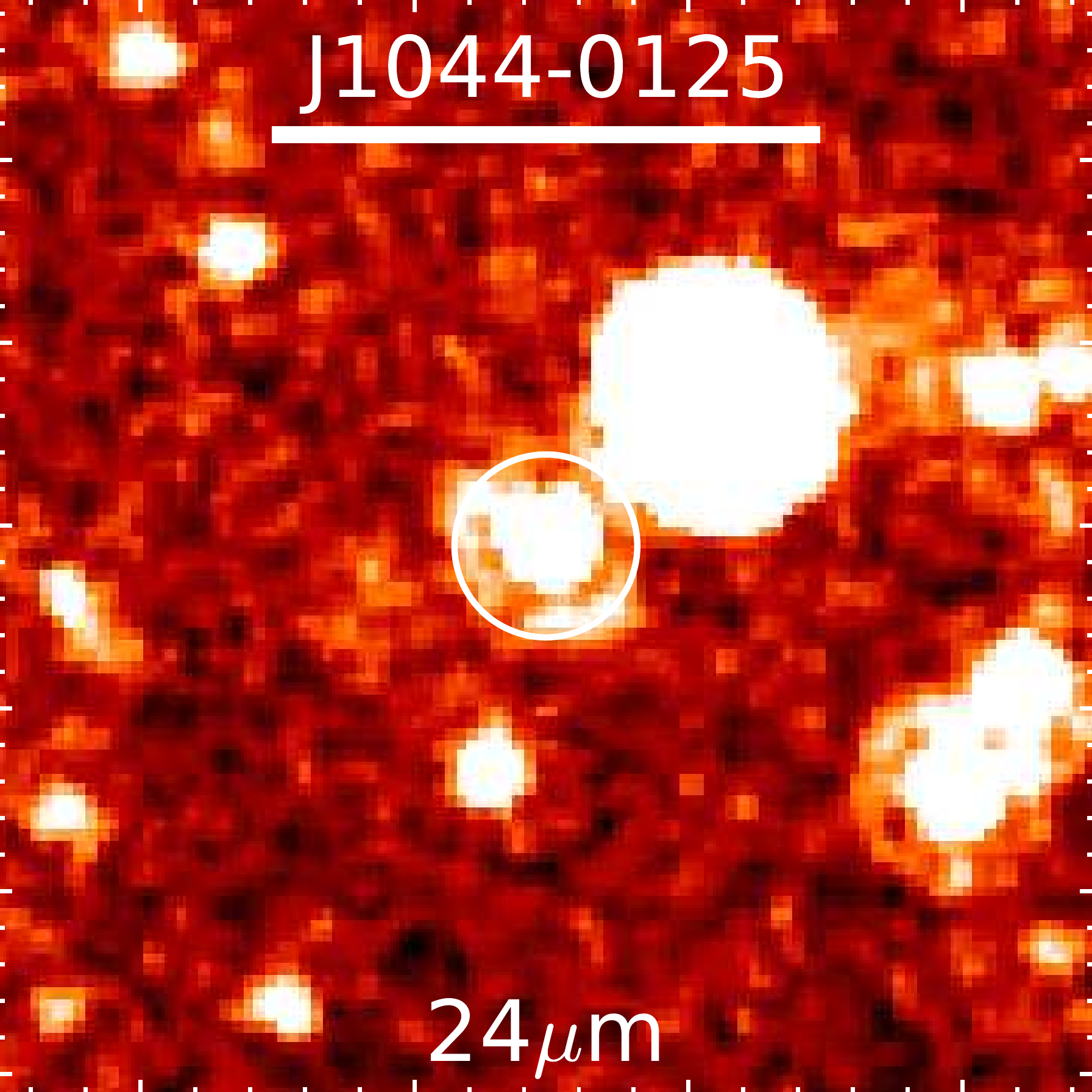}
\includegraphics[angle=0,scale=.23]{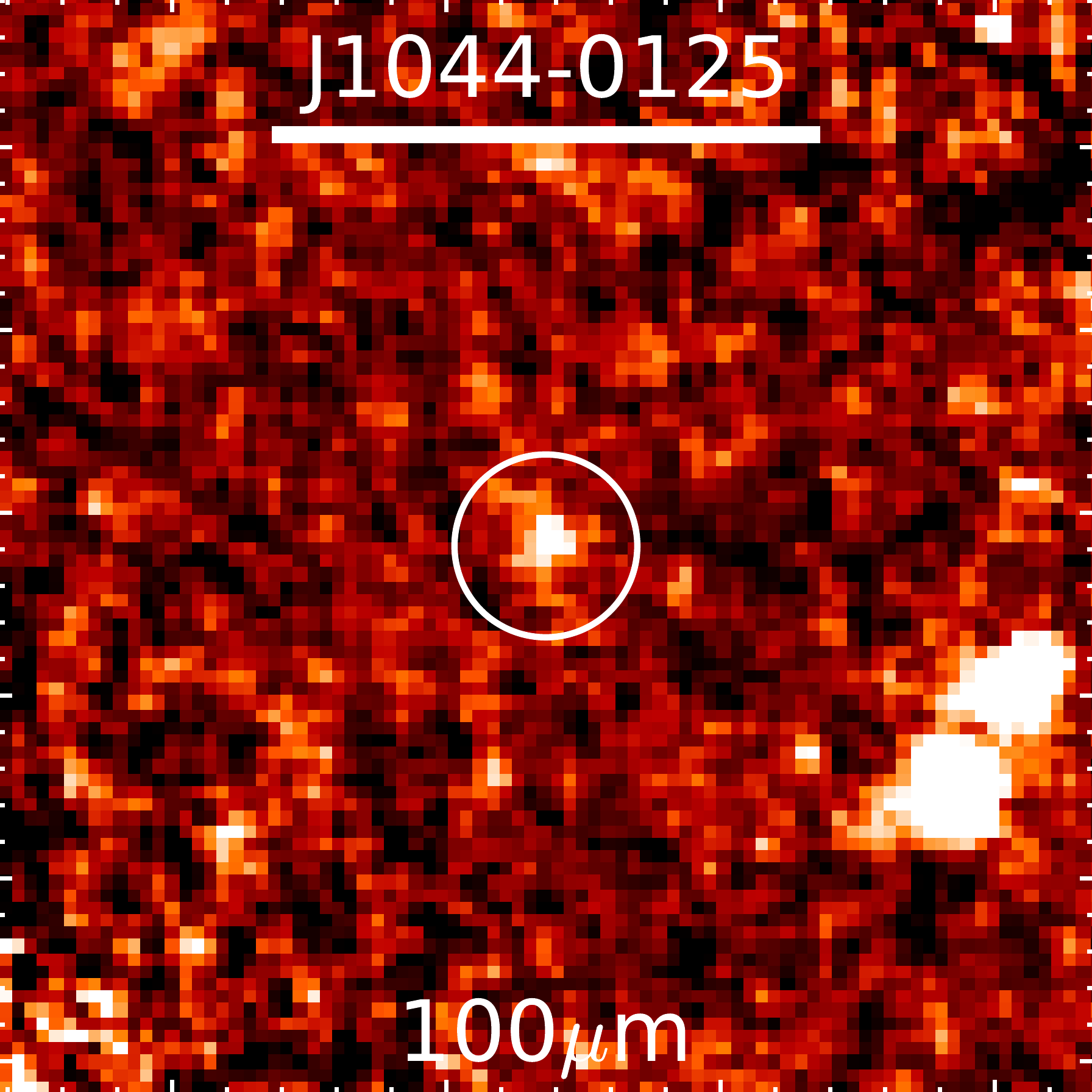}
\includegraphics[angle=0,scale=.23]{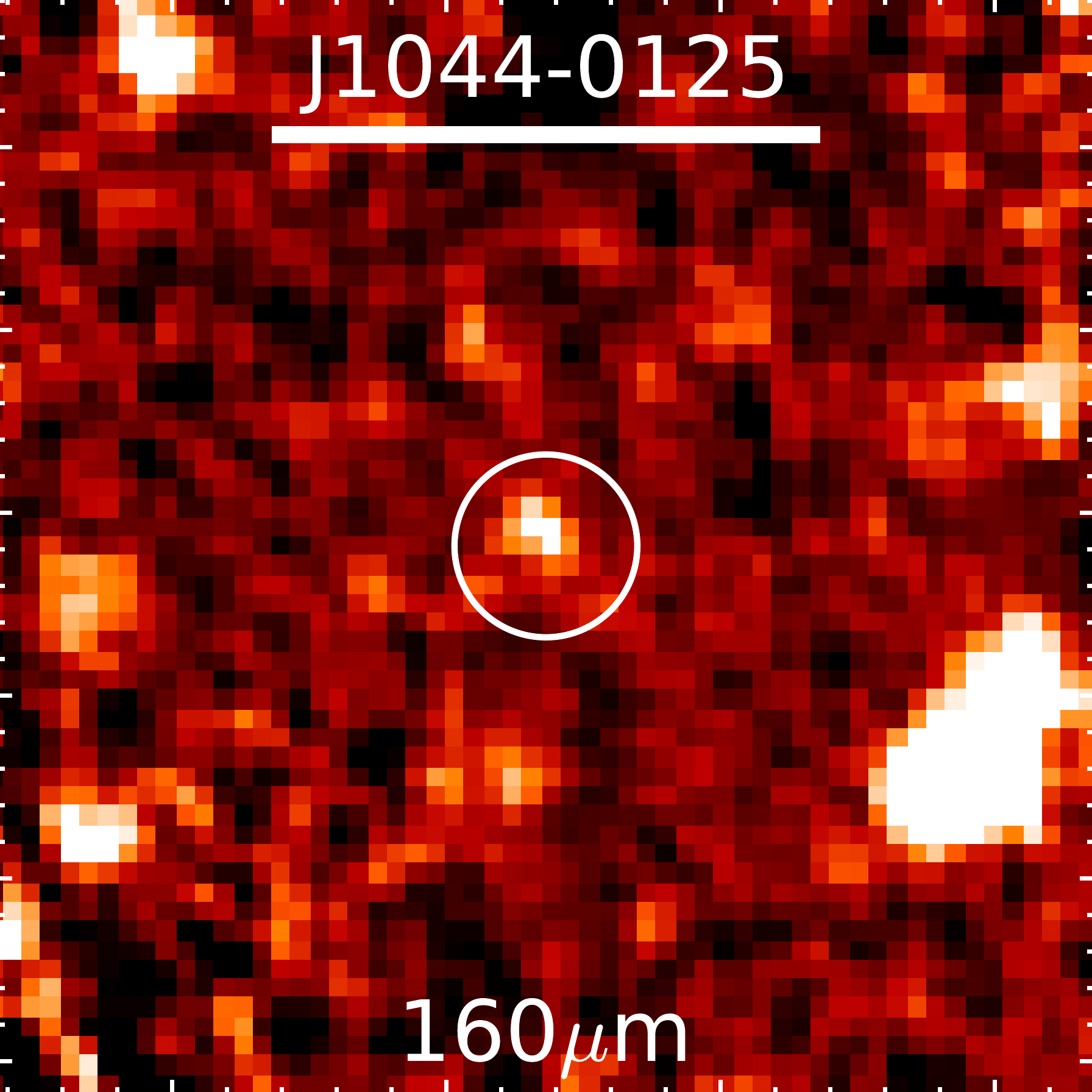}
\includegraphics[angle=0,scale=.23]{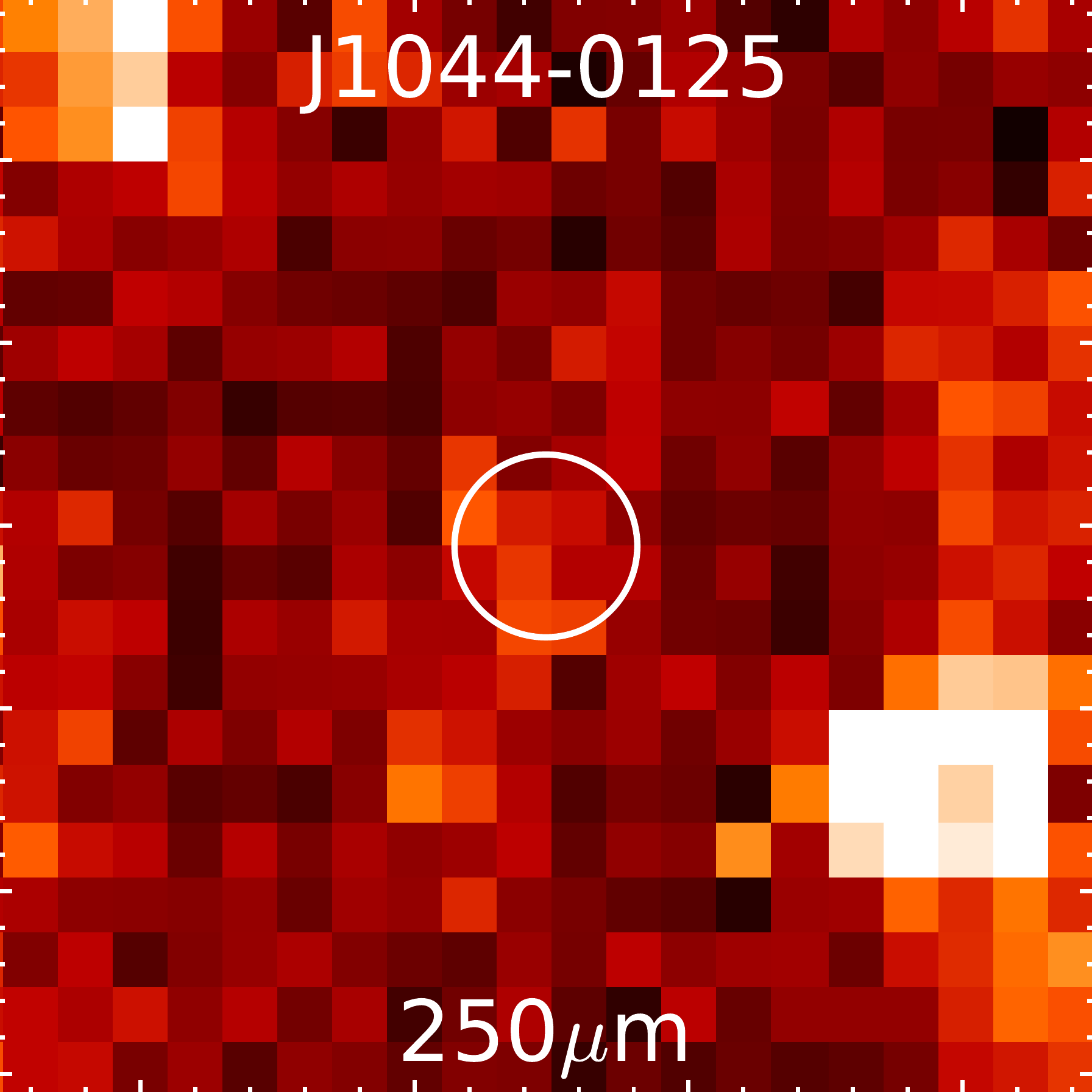}\\
\includegraphics[angle=0,scale=.23]{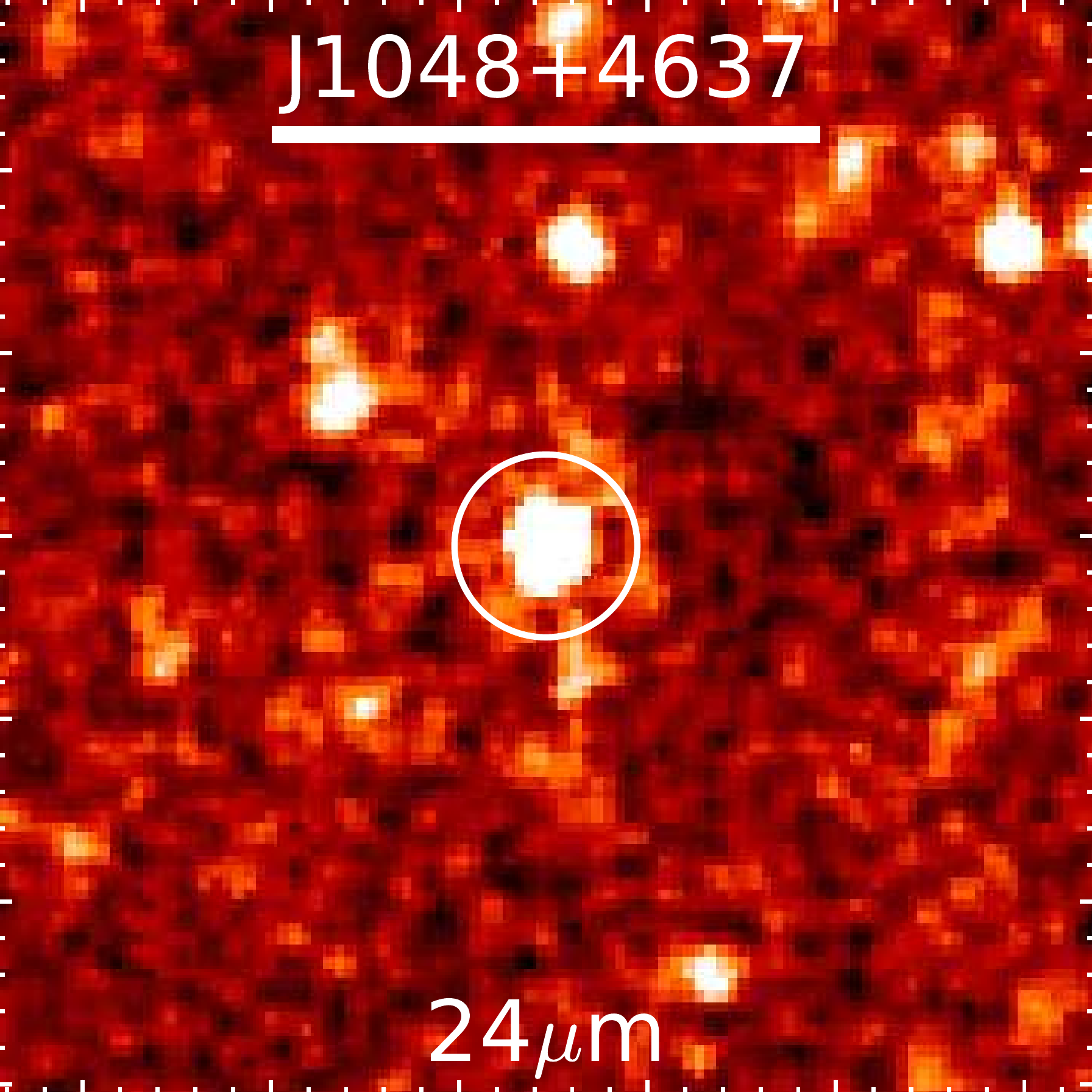}
\includegraphics[angle=0,scale=.23]{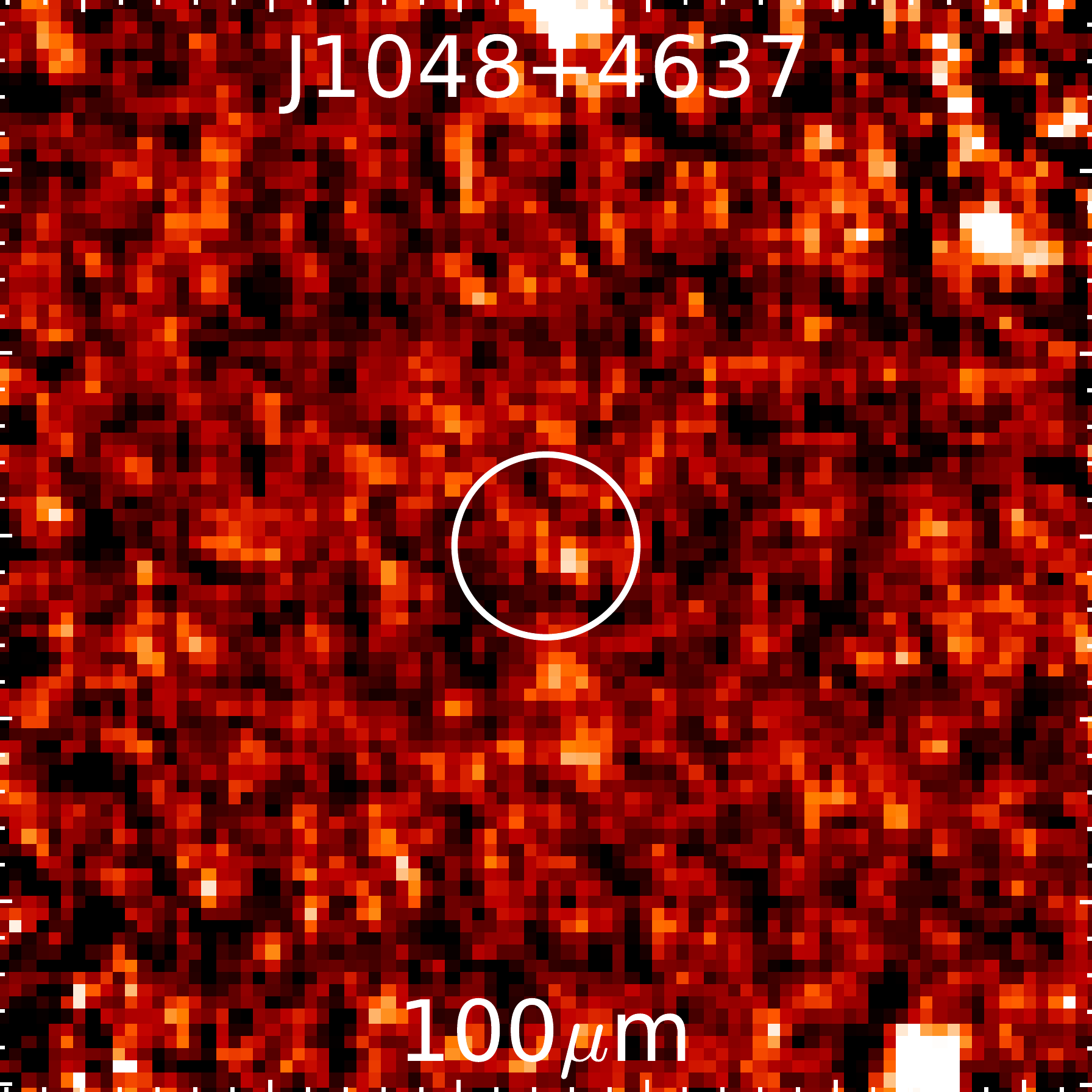}
\includegraphics[angle=0,scale=.23]{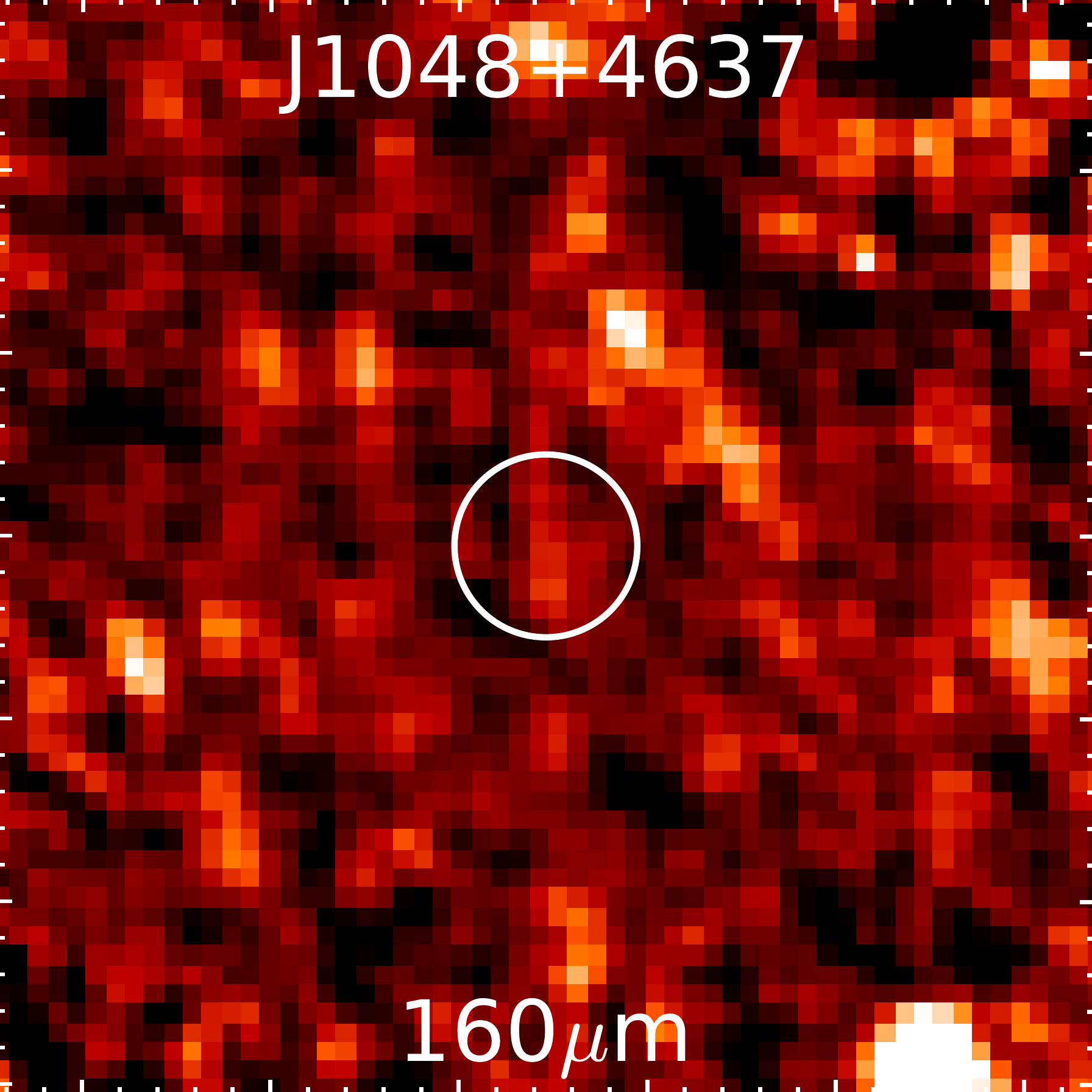}
\includegraphics[angle=0,scale=.23]{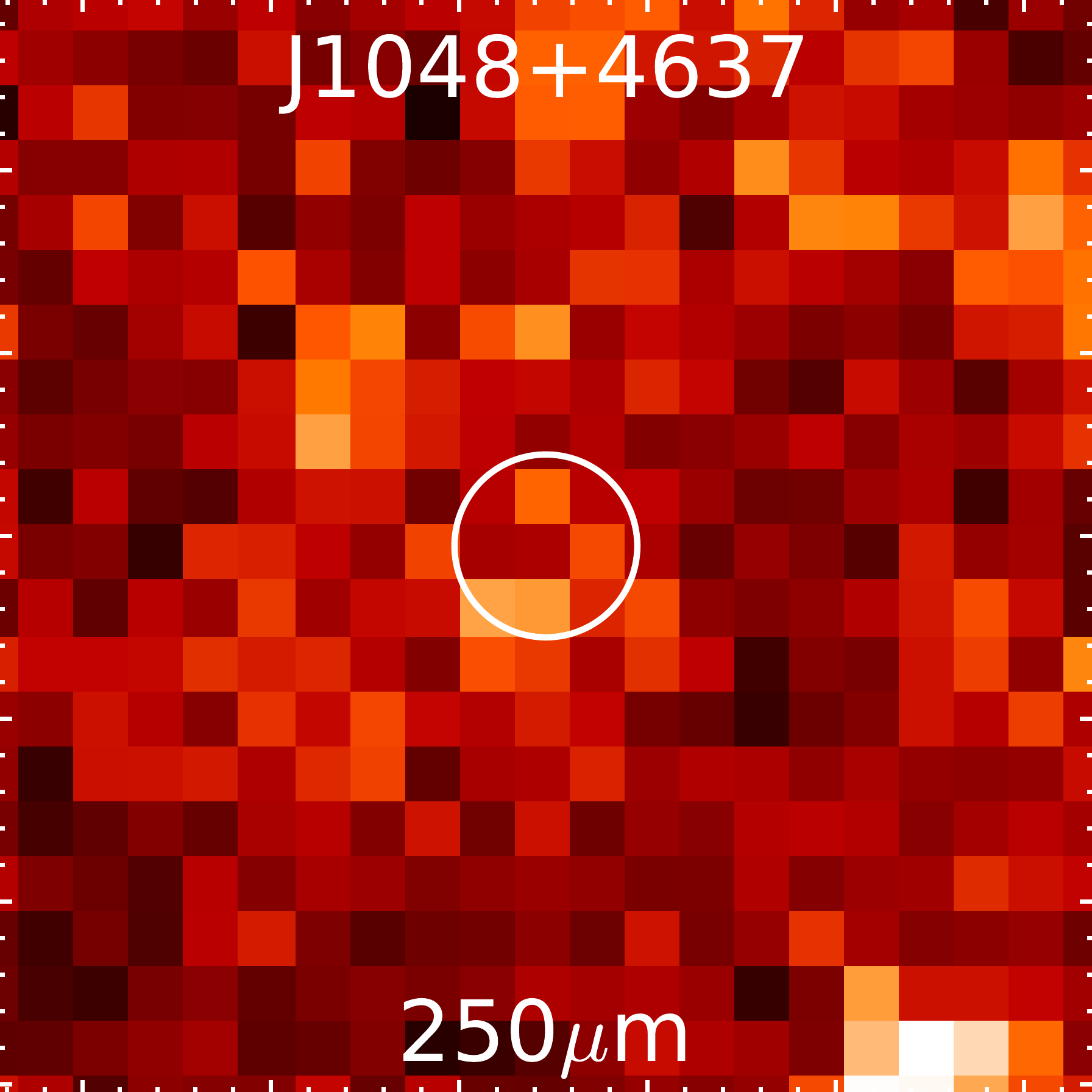}\\
\caption{{\it continued}}
\end{figure*}
\addtocounter{figure}{-1}
\begin{figure*}[t!]
\centering
\includegraphics[angle=0,scale=.23]{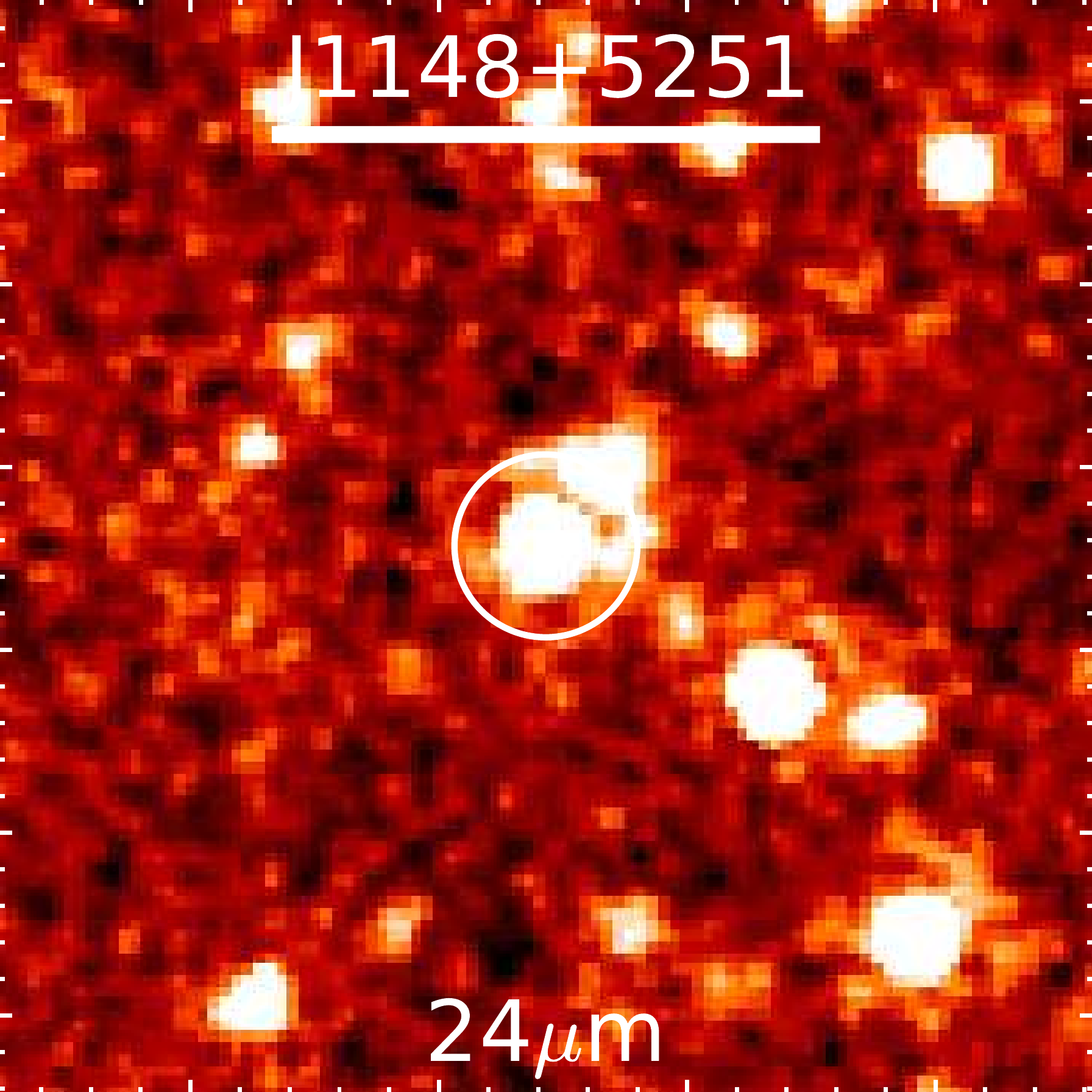}
\includegraphics[angle=0,scale=.23]{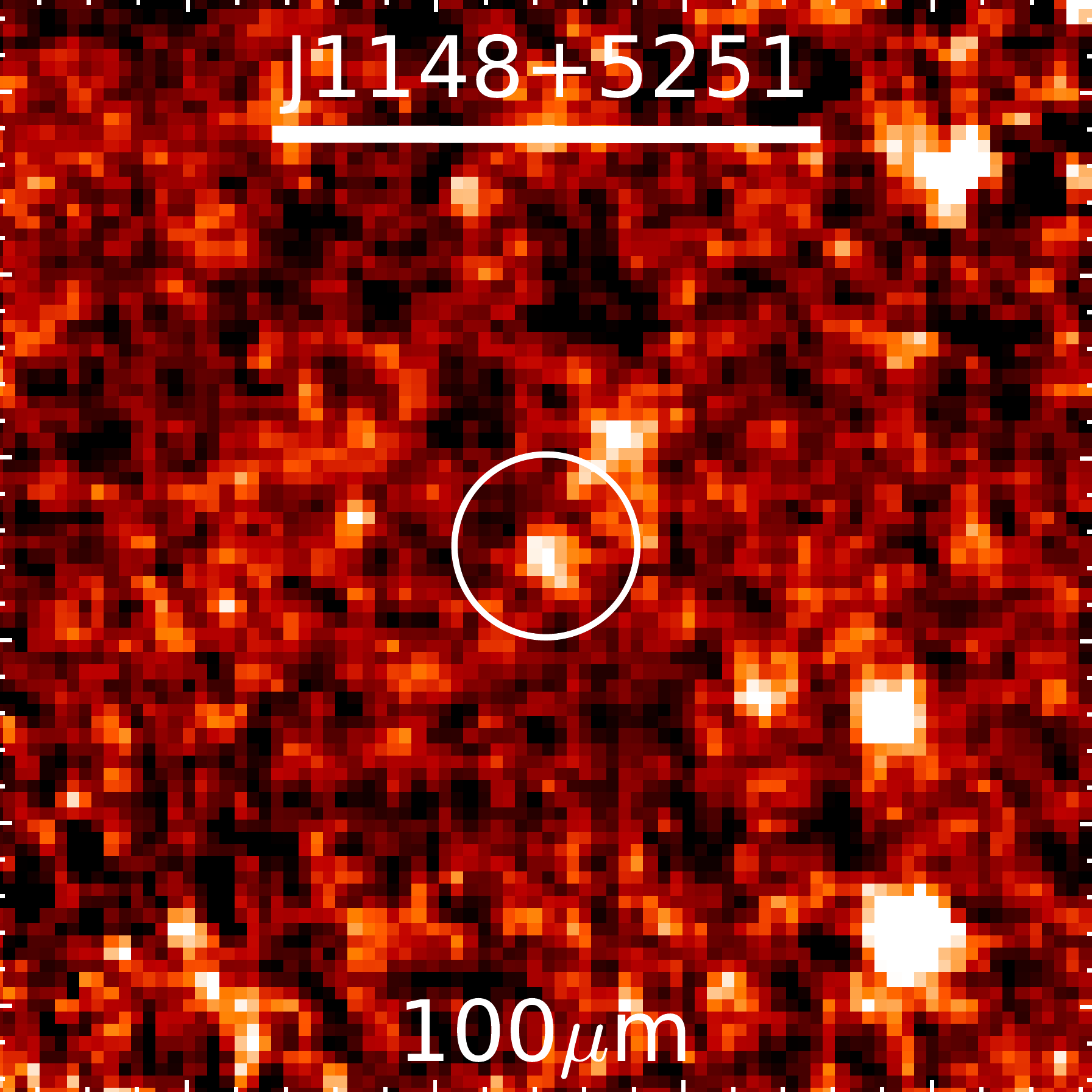}
\includegraphics[angle=0,scale=.23]{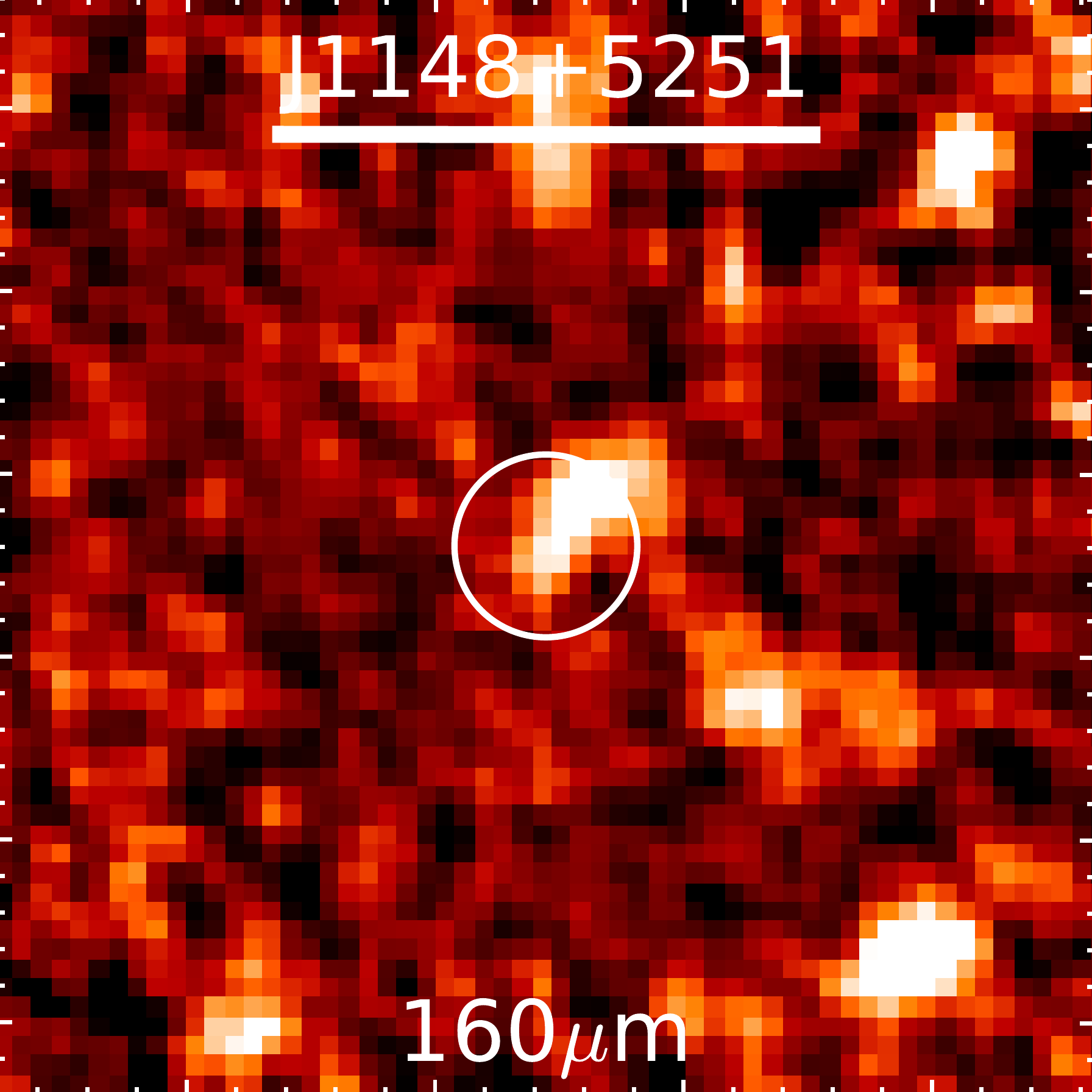}
\includegraphics[angle=0,scale=.23]{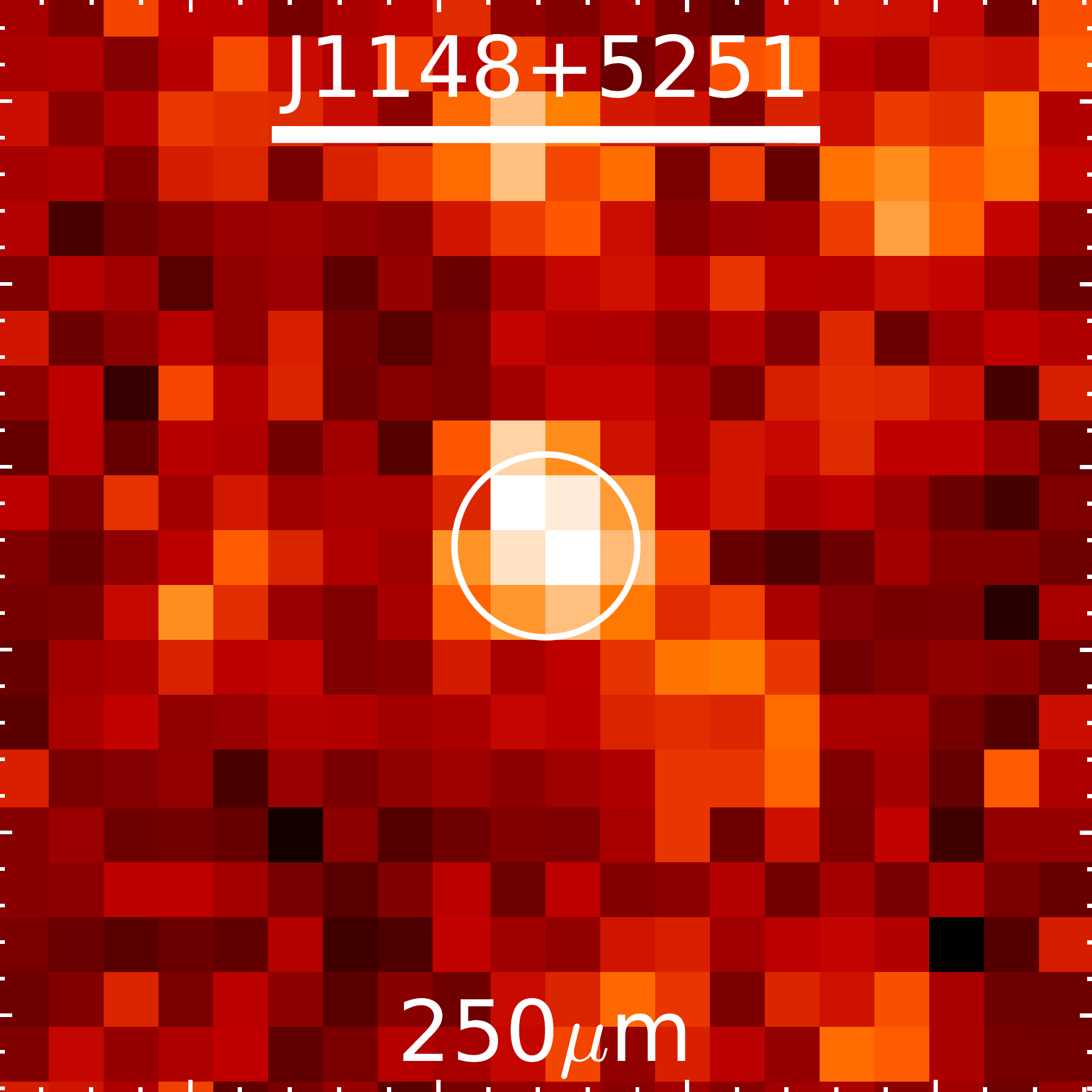}\\
\includegraphics[angle=0,scale=.23]{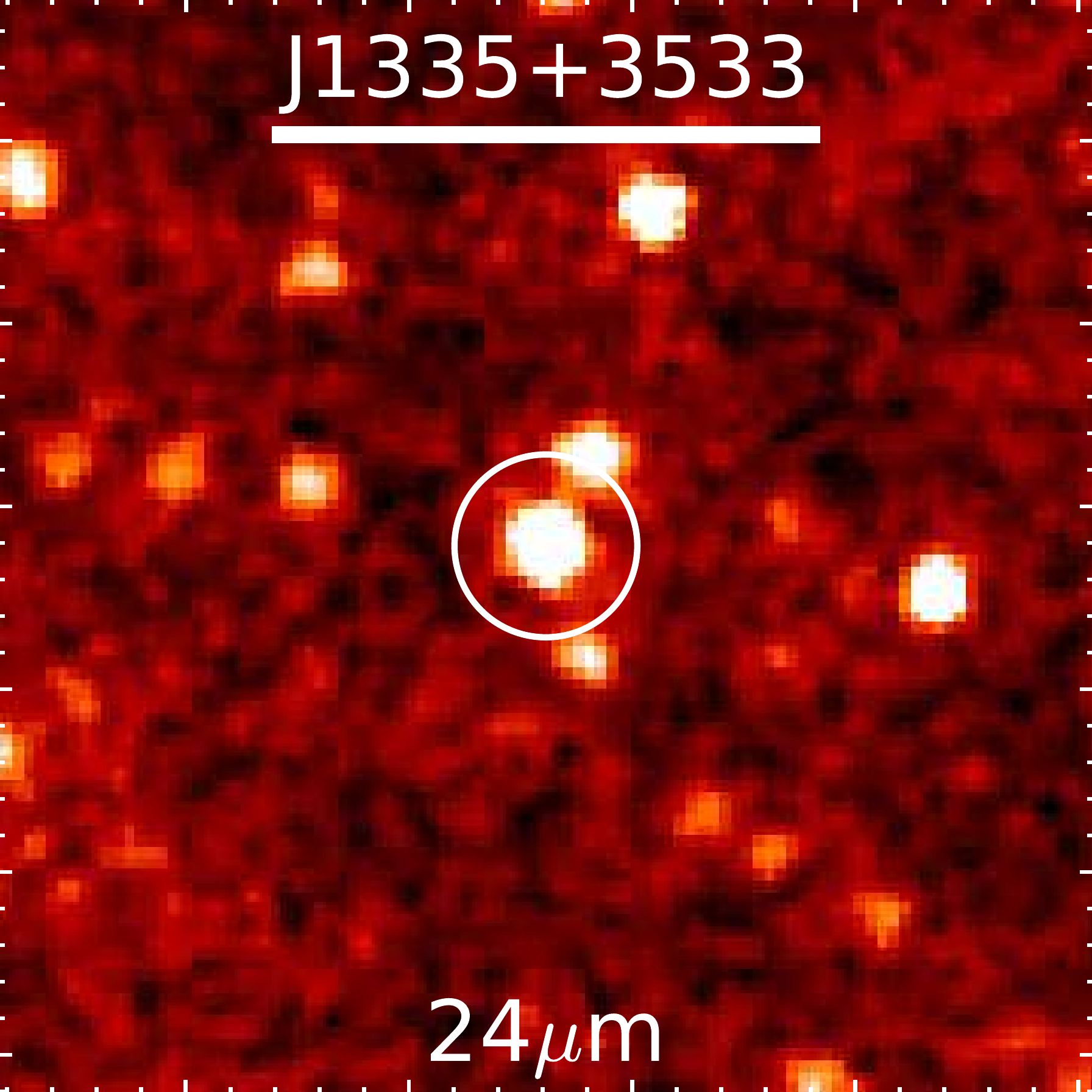}
\includegraphics[angle=0,scale=.23]{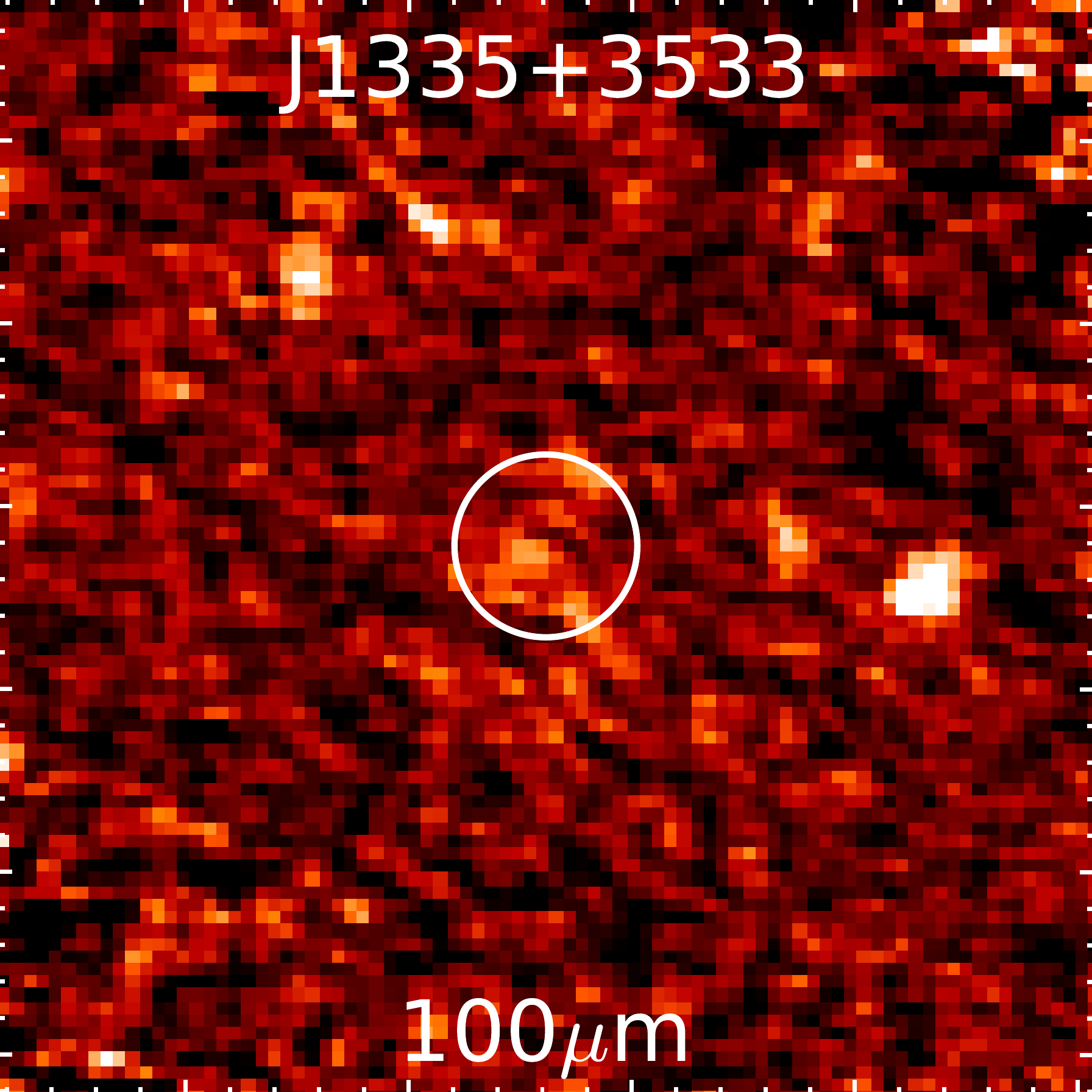}
\includegraphics[angle=0,scale=.23]{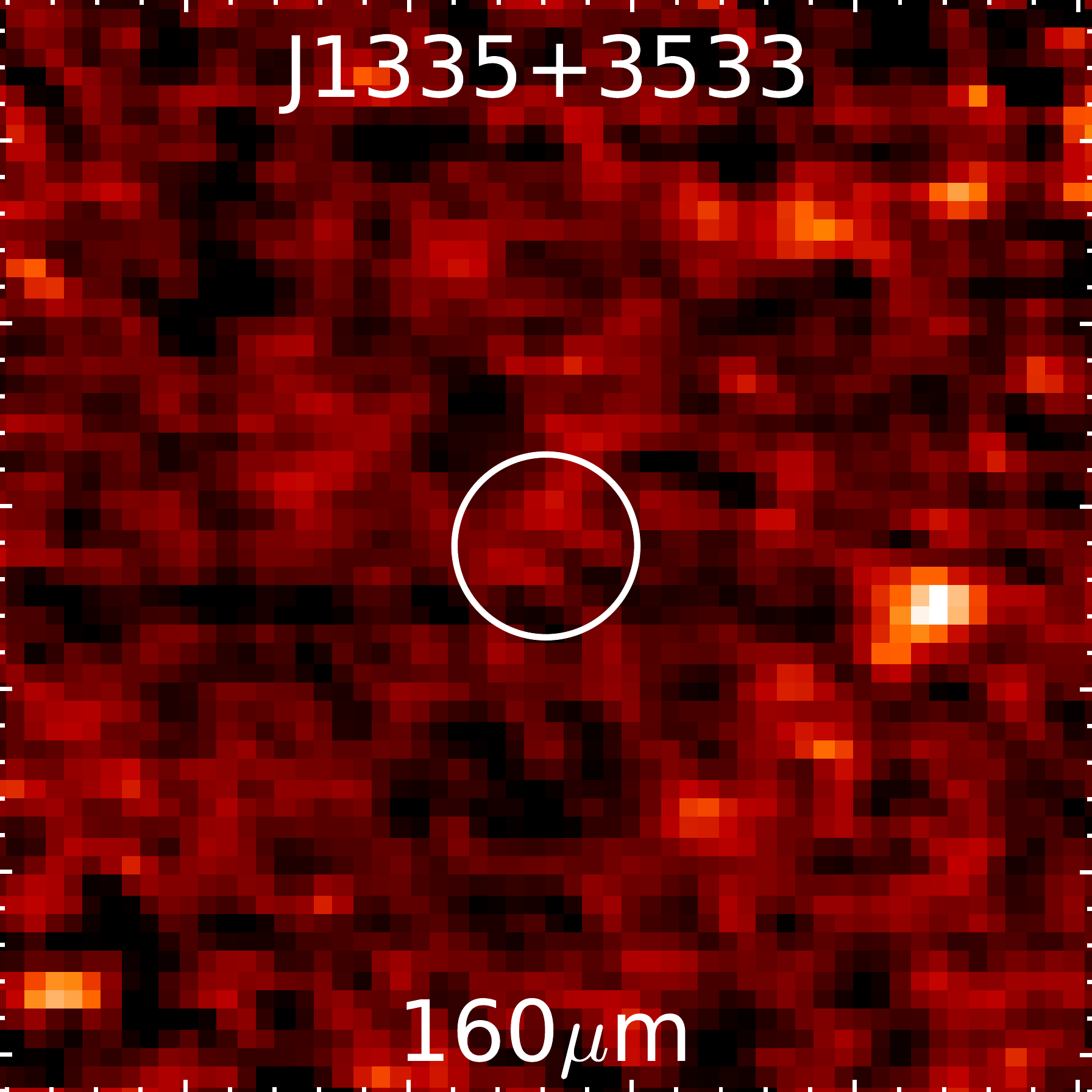}
\includegraphics[angle=0,scale=.23]{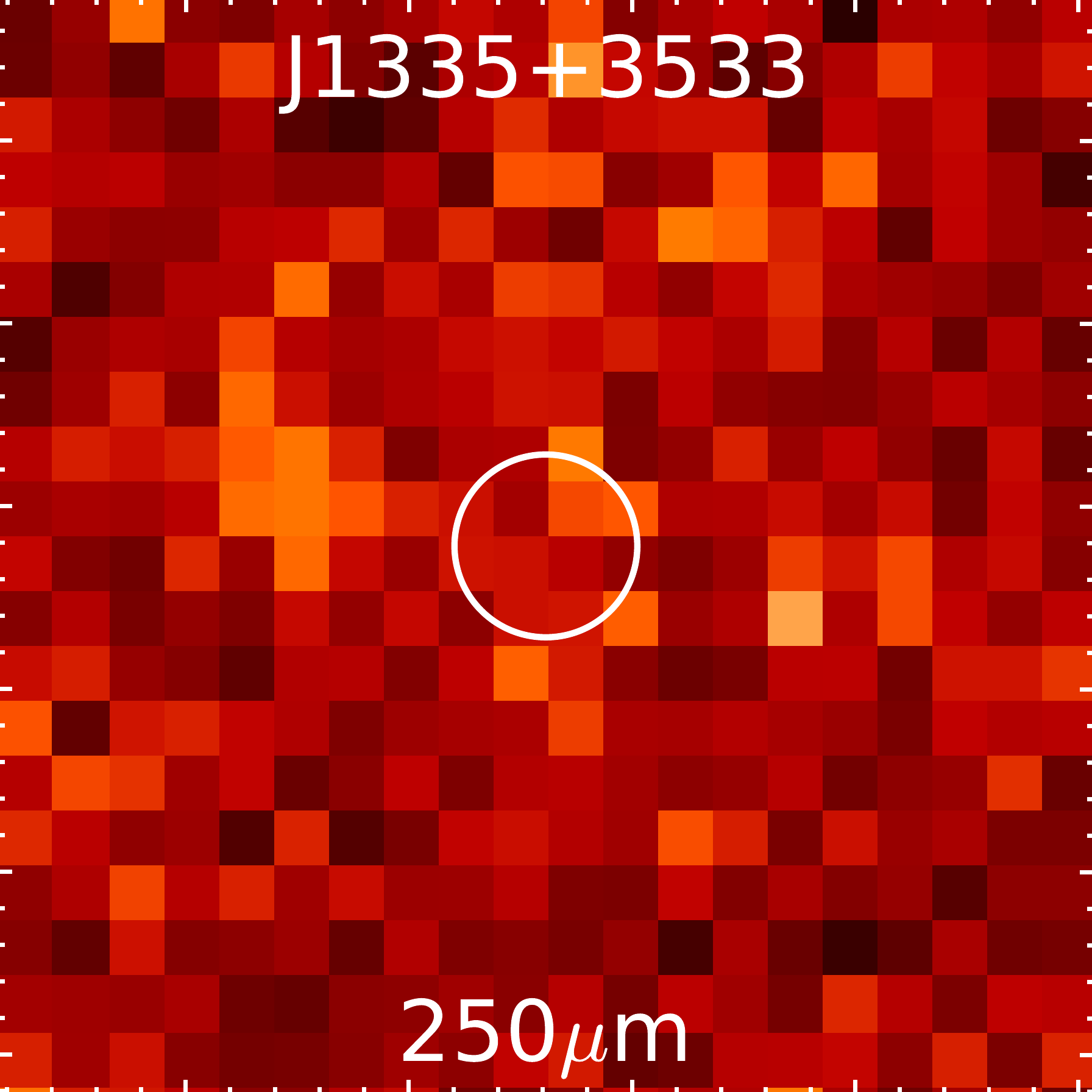}\\
\begin{minipage}[b][4.23cm][c]{4.23cm}{\vspace*{-0.4cm}\hspace*{1.5cm}\fbox{{\large no data}}}\end{minipage}
\includegraphics[angle=0,scale=.23]{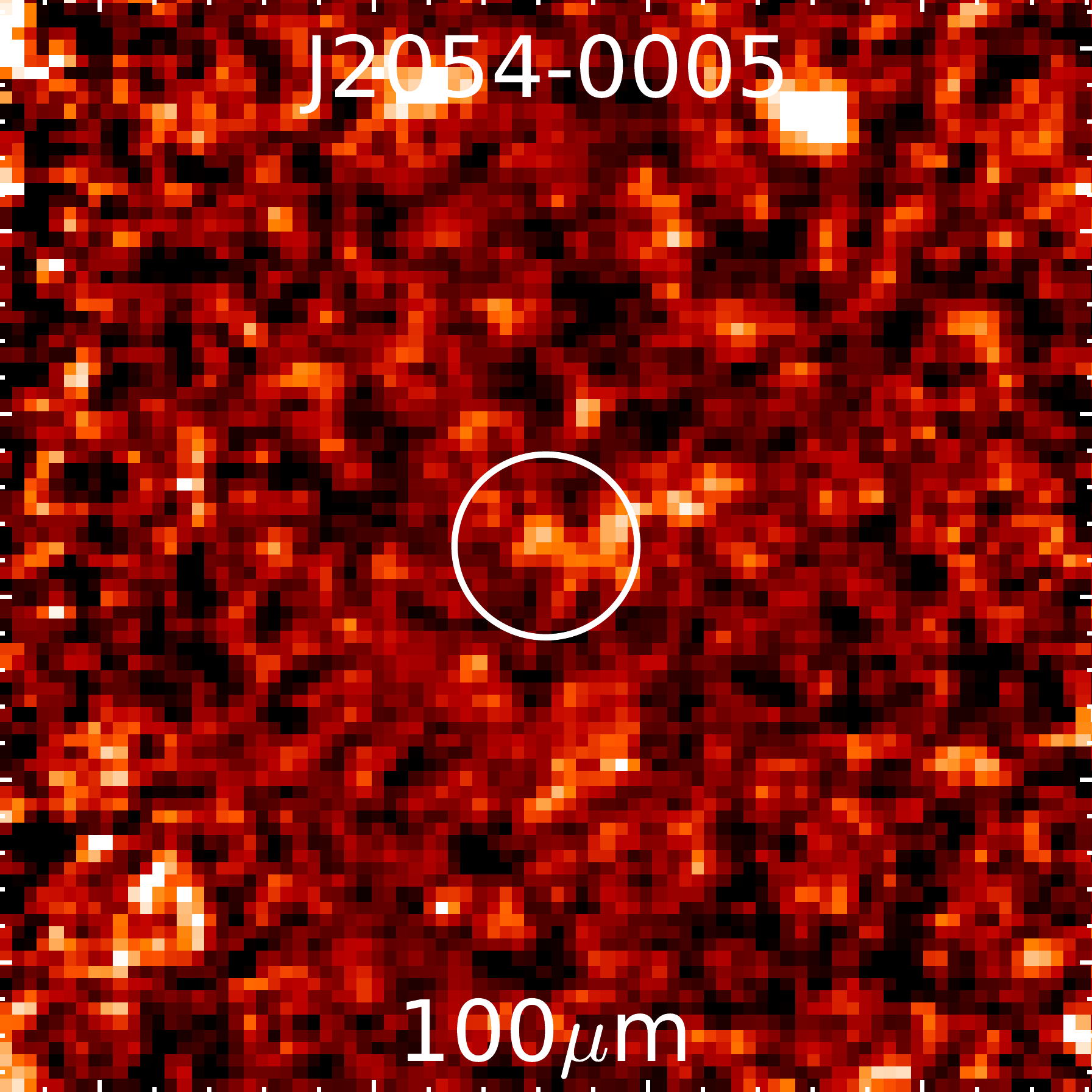}
\includegraphics[angle=0,scale=.23]{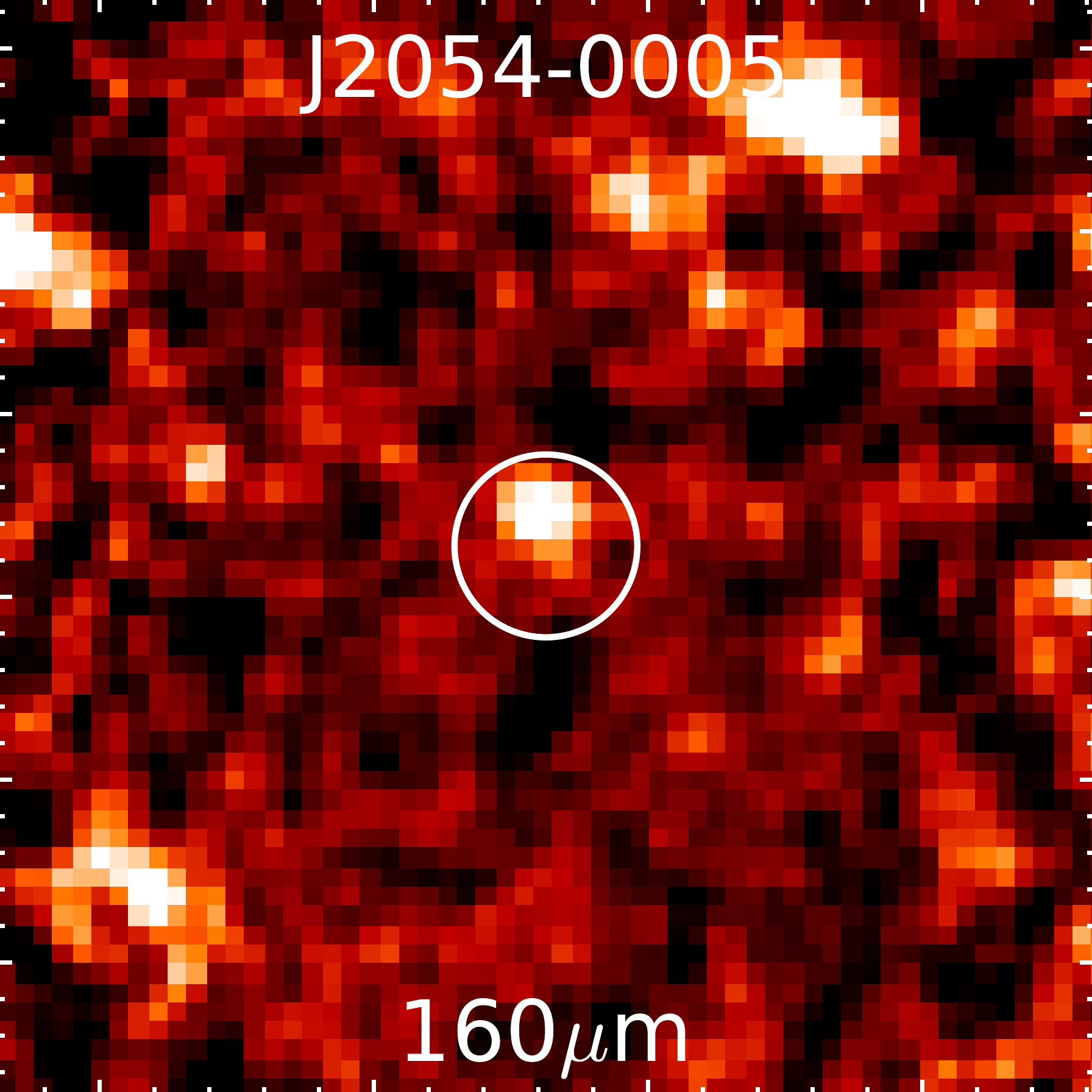}
\includegraphics[angle=0,scale=.23]{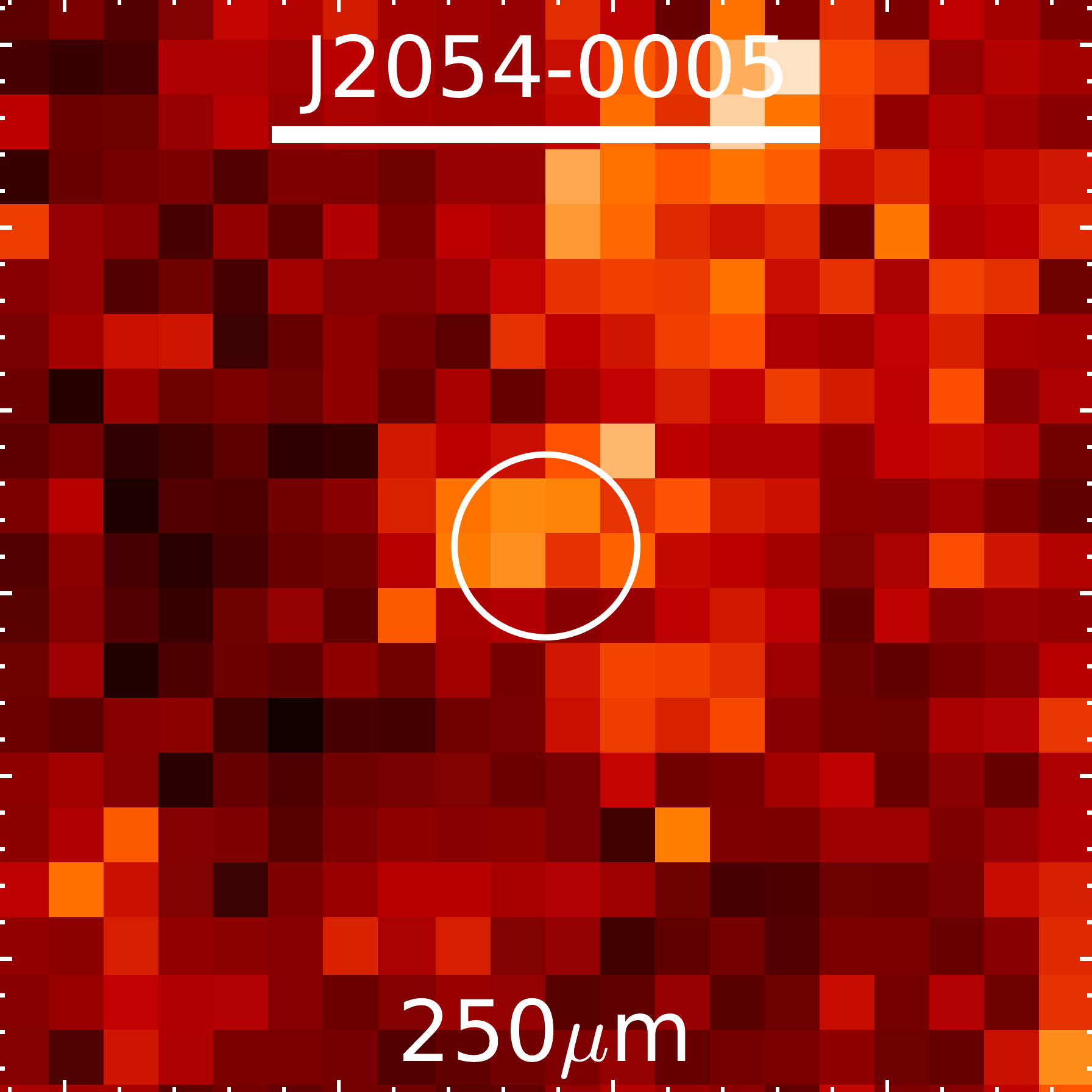}
\caption{{\it continued}}
\end{figure*}

\subsection{WISE}

The all-sky data release of the Wide-field
Infrared Survey Explorer \citep[WISE;][]{wri10} was queried for
photometry or upper limits in the  12\,$\mu$m band which can fill the
gap in the {\it Spitzer} photometry between 8 and 24\,$\mu$m. Only three 
quasars in this paper are detected at 12\,$\mu$m and their significance 
is low ($\lesssim$\,3.5$\sigma$, Tab.\,\ref{photometry}). Some objects (e.g. J2054$-$0005) are
not detected in any WISE band, and consequently no upper limits are
available in the
point-source catalogue. For such sources we performed aperture photometry 
on the WISE atlas images to determine upper limits, following the guidelines 
in the explanatory supplements to the WISE All-Sky Data Release Products 
\citep{cut12}. 

\subsection{Supplemental data}

Additional data from the literature was compiled, mainly in the observed 
NIR or mm regimes and often taken from the discovery papers. The latest data 
release of the UKIDSS survey \citep{law07} was also checked which yielded 
additional photometry in the NIR for six objects.

\section{Analysis}

\subsection{SED components}

\begin{figure*}[t!]
\centering
\includegraphics[angle=0,scale=1.0]{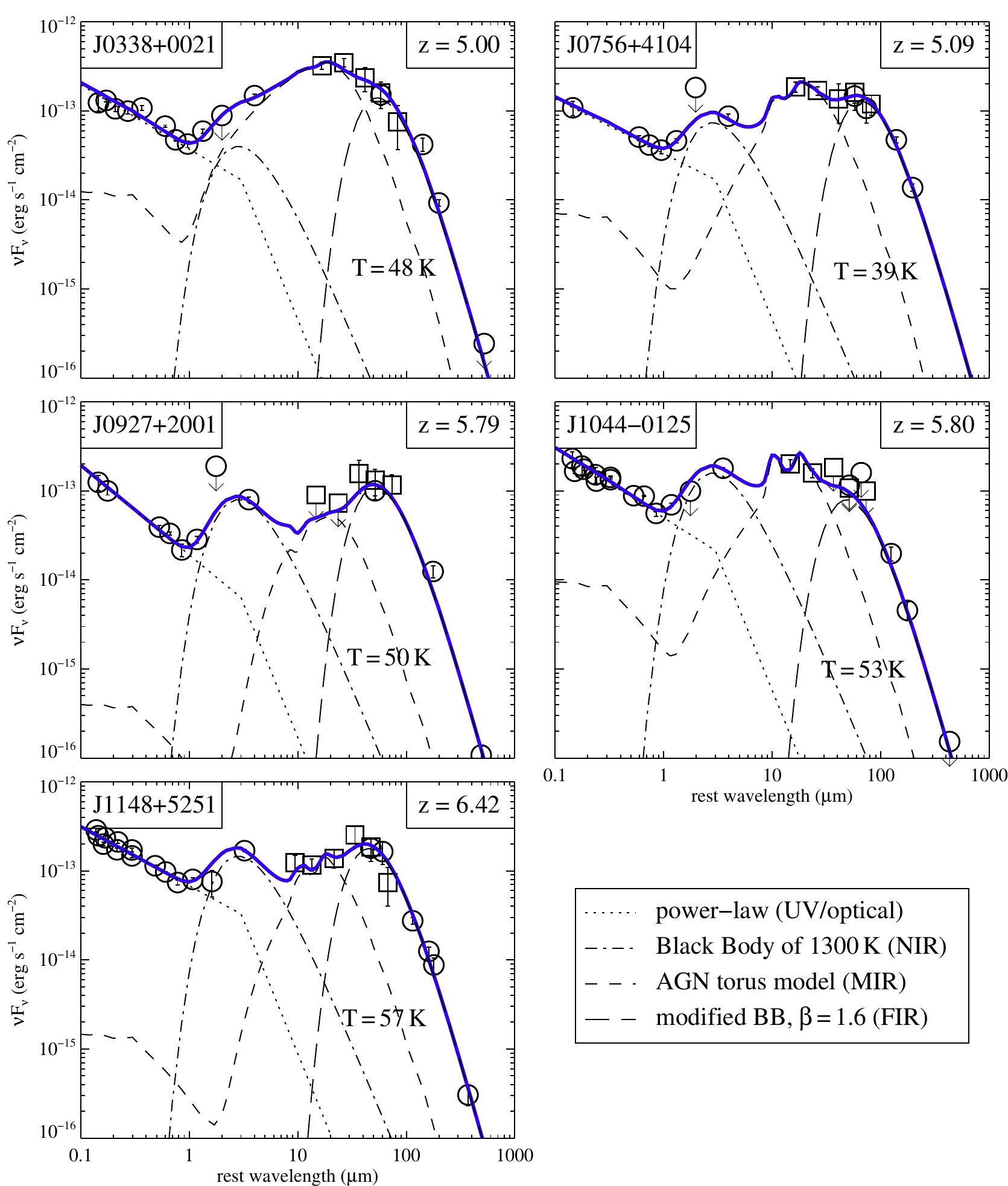}
\caption{Observed SEDs of mm-detected quasars with at least 
two {\it Herschel} detections; for these objects multi-component SED fits 
were carried out as outlined in section\,2. The SED fit is performed 
using a power-law in the UV/optical (dotted line), a 1300\,K black body in the
NIR (dot-dashed line), a torus model in
the NIR/MIR (short dashed line) and a modified 
black body in the FIR with emissivity index $\beta$ fixed to 1.6 (long
dashed line). 
 The blue solid line corresponds to the sum of the fitted
components which here represent the overall best fit. Thus, the 
temperature of the FIR component here may differ slightly from the 
overall mean temperature determined from all acceptable fits as 
presented in Tab.\,\ref{results}. The squares 
correspond to the new {\it Herschel} data. \label{sed_fits_detected}. }
\end{figure*}

\begin{figure*}[t!]
\centering
\includegraphics[angle=0,scale=1.0]{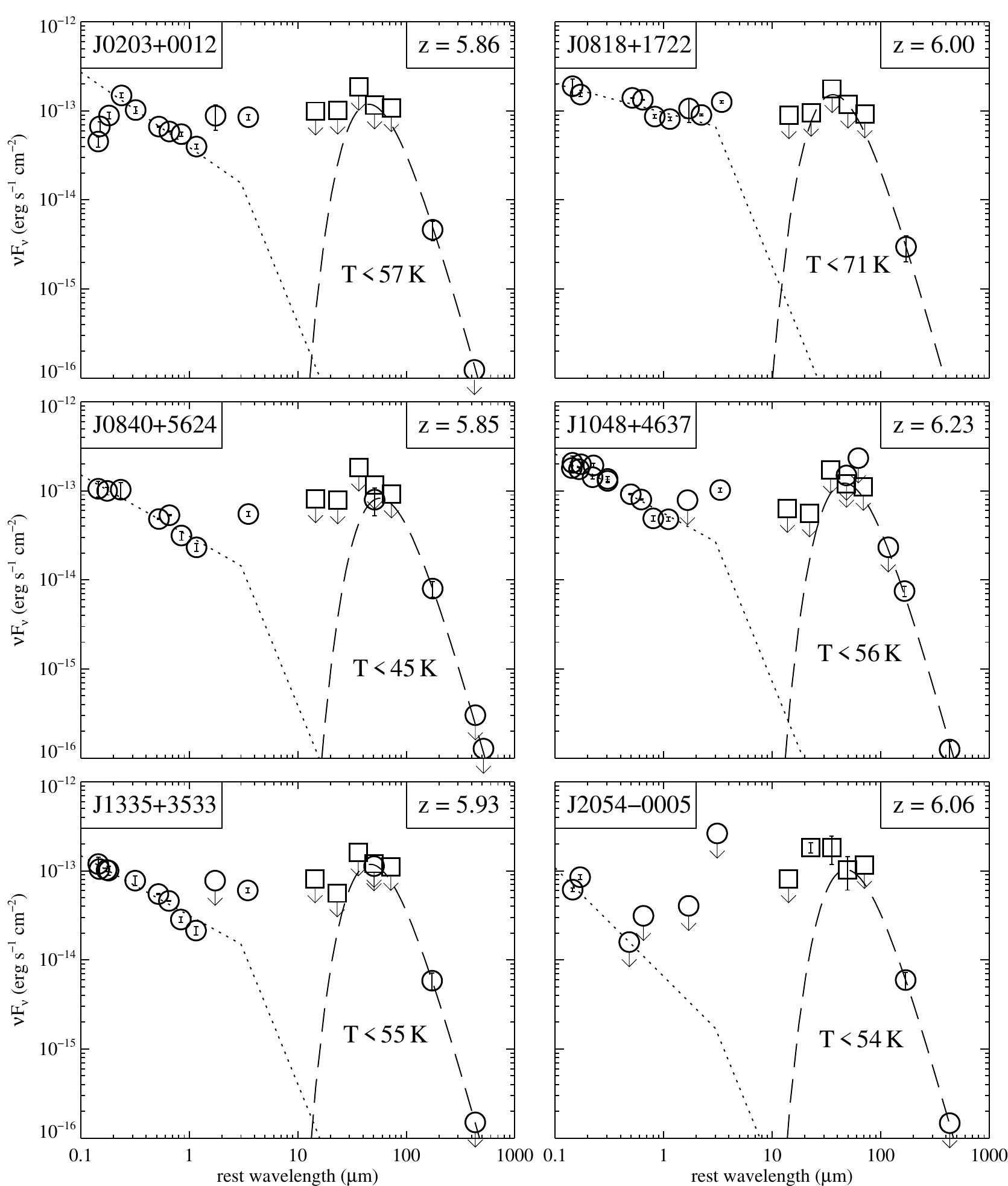}
\caption{Observed SEDs of mm-detected quasars without {\it Herschel} 
detections. Despite SPIRE detections, the object J2054$-$0005 is included 
here because the poor constraints on the rest frame optical through MIR 
SED prevents detailed SED fits.  \label{sed_fits_nondetected}}
\end{figure*}

In combination with other supplemental data from the literature (section 2), we
compile SEDs covering a rest frame wavelength range of typically 
0.1-400\,$\mu$m (see Figs.\,\ref{sed_fits_detected} and \ref{sed_fits_nondetected}). 
These SEDs are then fitted with a
combination of models to represent the different components contributing 
to the observed SED.  For this purpose we have divided our sample
into two groups, depending on the amount of data available to
constrain the fitted components. Objects detected in at least two
{\it Herschel} bands were subject to full SED fits (five sources), except 
J2054$-$0005, for which the lack of strong photometric constraints at 
rest frame wavelengths $<$\,10\,$\mu$m (no {\it  Spitzer} observations and only WISE 
upper limits) 
prevented a full SED fit. In our fits we consider four distinct components:

\begin{enumerate}
\item A power law in the UV/optical regime which represents the emission 
  from the accretion disk. We extend this component into the NIR and introduce 
  a break to a the Rayleigh-Jeans slope of 
$F_{\nu} \propto \nu^{2}$ at 3\,$\mu$m in the 
  rest frame \citep{hon10}.
  In the fitting, the power-law slope in the UV/optical and the overall 
  normalization are free parameters.

\item A blackbody of typically 1300\,K temperature, thus peaking in the 
restframe NIR. Empirically, such a component is often required to fit the optical 
through MIR SEDs  of 
luminous quasars \citep[e.g.][]{bar87,gal07,mor09,lei10b} and is generally 
interpreted as 
a signature of hot (graphite) dust close to the sublimation temperature.

\item A clumpy torus model from the library of \citet{hon10} to account for 
  the AGN heated dust from the ``dusty torus'' in the central parts of the 
  AGN. This component dominates the MIR and is important to disentangle, to first order, the 
  contributions from the nuclear dust to the rest frame FIR
  emission. Besides the choice of a particular model (see below), the 
  absolute scaling of the model is the only free parameter for this component.

\item A modified black body to account for possible FIR excess emission (over the 
AGN heated torus) which we here interpret as powered by star formation. For this component we fix 
the emissivity index $\beta$ to a value of 1.6 \citep[e.g.][but see section \ref{betafit}]{bee06,wan08a}. 
The temperature of the modified black body and the normalization are free 
parameters.

\end{enumerate}

\subsection{Fitting procedure}
\label{sec:fitting}

The torus models we consider here are available for seven different inclinations
(starting at 0\degr~and increasing in steps of 15\degr). For
each inclination, \citet{hon10} provide models of various
combinations of parameters (opening angle,  radial dust distribution,
etc.). Our fitting procedure takes one of the torus models and fits a linear
combination of the four components to the observed SED via chi-square
minimization using {\sc Mpfit} \citep{mar09} in IDL. We then cycle
through all torus models in the library and repeat the fitting for 
each of them. 
However, for efficiency we
limit the torus component to models with inclinations $\leq$\,45
degrees which seems reasonable given  that we observe luminous and
largely unreddened type-1 quasars. This leaves a total of  959
different torus models. We visually inspect the best-fitting 10\% of
the model combinations to confirm the fits.

With this approach we do not intend to develop a highly accurate model
for  the full SED emission in these objects. For the current work we
aim to  construct a physically motivated approximation that yields a
reasonable  description of the observed SEDs which allows us to
isolate excess FIR  emission and to account for (to first order)
contributions of the AGN  heated nuclear dust to the FIR
photometry. 

As outlined above, our fitting includes an additional, emprically
motivated  1300\,K black body in the NIR. We have also performed the
fits excluding this  component, only fitting a power-law, a torus
model and a FIR modified black body.   The comparison between both
cases shows that the fits including the NIR black body generally
represent the observed photometry better. This is  particularly
apparent in wavelength regions dominated by very hot dust
($\lambda_{\rm rest}$\,$\sim$\,1-3\,$\mu$m), in the  overlap region
between the torus and the FIR black body  ($\lambda_{\rm
rest}$\,$\sim$\,20-30\,$\mu$m) and in the fit at $\lambda_{\rm
rest}$\,$\gtrsim$\,100\,$\mu$m. The temperatures of the FIR dust
component also come out consistently lower (by about 5-10\,K) in fits
including the  additional NIR black body.

Recall that the additional NIR component contributes  significantly
(or dominantly) to the short infrared wavelengths. In cases where this
component is absent, torus models with a strong emphasis on emission
at  $\lambda$\,$\lesssim$\,10\,$\mu$m are favored in the fits to
accomodate (in particular)  the MIPS photometry. By design such torus
models contribute less flux at longer wavelengths
($\lambda$\,$\gtrsim$\,20\,$\mu$m) thus requiring  a hotter FIR
component to match the  {\it Herschel} photometry. This in turn
negatively affects the fit in the observed  sub-mm/mm regime. Including
the NIR component, more power in the torus component can shift  to
slightly longer wavelengths, allowing a cooler FIR component and
providing a better  overall fit to the data. The need to add an
additional hot component to torus models  when fitting type-1 AGN SEDs
has also been noted by e.g. \citet{mor12}. NIR reverberation mapping
observations \citep[e.g.][]{sug06} show that the size of the emitting
region of this very  hot dust is a factor of $\sim$20 smaller than the
torus as measured in the MIR via  interferometry (Burtscher et al., in
preparation), supporting this additional complexity in the
distribution of the AGN-heated dust.  The following results and
discussion will therefore be based on the fits including the
additional NIR black body.


\begin{figure}[t]
\centering
\includegraphics[angle=0,scale=0.49]{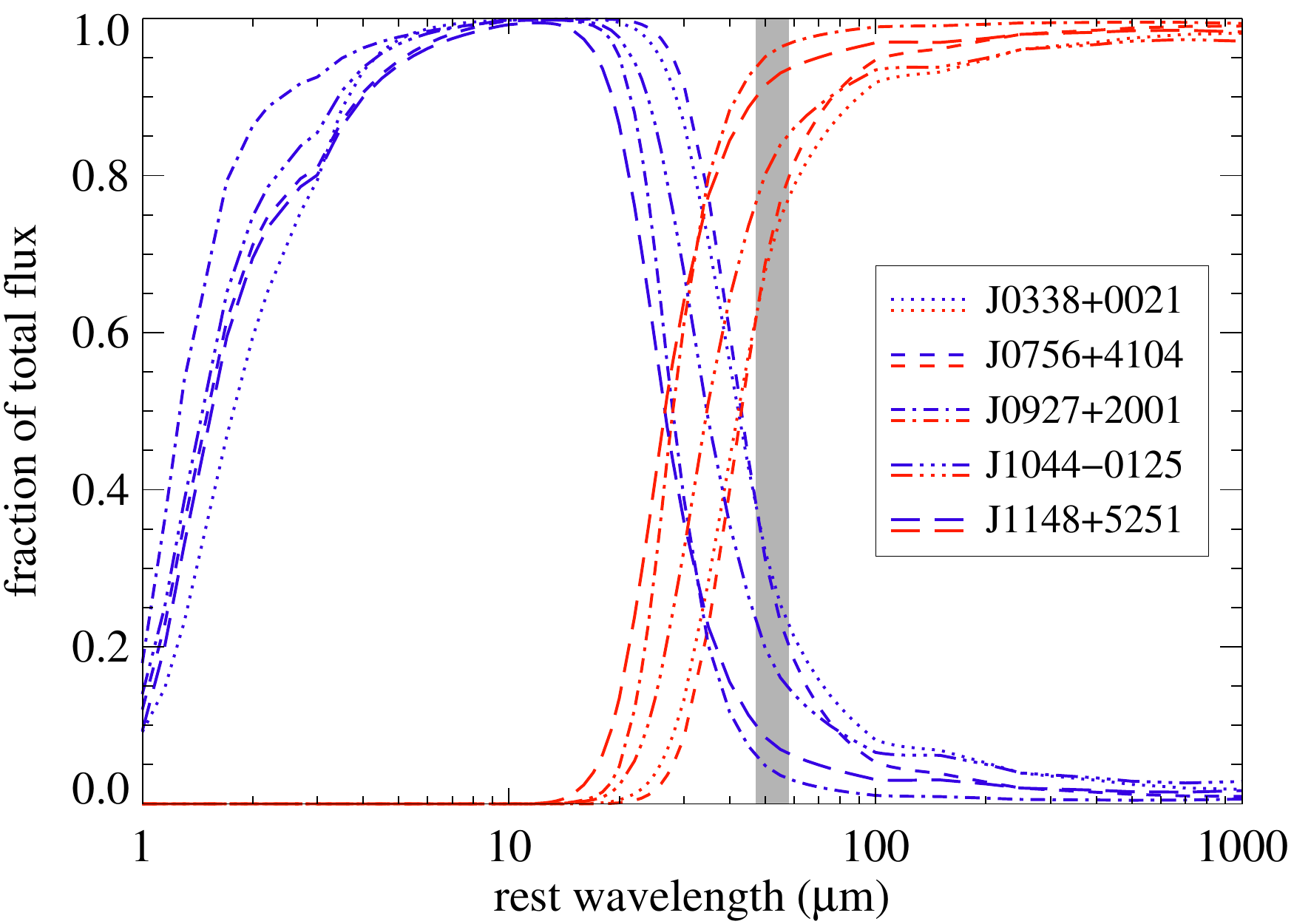}
\caption{Relative contributions of the NIR bump
  plus torus emission (presumably AGN-powered; blue) and the FIR black
  body (presumably star-formation powered ; red) 
  to the total SED fit as a function of wavelength. The vertical grey bar 
  indicates the wavelength range sampled by the observed 350\,$\mu$m band 
  for the redshifts of the sources in this plot ($z=5.0-6.4$).\label{flux_ratios}}
\end{figure}

\section{Results}

\subsection{Detection rates}

In our new {\it Herschel} observations we detect six out
of eleven sources (Tab.\,\ref{photometry}). Typically, the quasars are 
either detected in all five bands or not at all with {\it Herschel}. 
This is an 
important point which will be revisited in section \ref{sec:stacking}.
The exceptions are J0927+2001 which is detected with SPIRE but not 
with PACS and 
J1044$-$1025 for which the opposite is the case. 

From the ten objects 
observed with {\it Spitzer}, all are detected in all bands. With 
WISE at 12\,$\mu$m only three objects are detected and at low significance. 
More detailed information on the individual objects can be found in 
the Appendix. 

\subsection{The temperature of the FIR dust}
\label{sec:t_fir}

Previous studies of star formation in high-redshift ($z>5$) quasars 
often had to rely on single band mm emission as a tracer for the
starburst heated dust  \citep[e.g.][]{ber03a,wan08a}.  Far-infrared
luminosities were determined by fitting a single modified black body
to the mm photometry and integrating under this component. Since no
knowledge about the temperature of the dust was at hand for most
cases, typical values found for FIR bright quasars at lower redshift
($z$\,$\sim$\,$2-4$)  were assumed (e.g. T = $47$\,K, $\beta=1.6$;
\citealt{bee06}). This approach has been tentatively supported for some 
high-redshift quasars by ground-based observations at 350\,$\mu$m
\citep{wan08b,wan10}.

The new multi-wavelength FIR photometry now allows us to estimate the
temperature of the FIR emitting dust directly while simultaneously
accounting for the contributions from AGN heated nuclear dust to the
infrared. In Figure \ref{sed_fits_detected} we present the SEDs of the
five objects which are detected in at least two {\it Herschel}
bands and their accompanying best fits.

 As explained above, details of the SED of the AGN heated dust torus affects 
the shape (and temperature) of the FIR component. We have taken this into account 
when calculating the uncertainties of the dust temperature (see
section \ref{sec:error_estimates}). The values of 
T$_{\rm FIR}$ we obtain here for these five objects are reported in Table
\ref{results}. We clearly see a range of temperatures among these 
objects, spanning almost 20\,K. 

One important result from the SED fits in Figure
\ref{sed_fits_detected} is that  the flux at  $\lambda$$_{\rm
rest}$\,$\gtrsim$\,50\,$\mu$m is usually dominated  by the FIR excess
emission with only minor contributions from the AGN heated torus. 
  This is further illustrated in  Fig.\,\ref{flux_ratios} where we show the
fractional contribution of the hot dust plus dusty torus component
(both presumably powered by the AGN) compared with the contributions
from the FIR excess component, which may be powered by star
formation. As indicated in this figure, the SPIRE
350\,$\mu$m band is typically dominated  by emission from the FIR
black body for the redshifts considered here. In fact,  fitting only a
single modified black body ($\beta = 1.6$) to the photometry at
$\lambda$\,$\geq$\,350\,$\mu$m  gives very similar dust temperatures
compared to the full SED fits.

This result can immediately be utilized for an estimate  on the dust
temperature for objects which do not qualify for full SED fits (see
Figure \ref{sed_fits_nondetected}).  In these cases we use the
250\,GHz detection as an anchor for the (modified) black body while
the FIR upper limits at $\lambda$\,$\geq$\,350\,$\mu$m (mostly from
SPIRE)  allow us to constrain the maximum permitted temperature of
this component.  Because of our findings from the full SED fits
(Fig.\,\ref{flux_ratios}) we can assume that our upper limits to the
dust temperature are reasonably robust when limiting the fits to
$\lambda$\,$\geq$\,350\,$\mu$m, even without suitable constraints  on
the AGN dust emission in these objects. We determine dust temperatures
of $\leq$57\,K, with J0818+1722 being the only exception
(T$_{FIR}$\,$\leq$\,71\,K). However, the latter value has to be taken
with caution because the 250\,GHz photometry  of this source could be
contaminated by the emission from the nearby galaxy that is detected
in most infrared bands (see notes on individual objects in the
Appendix and Figure \ref{images}).


\subsection{The emissivity index $\beta$}
\label{betafit}

 In the previous fits, the emissivity index $\beta$ was fixed to
a value of 1.6 to enable the comparison with earlier
literature studies. The good photometric coverage in the FIR and 
sub-mm for some of our objects now allows us to explore in how far $\beta$ 
can be constrained using the quasar SEDs. For such a
study, additional photometry at lower frequencies (typically around
90\,GHz in the observed frame) is very important as it helps
to further constrain the Rayleigh-Jeans tail of the fitted dust
component. In fact, considering only photometry at
$\nu_{\rm obs}$\,$>$\,250\,GHz does not provide good constraints on $\beta$ (or
the FIR dust temperature) if both parameters are kept free during
fitting. Photometry at $\sim$\,90\,GHz in combination with {\it
Herschel} FIR detections is available for four objects in our
sample. Re-fitting the SEDs as outlined  previously but now keeping
$\beta$ as a free parameter, we find relatively  high $\beta$ values
($2.0-2.7$) combined with relatively low temperatures  ($\sim$\,33\,K,
but still 54\,K for J1148+5251). The integrated luminosity  of the FIR
dust when determined from these new fits remains virtually unchanged
as compared to a fixed $\beta$ approach. We caution, however, that a
reliable measure of $\beta$ is hard to obtain in these objects because
the peak of the dust emission is not well defined (or isolated) in the
SED due to the strong nuclear dust emission from the torus. Therefore,
the peak wavelength (and temperature) of the FIR black body depends on
the choice of the torus model which adds additional uncertainty
in the determination of $\beta$.


\begin{table*}[t]
\begin{center}
\caption{Physical parameters derived from the fitted components.\label{results}}
\begin{tabular}{lcccccccc}
\tableline\tableline
name                          & $\alpha$$_{\rm UV/opt}$ & L$_{\rm UV/opt}$        & L$_{\rm NIR/MIR}$ & T$_{\rm FIR}$        & L$_{\rm FIR}$          & \% of & SFR                   & M$_{\rm dust}$ \\
                              &                        & $10^{46}$erg\,s$^{-1}$ & $10^{46}$erg\,s$^{-1}$    & K                   & $10^{13}$\,L$_{\odot}$ & L$_{\rm FIR}$ & $10^3$\,M$_{\odot}$yr$^{-1}$  & $10^{8}$\,M$_{\odot}$ \\
(1)                           & (2)                    & (3)            & (4)              & (5)                 & (6)                   & (7)     & (8)                  & (9)      \\
\tableline                       
J0203+0012                    & $-0.16 \pm 0.02$ & $ 9.6 \pm 0.2$ & \hspace*{-3mm}$<$10.5  &\hspace*{-2mm} $<$57 &  \hspace*{-3mm}$<$1.3 & \nodata & \hspace*{-3mm}$<$2.2 & \nodata \\
J0338+0021                    & $-0.39 \pm 0.03$ & $ 4.8 \pm 0.3$ & $18.1 \pm 0.9$        &          $47 \pm 4$ &         $1.1 \pm 0.4$ & 24      & $1.8 \pm 0.6$        & $6.8 \pm 2.0$ \\
J0756+4104                    & $-0.42 \pm 0.03$ & $ 4.1 \pm 0.3$ & $10.3 \pm 0.3$        &          $40 \pm 2$ &         $1.1 \pm 0.2$ & 35      & $1.9 \pm 0.3$        & \hspace*{-1.4mm}$15.9 \pm 1.6$ \\
J0818+1722\tablenotemark{a}   & $-0.68 \pm 0.02$ & $12.5 \pm 0.2$ & \hspace*{-3mm}$<$8.3  & \hspace*{-2mm}$<$71 &  \hspace*{-3mm}$<$1.8 & \nodata & \hspace*{-3mm}$<$3.1 & \nodata \\
J0840+5624                    & $-0.31 \pm 0.02$ & $ 6.1 \pm 0.2$ & \hspace*{-3mm}$<$8.8  & \hspace*{-2mm}$<$45 &  \hspace*{-3mm}$<$0.9 & \nodata & \hspace*{-3mm}$<$1.6 & \nodata \\
J0927+2001                    &  $0.00 \pm 0.03$ & $ 5.5 \pm 0.2$ & \hspace*{-3mm}$<$7.3  &         $ 50 \pm 2$ &        $ 1.3 \pm 0.2$ & 62      & $ 2.1 \pm 0.3$       & $5.3 \pm 0.4$ \\
J1044$-$0125\tablenotemark{b} & $-0.33 \pm 0.03$ & $ 9.9 \pm 0.3$ & $19.3 \pm 0.5$        & \hspace*{-2mm}$<$53 &  \hspace*{-3mm}$<$1.2 & $<$24   & \hspace*{-3mm}$<$2.1 & \nodata \\
J1048+4637                    & $-0.33 \pm 0.02$ & $12.4 \pm 0.2$ & \hspace*{-3mm}$<$11.5 & \hspace*{-2mm}$<$56 &  \hspace*{-3mm}$<$1.6 & \nodata & \hspace*{-3mm}$<$2.7 & \nodata \\
J1148+5251                    & $-0.35 \pm 0.03$ & $15.3 \pm 0.2$ & $16.6 \pm 0.9$        &          $59 \pm 3$ &         $2.7 \pm 0.3$ & 54      & $4.6 \pm 0.5 $       & $4.7 \pm 0.6$ \\
J1335+3533                    & $-0.33 \pm 0.02$ & $ 6.3 \pm 0.2$ & \hspace*{-3mm}$<$7.3  & \hspace*{-2mm}$<$55 &  \hspace*{-3mm}$<$1.4 & \nodata & \hspace*{-3mm}$<$2.4 & \nodata \\
J2054$-$0005                  & $-0.22 \pm 0.02$ & $ 3.0 \pm 0.2$ & \hspace*{-3mm}$<$22.5              & \hspace*{-2mm}$<$54 &  \hspace*{-3mm}$<$1.3 & \nodata & \hspace*{-3mm}$<$2.2 & \nodata \\
\tableline
\end{tabular}
\tablecomments{(2) Power-law slope in the UV/optical ($F_{\nu}\sim{\nu}^{\alpha}$); 
(3) Integrated luminosity between 0.1 and 1\,$\mu$m of the power-law component;
(4) Integrated luminosity between 1 and 1000\,$\mu$m of (presumbly) AGN powered dust emission (NIR black body and 
torus component combined); 
(5) Temperature of the modified black body fitted in the FIR;
(6) Integrated luminosity between 8 and 1000\,$\mu$m of the star-fromation powered FIR component; 
(7) Percentage contribution of star formation to total FIR luminosity between 8 and 1000\,$\mu$m; (8) 
Star-formation rate derived from L$_{\rm FIR,SF}$ using \citet{ken98}; (9) Dust mass derived from the 
star-formation powered FIR component following equation 1.}
\end{center}
\end{table*}

\begin{figure*}[t!]
\centering
\includegraphics[angle=0,scale=.33]{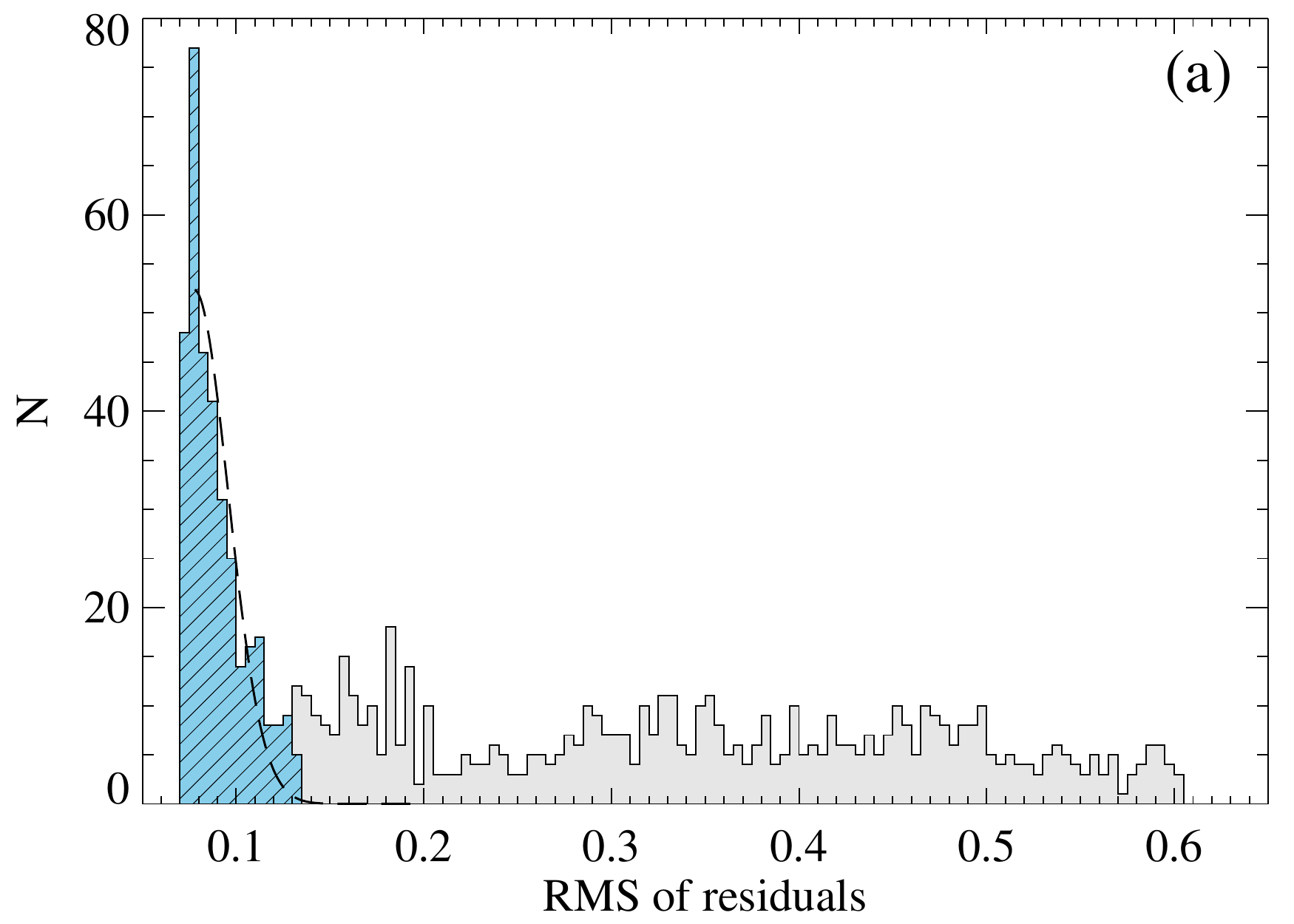}
\includegraphics[angle=0,scale=.33]{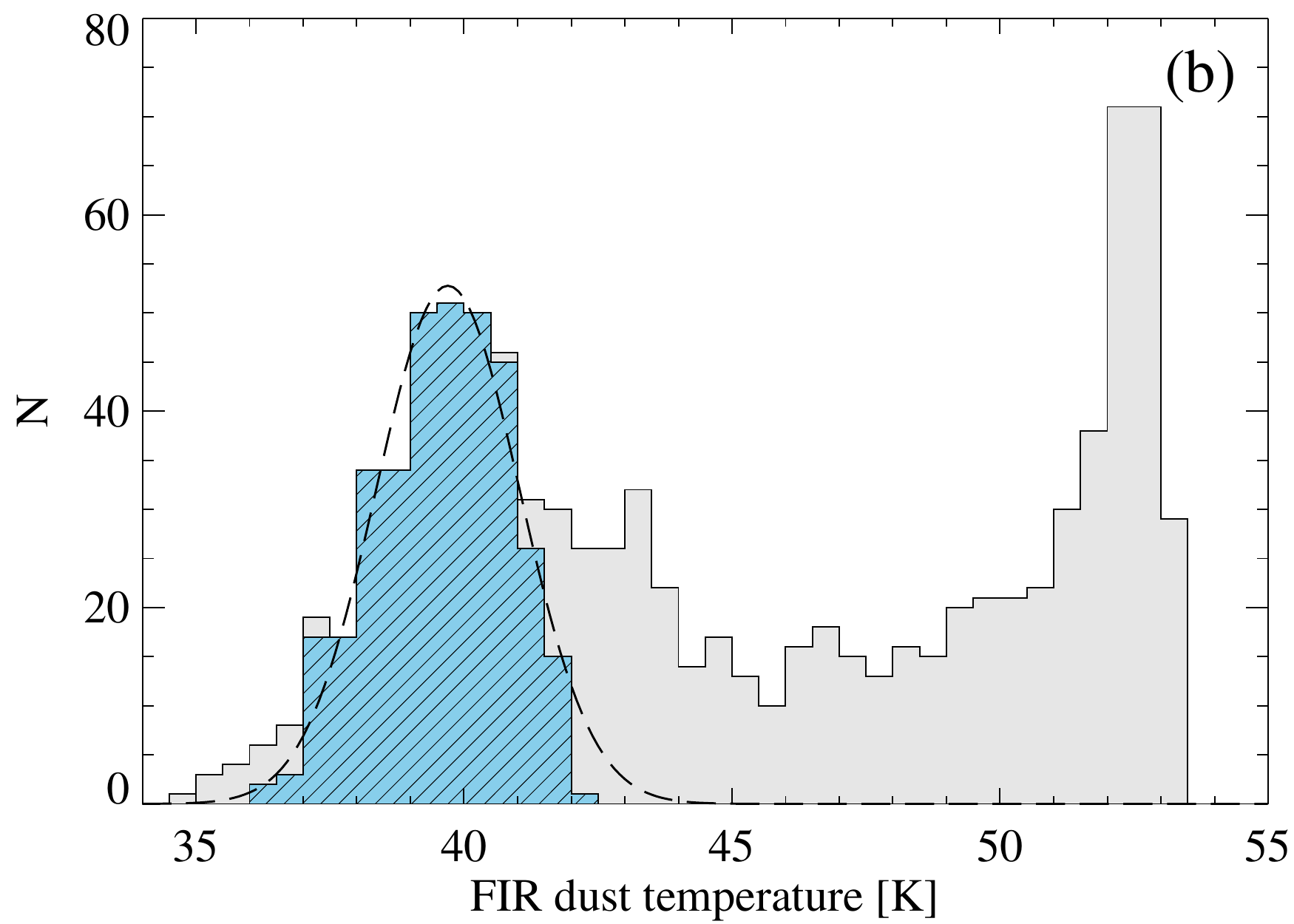}
\includegraphics[angle=0,scale=.33]{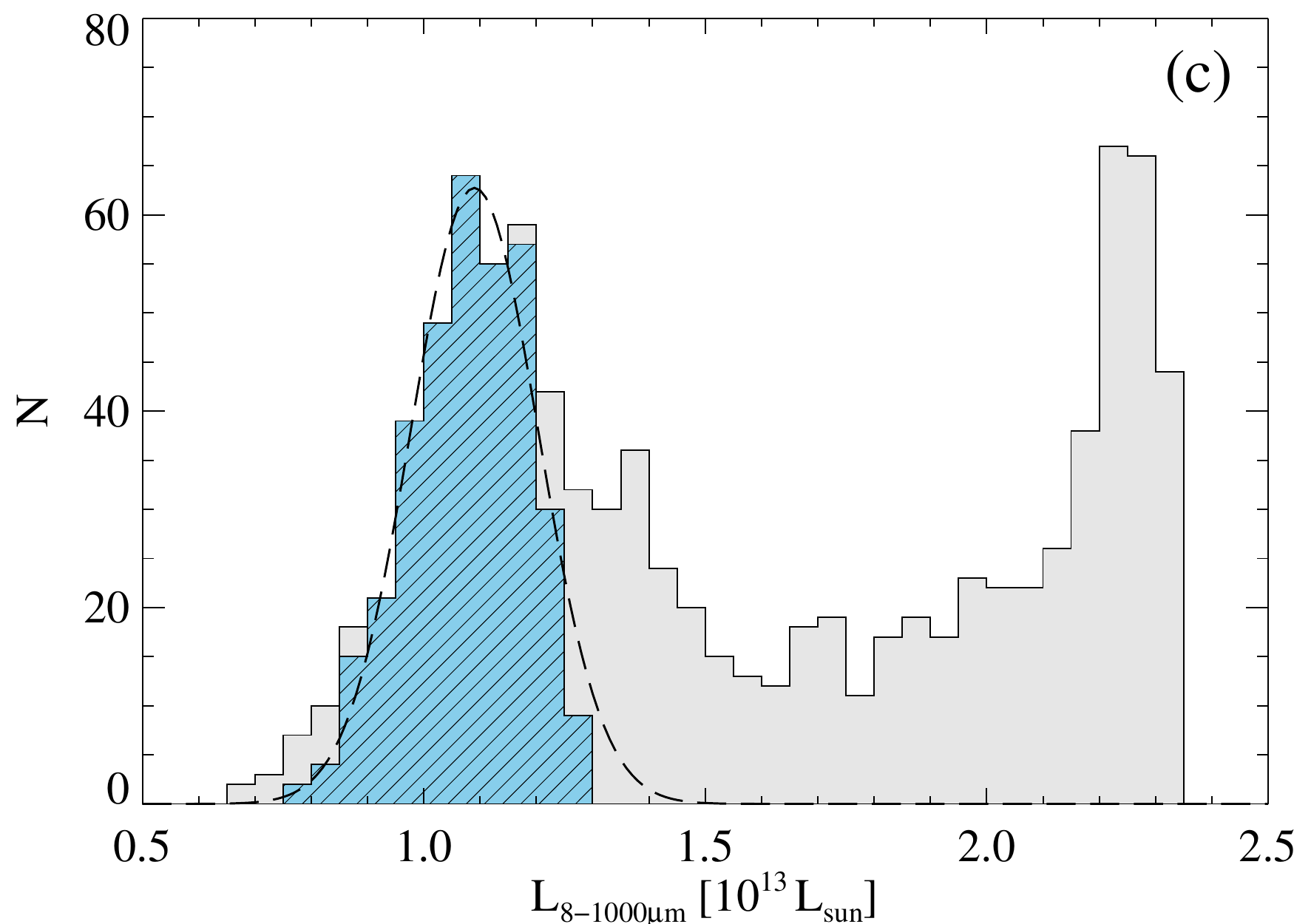}\\
\includegraphics[angle=0,scale=.33]{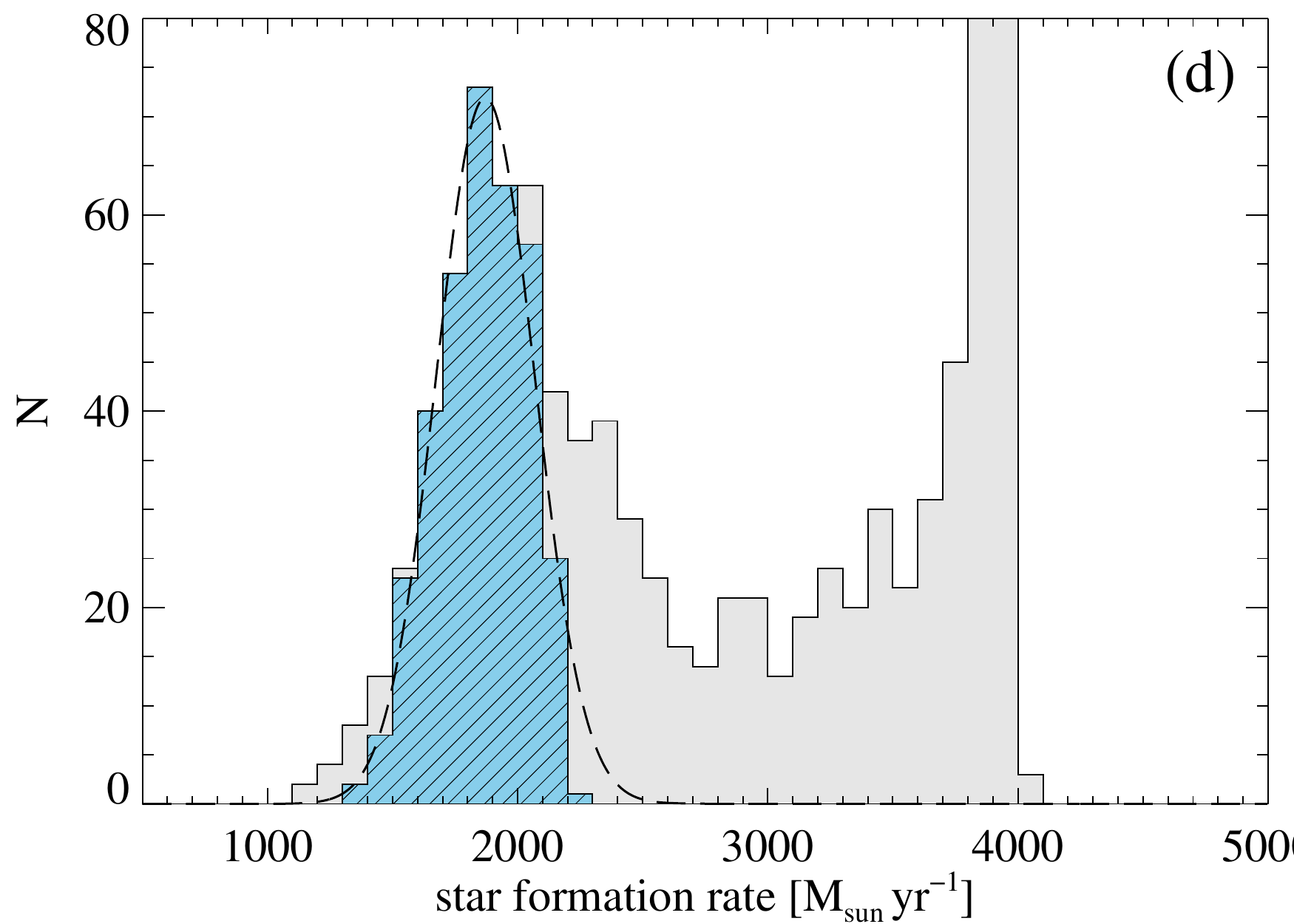}
\includegraphics[angle=0,scale=.33]{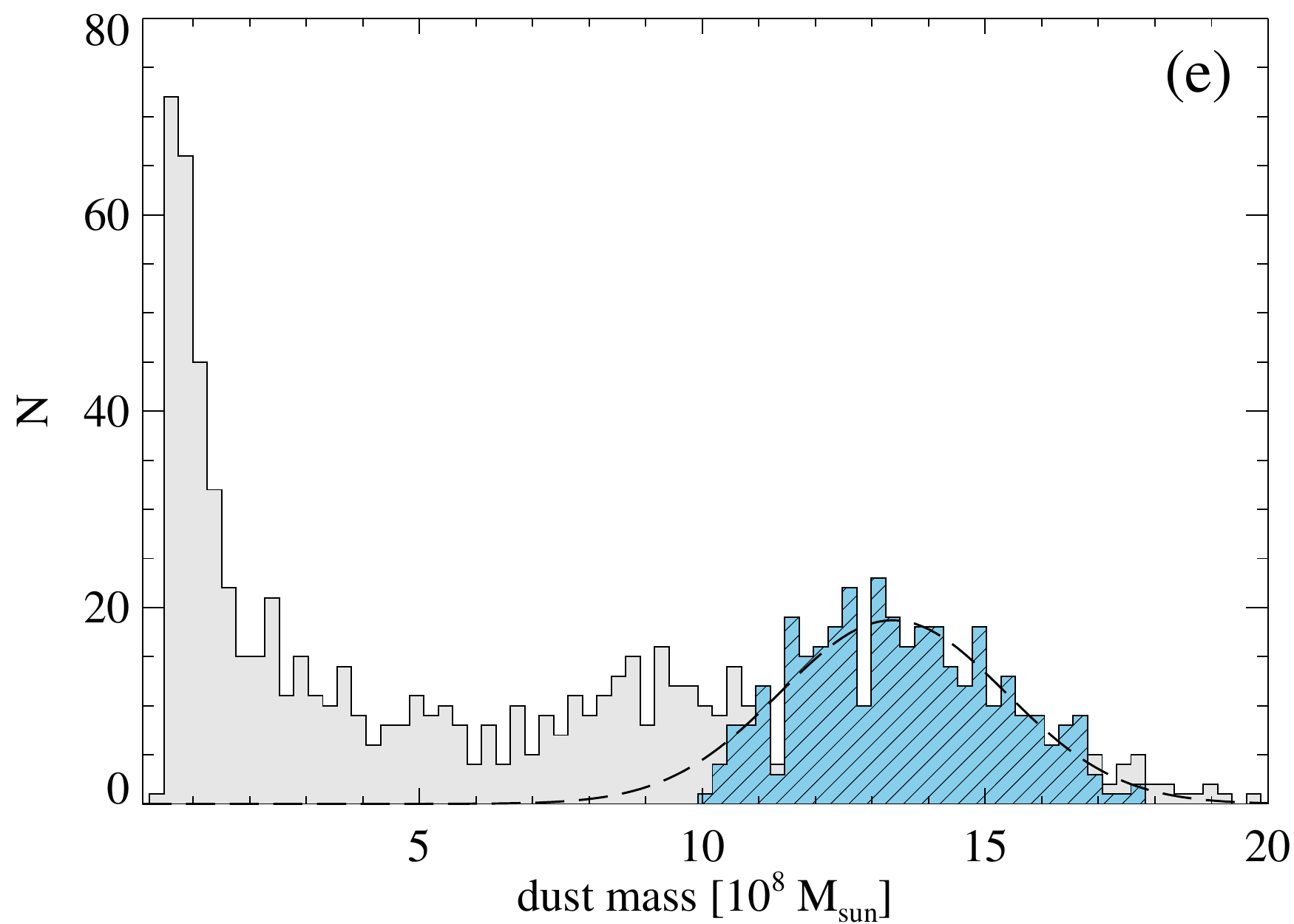}
\caption{Histograms of derived values from the SED fits of the quasar J0756+4104. 
The grey shaded 
area in each panel shows the N\,$\sim$\,1000 solutions from all fits, while 
the blue and hashed region identifies values based on acceptable fits (see text). 
The dashed lines in each panel represent Gaussian fits to the data from the 
acceptable fits. Panels (d) and (e) rely on the assumption that the FIR emission 
is powered by star formation.\label{uncertainties} }
\end{figure*}

\subsection{Far-infrared luminosities, star-formation rates and dust masses}
\label{sec:l_fir}

Integrating the emission of the FIR excess component only
(i.e. the modified black body) between 8 and 1000\,$\mu$m in the rest
frame allows us to determine its luminosity L$_{\rm FIR}$. 
The FIR luminosities we derive are of the order of
$\sim$\,10$^{13}$\,L$_{\odot}$ (see Tab.\,\ref{results}), which 
corresponds to $\sim$25-60\% of the bolometric FIR luminosity in these 
objects. For sources with  full SED fits, the values for L$_{\rm FIR}$ do 
not change significantly ($\lesssim 10$\%) if we fit only a single 
modified black body to the photometry at 
$\lambda_{\rm obs}$\,$\geq$\,350\,$\mu$m.

Assuming that the FIR emission is powered by star formation, we converted 
the FIR luminosities into star-formation rates using the formula 
in \citet{ken98}. The results are reported in Table \ref{results}. 
Using our fits of the FIR excess component we can also derive an 
estimate for the dust masses in the star forming regions:

\begin{equation}
M_{dust} = \frac{S_{250{\mu}m} D_{L}^2}{\kappa_{250{\mu}m} B_{\nu}(250{\mu}m,T_{FIR})}
\end{equation}

where $S_{250{\mu}m}$ is flux level at a rest frame wavelength
of 250\,$\mu$m as determined from the {\it fit}, $ D_{L}$ is the
luminosity distance,  $\kappa_{250{\mu}m}$ is the dust absorption
coefficient at 250\,${\mu}$m as determined from the models of
\citet{dra03}, and $B_{\nu}(250{\mu}m,T_{FIR})$ is the value of the Planck
function with temperature $T_{FIR}$ at a wavelength of 250\,${\mu}$m.
The results are also reported in Table\,\ref{results}.

\subsection{Error estimates on physical parameters}\label{sec:error_estimates}

In order to estimate uncertainties in the derived parameters we 
studied the distribution of their values resulting from 
all the fitted models  (N\,$\sim$\,1000; as outlined in section 
\ref{sec:fitting} we only include torus models with  inclinations of 
$\leq$45 degrees). This also allows us to account for the influence the choice 
of a particluar torus model has on these parameters. 

As a first step, we calculated for all the fitted models the residuals 
between the global fit and the observed data in the infrared  
($\lambda_{\rm rest}>1\,\mu$m) and determined the error-weighted RMS for 
these points.  A typical
distribution of the resulting RMS values is presented in panel '(a)'
of  Figure \ref{uncertainties} (we here use the quasar J0756+4104 as 
an example to demonstrate our approach). In this
figure we see a clear peak at low RMS values representing a family of
good fits, with an extended tail to large RMS values corresponding to
increasingly worse model representations of the observed SED. We then fitted 
the right side of the RMS peak with a Gaussian (dashed line). All fits with 
an RMS value within 3$\sigma$ of the centroid value of the Gaussian are 
identified as acceptable model fits. The values corresponding to these 
fits are marked as blue and hashed regions in all panels of Figure 
\ref{uncertainties} and are used for estimating uncertainties on 
the derived values.

Each of the N\,$\sim$\,1000 model fits provides a value for the 
temperature of the modified black body in the FIR (T$_{\rm FIR}$, 
panel '(b)' in Fig.\,\ref{uncertainties}). From the temperature and the
normalization of this component we can then calculate 
(see section \ref{sec:l_fir}) L$_{\rm FIR}$
(panel '(c)'), a star-formation rate  (panel '(d)'), and a dust
mass M$_{\rm dust}$ (panel '(e)'). In each histogram we fit the 
parameter values obtained from the acceptable fits (as determined from the 
residual RMS distribution; blue and hashed regions) with a Gaussian 
(dashed line). This gaussian fit provides us with a mean parameter value 
(centroid) and an uncertainty ($\sigma$) for each fitted object. These 
results are reported in Tab.\,\ref{results}.

\begin{figure*}[t!]
\centering
\includegraphics[angle=0,scale=.18]{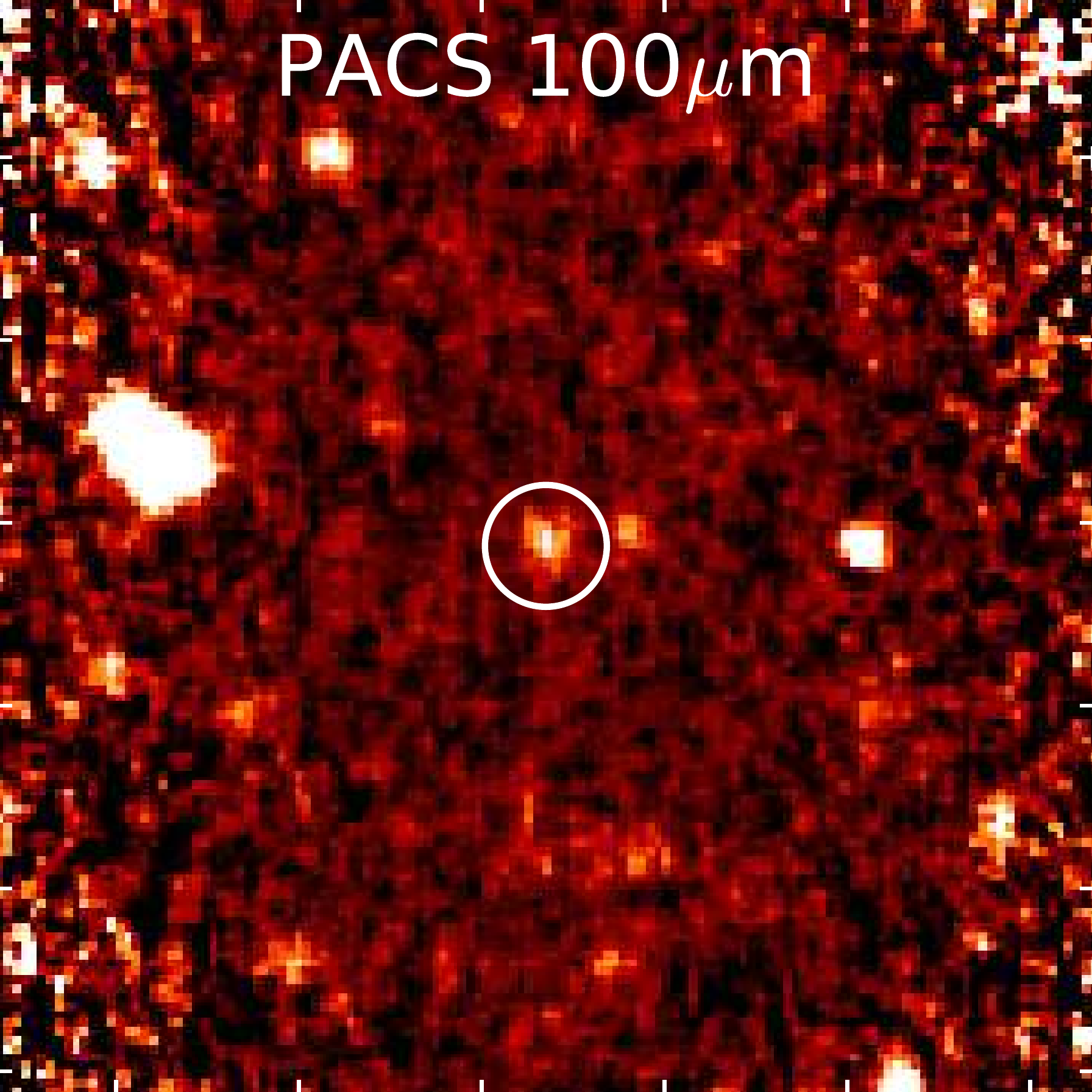}
\includegraphics[angle=0,scale=.18]{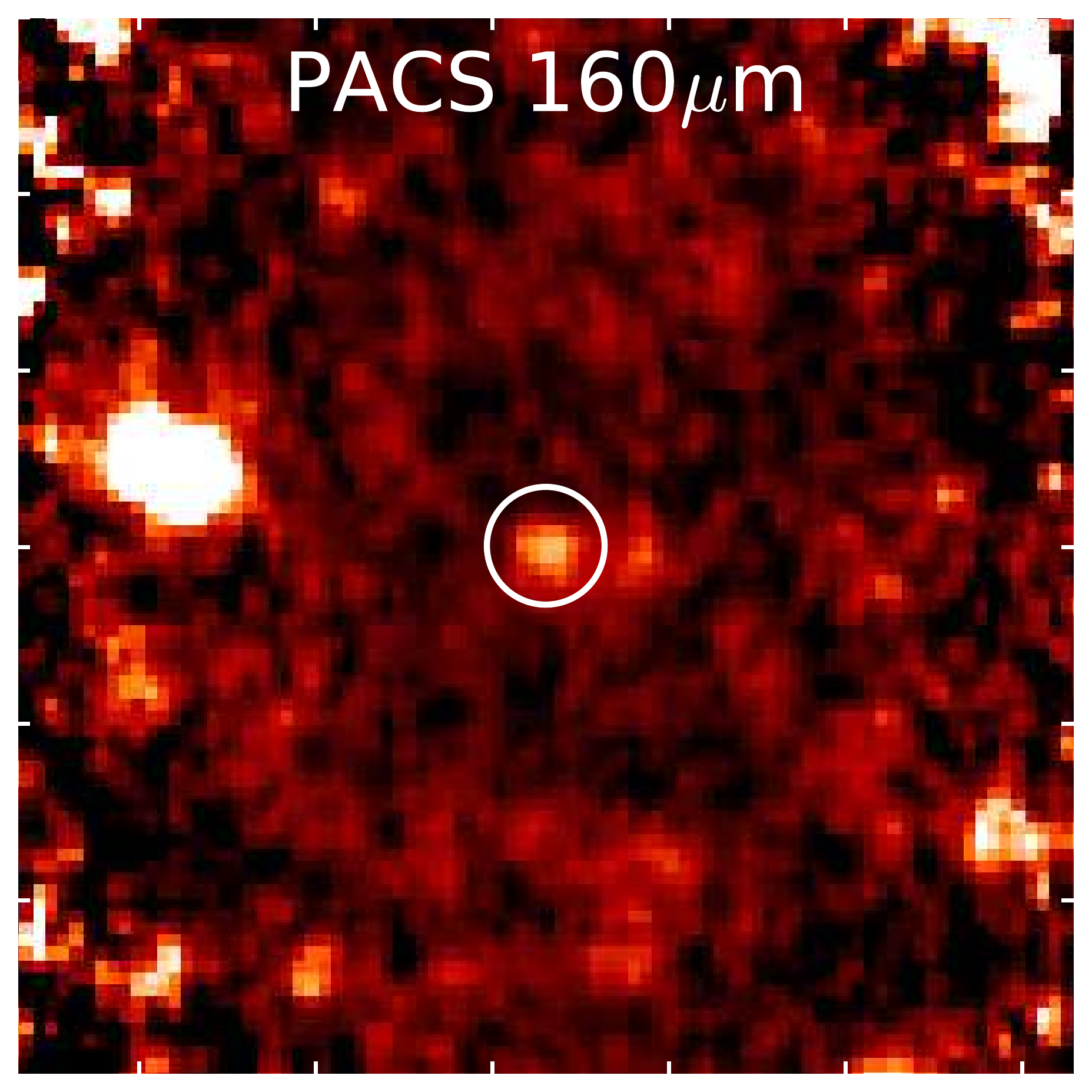}
\includegraphics[angle=0,scale=.18]{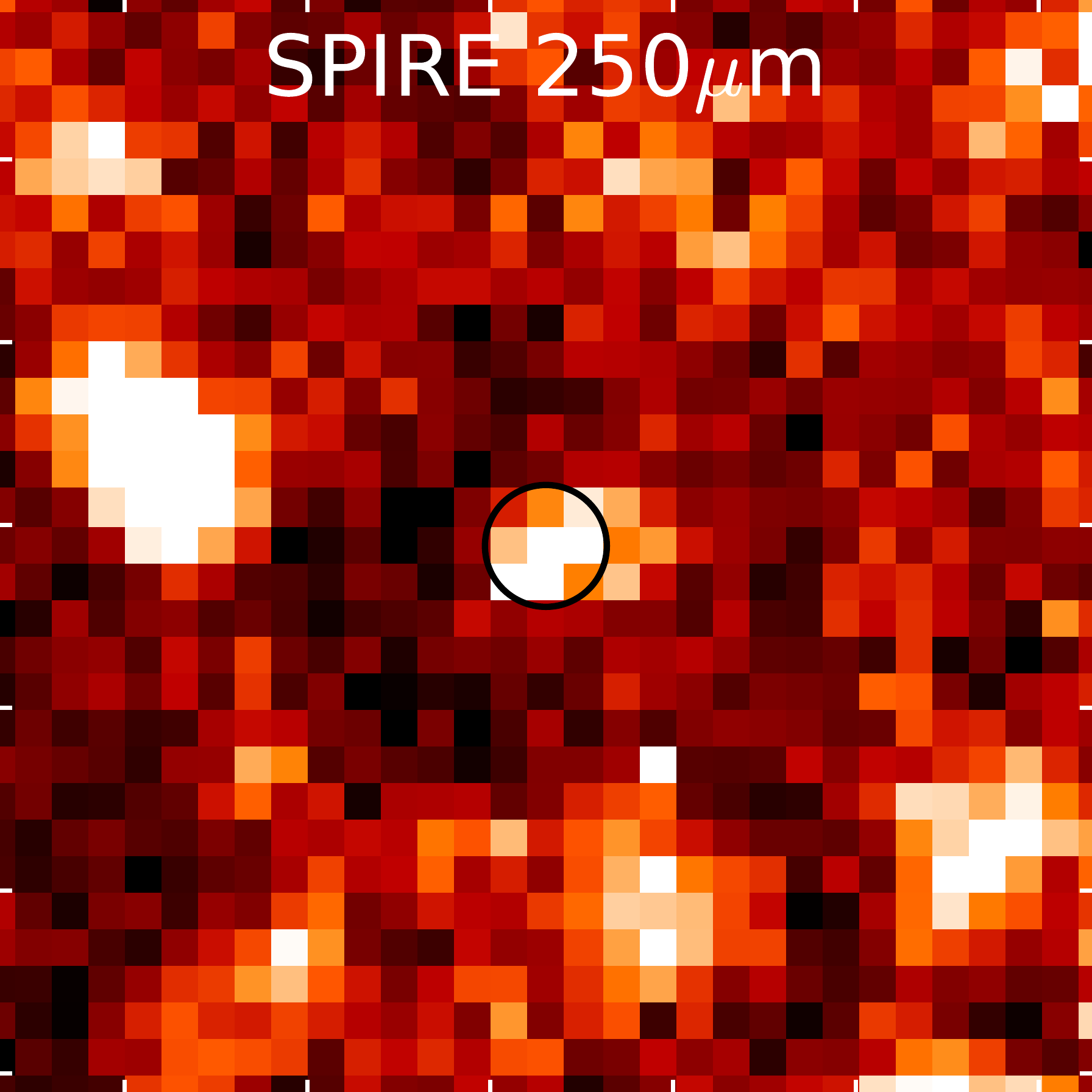}
\includegraphics[angle=0,scale=.18]{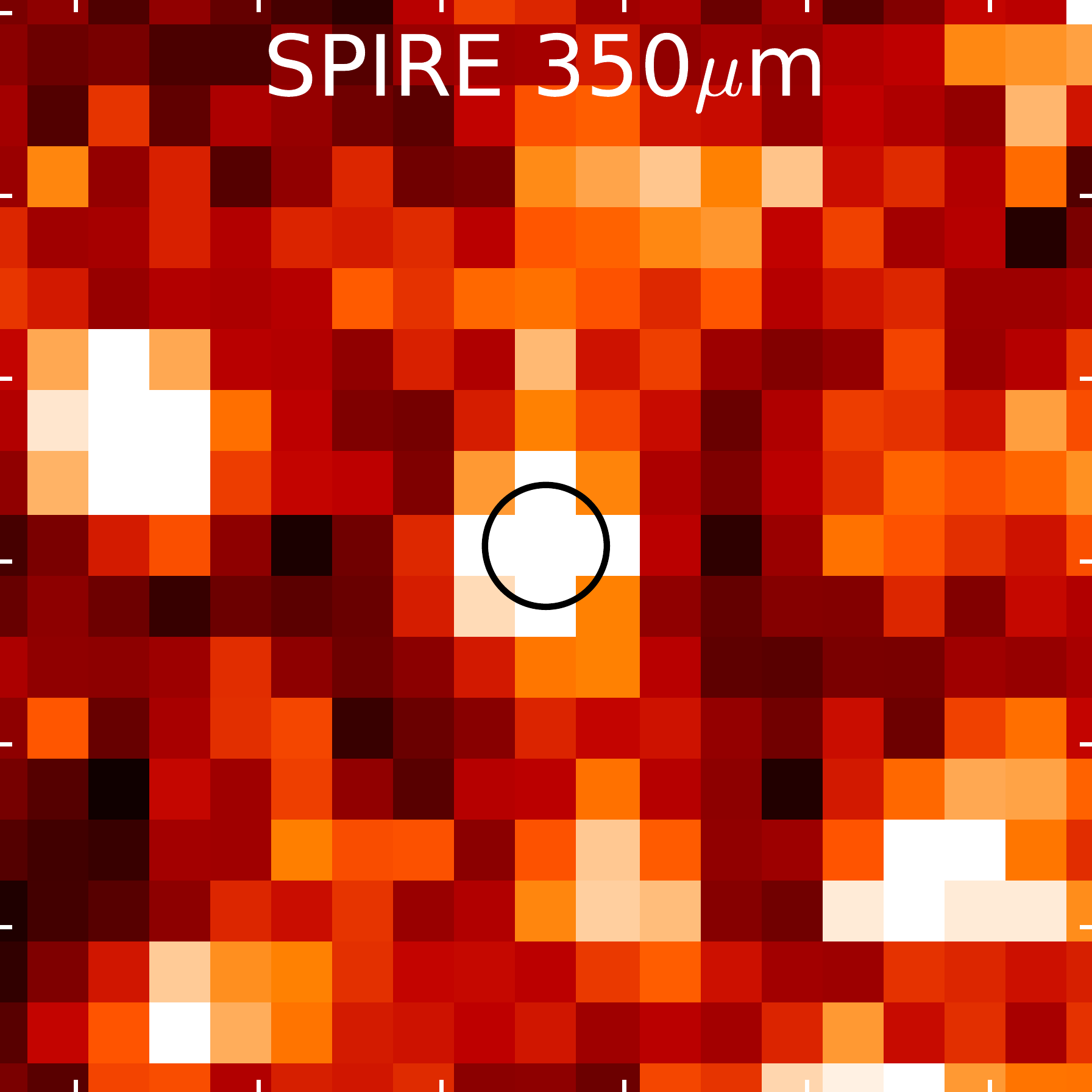}
\includegraphics[angle=0,scale=.18]{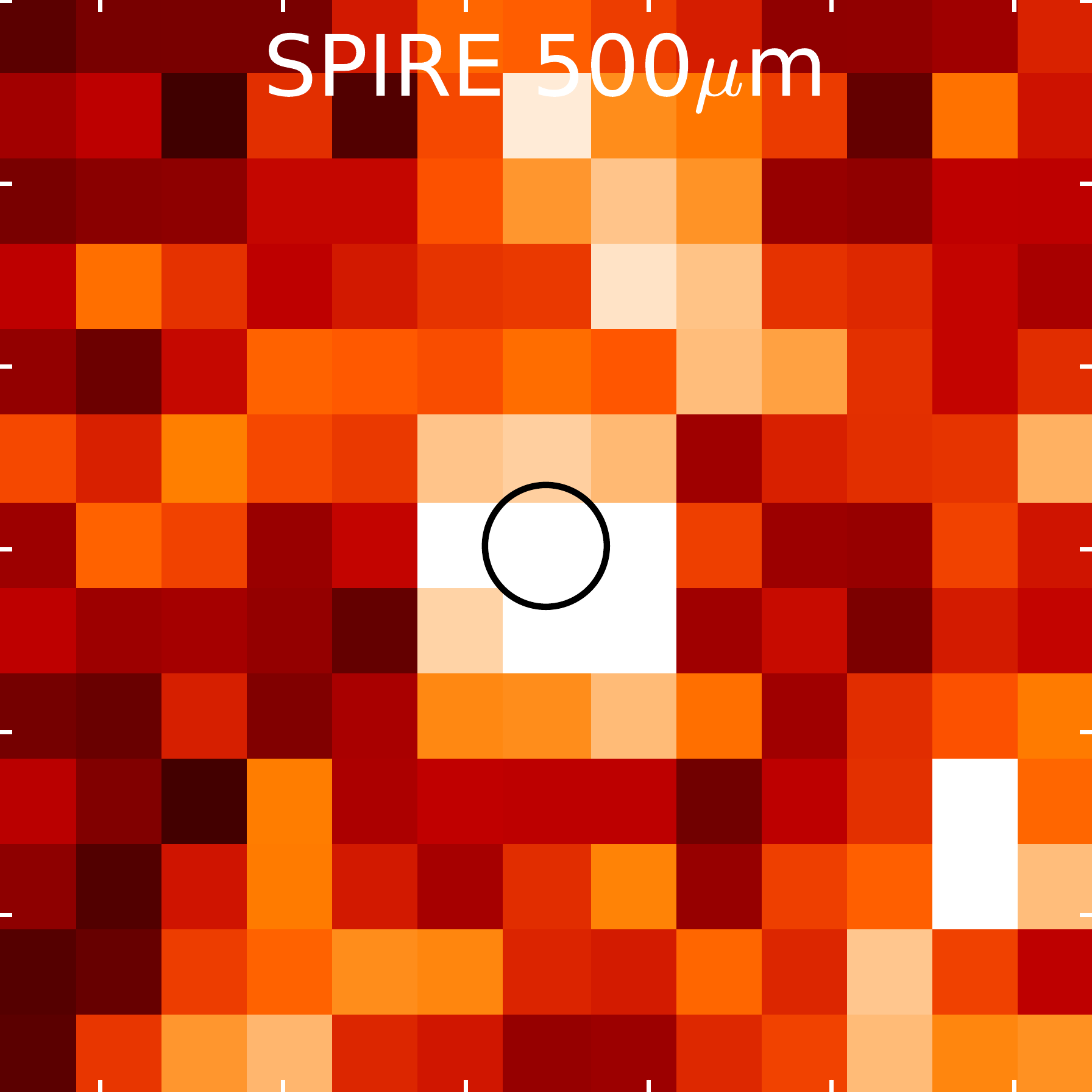}\\
\includegraphics[angle=0,scale=.18]{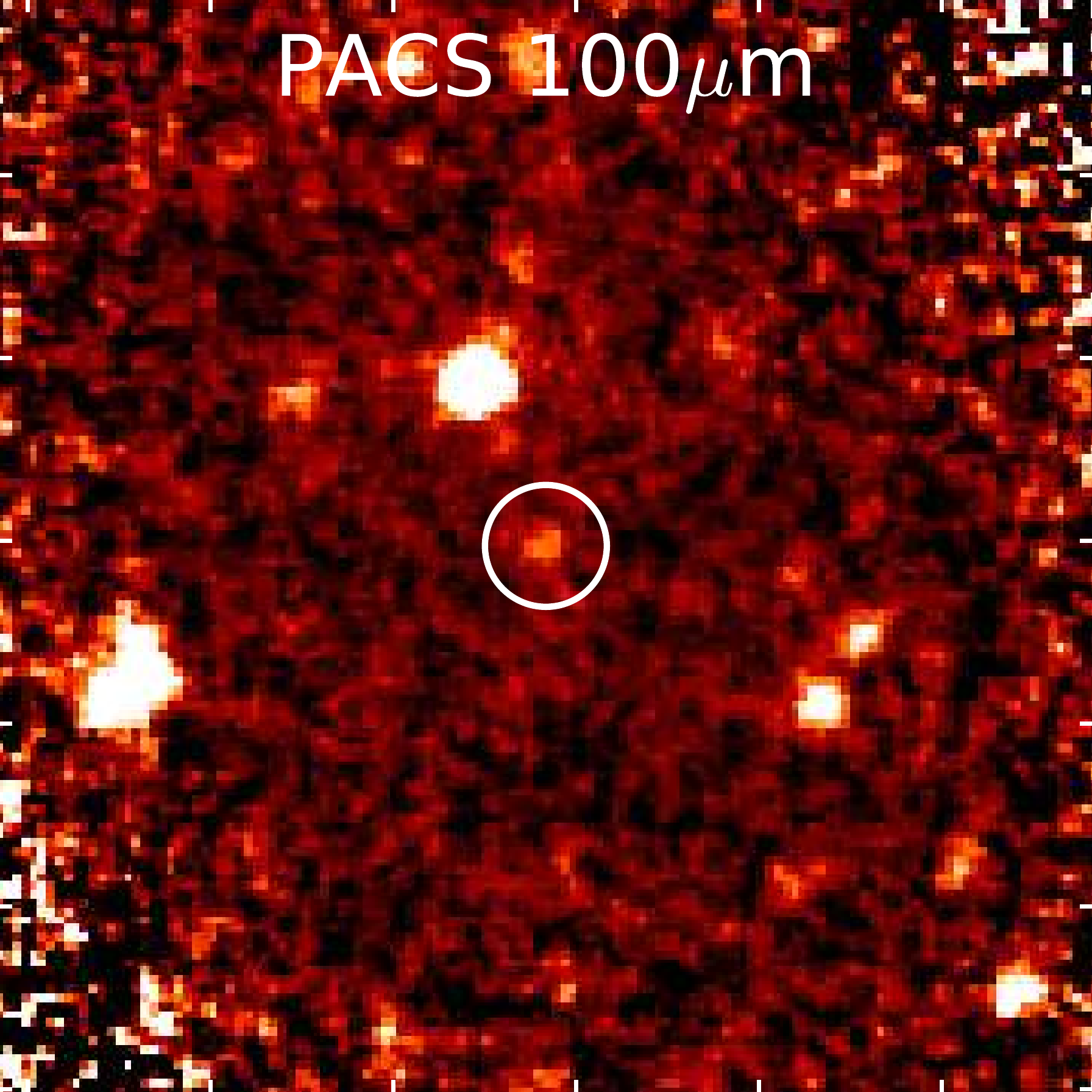}
\includegraphics[angle=0,scale=.18]{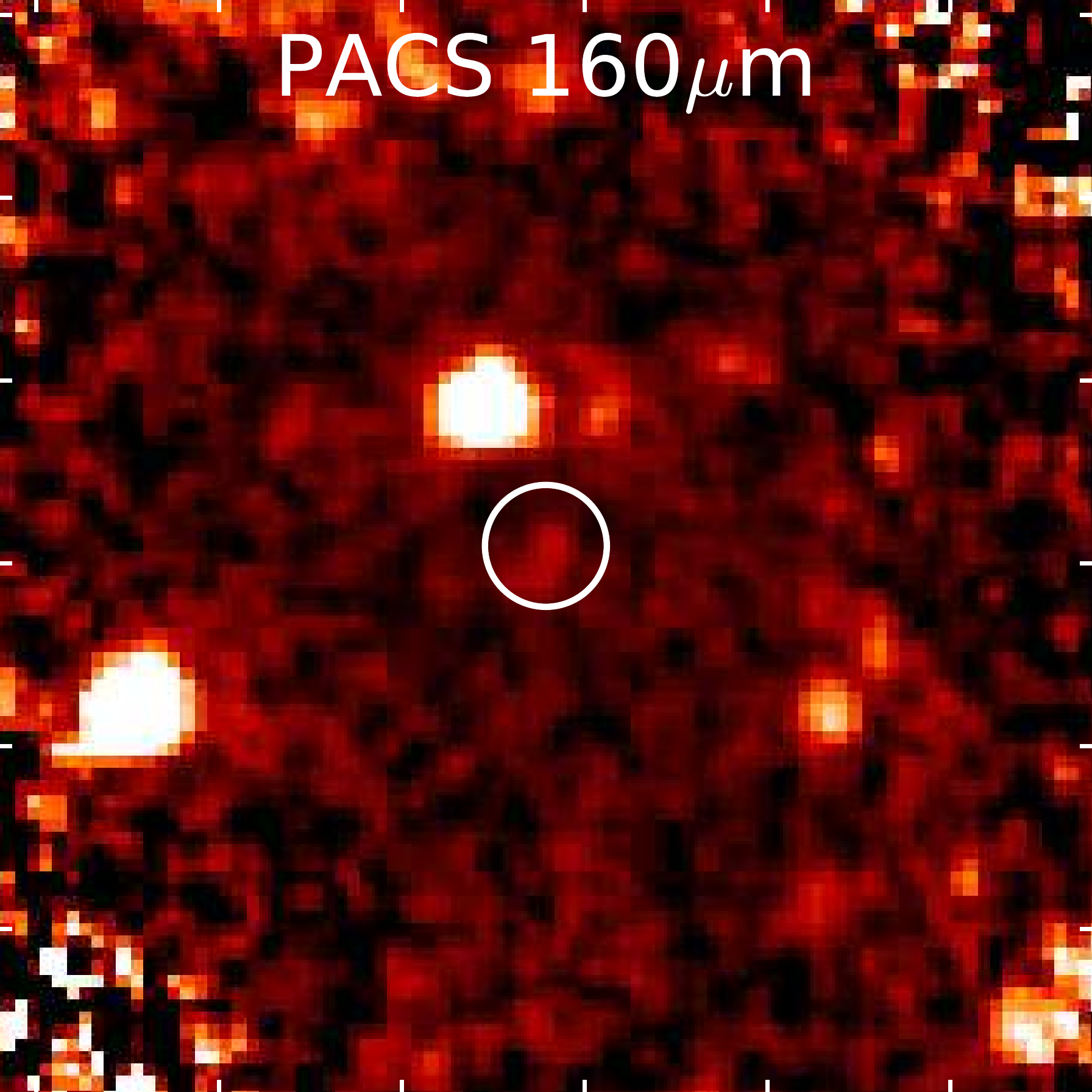}
\includegraphics[angle=0,scale=.18]{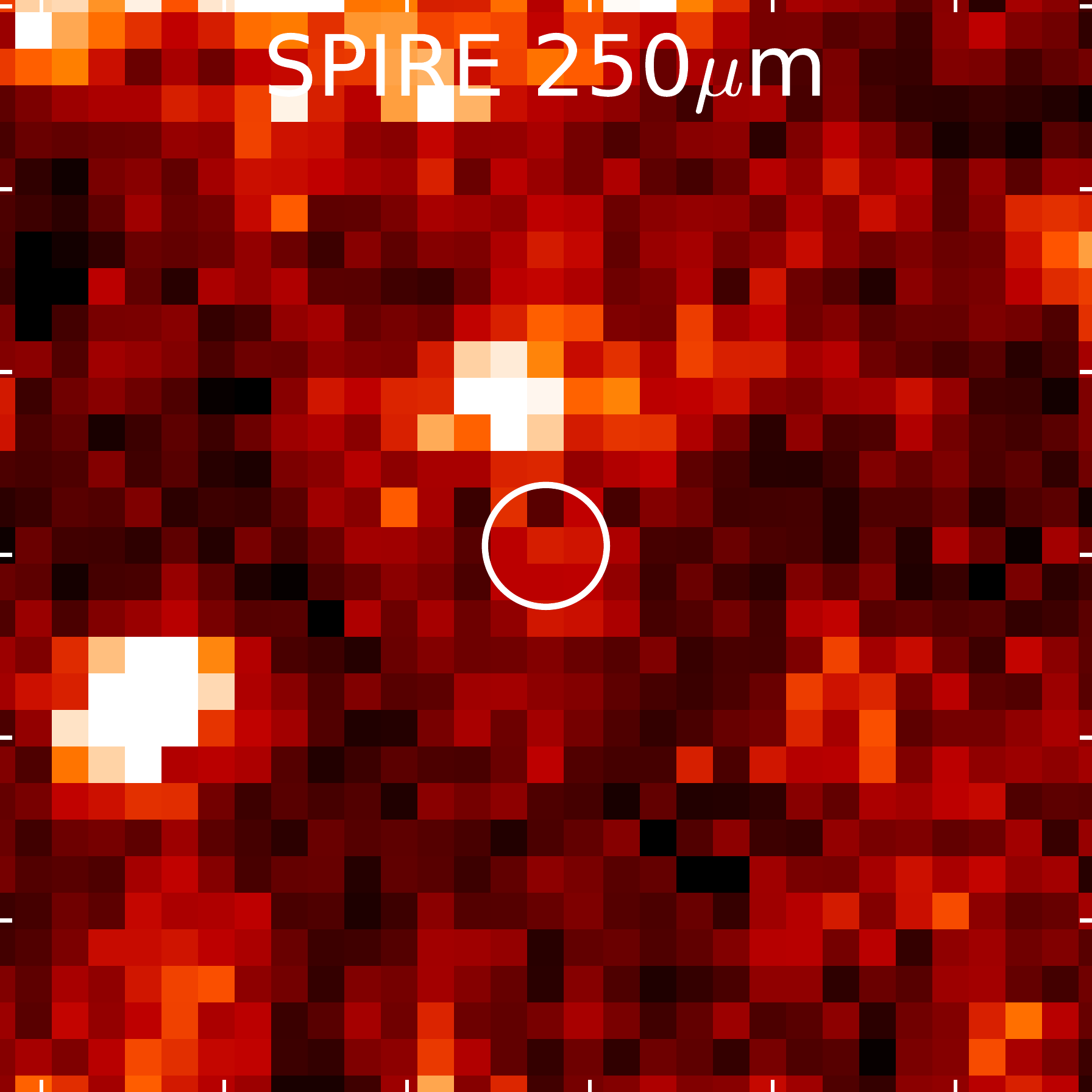}
\includegraphics[angle=0,scale=.18]{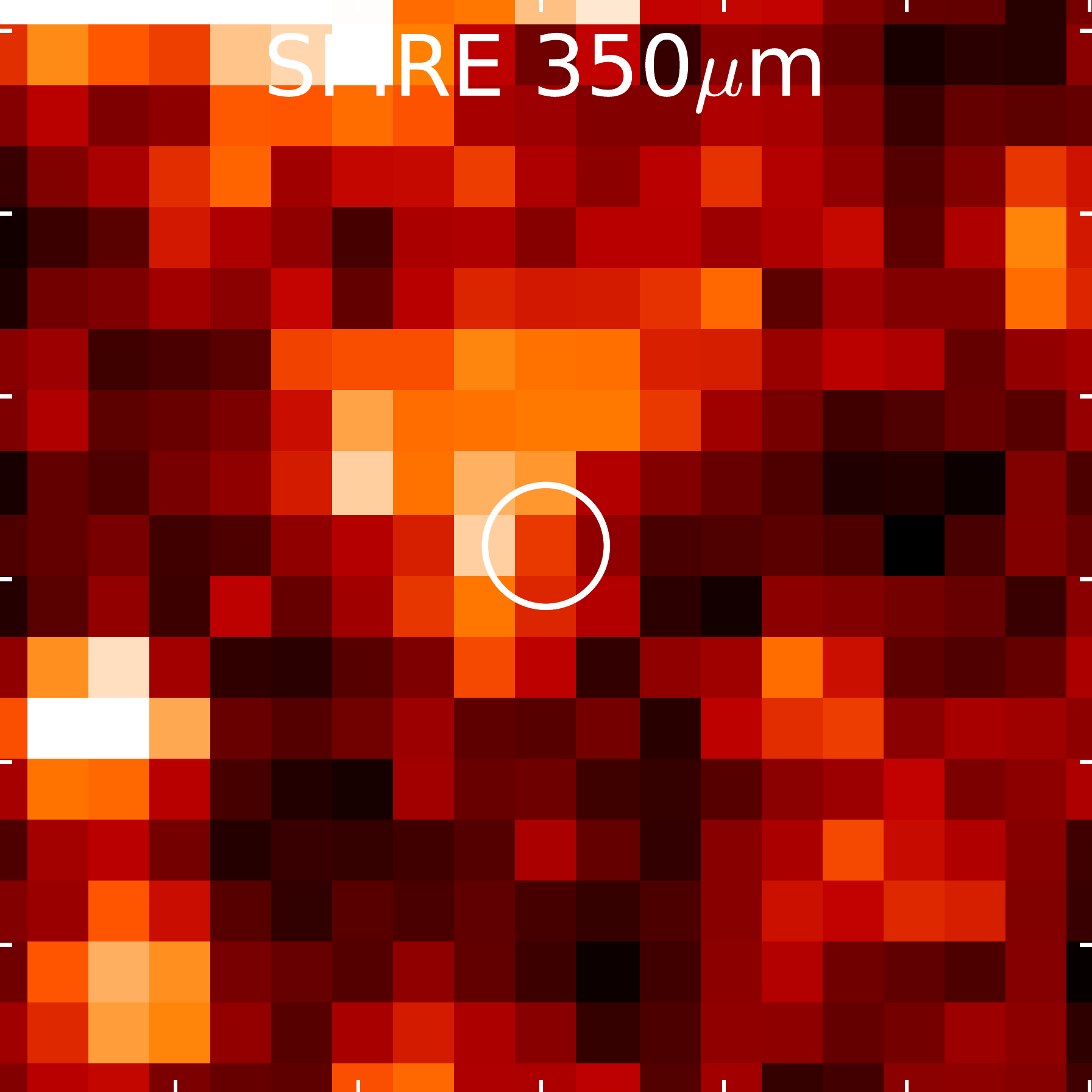}
\includegraphics[angle=0,scale=.18]{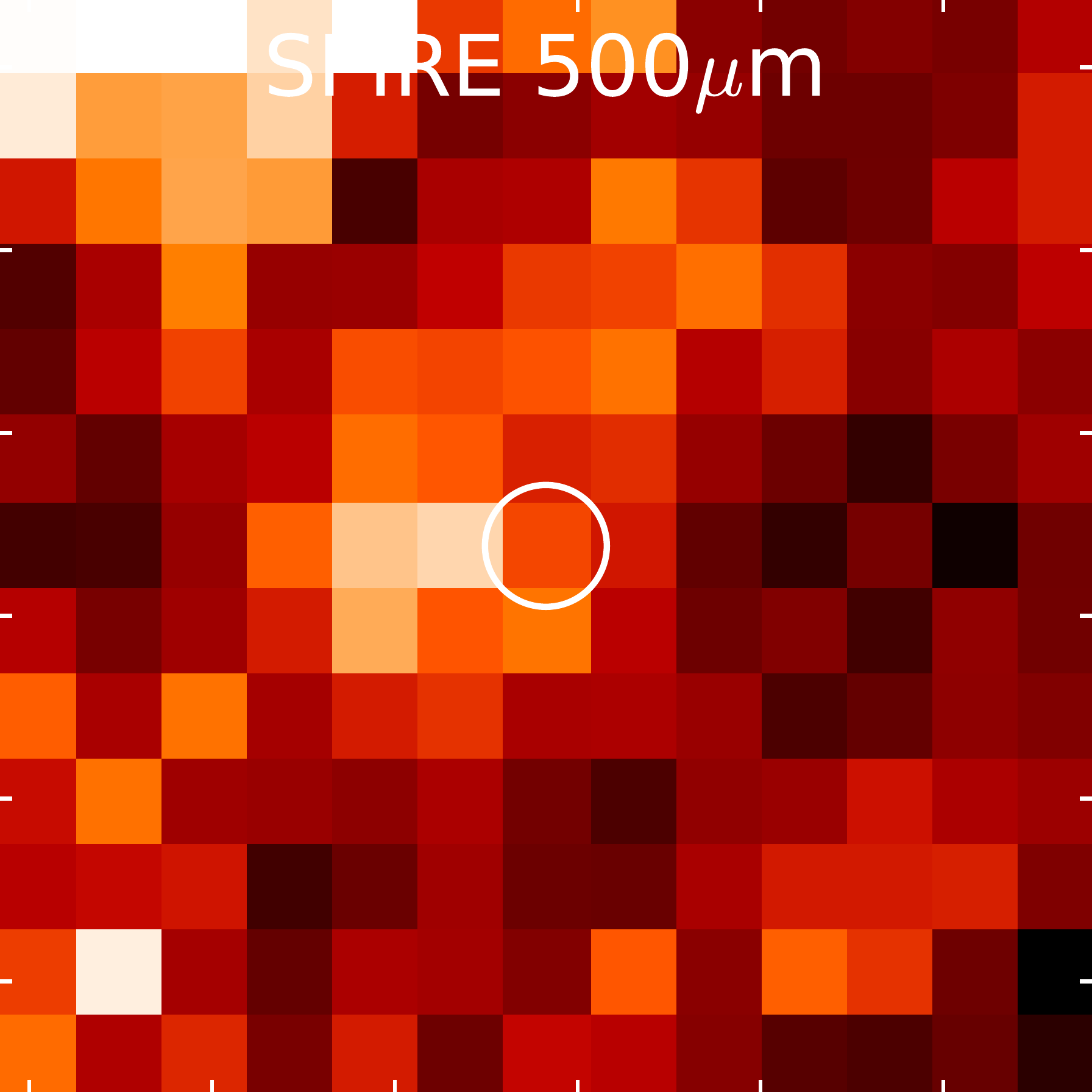}
\caption{Stacked {\it Herschel} images of sources with individual 
SPIRE detecions (top row) and without individual SPIRE detections 
(bottom row). The images are 3\arcmin~on a side and the the circle 
indicates the center of the stack and has a diameter of 20\arcsec.
\label{stacked_images} }
\end{figure*}

\section{Discussion}

\subsection{Comparison with previous studies}

The average temperature of the modified black body used to model the
FIR emission is comparable to the $\sim$\,47\,K measured
for lower redshift FIR bright quasars \citep{bee06}. This finding 
is confirmed  by the average SEDs presented in section
\ref{sec:stacking} below.

While this is true on average, we see a significant spread in dust
temperature ($\sim$\,20\,K) between individual objects, even for
comparable FIR luminosities.  Despite the low number of objects for
which such fits can be performed, this finding highlights that the
choice of the dust temperature can add uncertainty to the estimate
of $L_{\rm FIR}$ and M$_{\rm dust}$, in particular for objects with
only single photometric measurements and thus no individual
constraints on $T_{\rm FIR}$.

A related issue is that of possible AGN contributions to the heating
of the  FIR dust which will be discussed briefly in the following
section.

We also find that in our modeling strategy the FIR component can be 
isolated from the torus component if only data at 
$\lambda$$_{\rm rest}$\,$\gtrsim$\,50\,$\mu$m are considered (see section 
\ref{sec:t_fir} and Fig.\,\ref{flux_ratios}). Consequently, 
single-component fits to data at these wavelengths yield estimates of the FIR
luminosity, temperature and dust mass that match the values based on the full SED
fits. This result validates the approach in previous studies of high-$z$ 
quasars \citep{ber03a,bee06,wan08a,wan08b} in which single-component fits to
(ground-based) photometry at $\lambda_{\rm obs}$\,$\geq$\,$350\,\mu$m
was used to derive physical parameters (with the caveat of unknown dust temperature 
in some of these studies). It also adds further significance to the upper 
limits on $T_{FIR}$, $L_{FIR}$, and SFR we determine for the remainder 
of our sample where we are limited to single component fits at 
$\lambda_{\rm obs}$\,$\geq$\,$350\,\mu$m. 

The strong overlap between the torus and the FIR components in our fits 
does not provide good constraints on the emissivity index $\beta$ 
of the latter component. The difficulty of determining reliable $\beta$ 
estimates in objects with a strong AGN is also apparent from the literature: 
\citet{pri01} find a high $\beta$ value of $\sim$\,2 (with a FIR dust 
temperature of 41\,K) by combining the available photometry for a number 
of $z$\,$\sim$\,4 quasars into a single SED and using this global SED to 
constrain a modified black body fit. On the other hand, using a
similar approach (and much of the same data), \citet{bee06} find
$\beta$\,$\sim$\,1.6  and T$_{\rm FIR}$\,$\sim$\,47\,K for a sample of
quasars with $z$\,=\,$1.8-6.4$.

\subsection{Stacking of the FIR data}\label{sec:stacking}

In order to better constrain the FIR emission of the SPIRE
non-detected objects, we stacked the individual  SPIRE observations at
the nominal position of the quasar (excluding J0818+1722 due to
possible confusion issues)\footnote{We also excluded J1044$-$0125 from this 
stack. While the source is also not detected with SPIRE, its PACS detection 
would influence the average PACS flux of this subsample significantly.}.  
Even in the stacked images no significant
detection was achieved (see Fig.\,\ref{stacked_images}).  Stacking the 
corresponding PACS data we recover a faint ($\sim$\,3\,$\sigma$) 
average signal 
at 100 and 160\,$\mu$m. We iterated during the stacking, leaving a 
different source out of the stack for 
every iteration to verify that the result is not biased by any individual
object. Differences between these stacks were usually smaller than
the uncertainty on the photometry of the total stack. The photometry
was performed in  an identical manner to the individual frames and as
outlined in section\,2.

\begin{figure}[t!]
\centering
\includegraphics[angle=0,scale=.5]{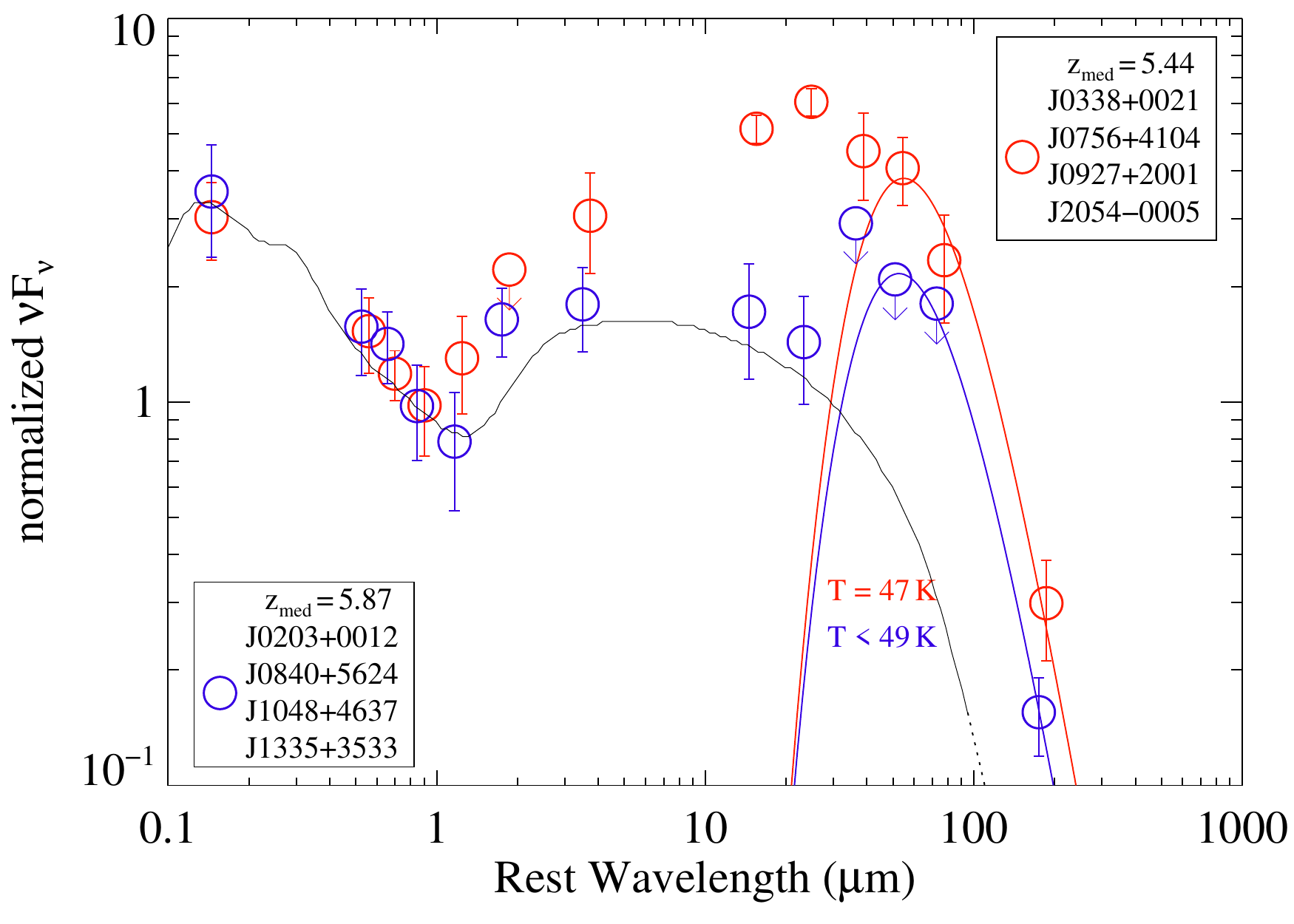}
\caption{Average SEDs of mm-detected quasars with individual FIR detections 
({\it red}) and without individual FIR detections ({\it blue}).
The solid colored lines represent modified black body fits ($\beta=1.6$) 
to the 
observed photometry at $\lambda$\,$>$\,350\,$\mu$m. The solid 
black line shows the mean SDSS quasar template presented in \citet{ric06}, 
matched in the UV/optical and extended by a power law of $F_{\nu} \propto {\nu}^{-2}$ at the longest 
wavelengths. All SEDs shown here were normalized at a rest frame wavelength of 1\,$\mu$m. 
\label{average_seds} }
\end{figure}

Using the stacked {\it Herschel} fluxes and combining them with 
stacked WISE data 
as well as averaged {\it Spitzer} and mm photometry, we can
produce an average SED for the FIR non-detected objects which is presented in 
Fig.\,\ref{average_seds}. Performing the same stacking/averaging
procedure for the objects detected in the FIR individually (excluding
J1148+5251 due to possible confusion issues) also provides us with an
average SED for these objects. 

Fitting the average FIR SEDs with a modified black body 
(at $\lambda_{\rm obs}$\,$\geq$\,350\,$\mu$m, $\beta$\,=\,1.6) 
reveals dust temperatures of 47\,K for the stack of the individually 
detected objects, as expected considering the individual FIR dust 
temperatures of the objects in this 
stack (Tab.\,\ref{results}). For the average SED of the objects 
individually undetected in SPIRE we determine an upper limit of 49\,K 
which is lower than the upper limits determined individually 
(Tab.\,\ref{results}). These temperatures are very similar to the values 
commonly adopted for mm detected quasars with only few or single photometric 
data points \citep[e.g.][]{bee06,wan08b}. They also agree well with the dust
temperatures observed for sub-mm galaxies of comparable FIR luminosity
\citep[][]{mag12}. 

In Fig.\,\ref{average_seds} we also compare our average SEDs with the
SDSS quasar template of \citet{ric06}. When normalized at 1\,$\mu$m 
(using a power-law fit to the rest frame UV/optical emission), 
we see that below $\sim$30\,$\mu$m the 
average SED of the SPIRE non-detections (blue points in 
Fig.\,\ref{average_seds}) looks similar to the typical
SDSS quasar template. At longer wavelengths we see an 
additional cool dust component for our sources, as indicated by 
the available mm observations. Not considering these longer wavelengths, 
 we see AGN powered emission with possibly increased emphasis on the 
hottest dust (NIR) when compared to the template.


In contrast, for a similar scaled UV/optical emission, the average 
infrared SED of the
SPIRE detected quasars (red points in Fig.\,\ref{average_seds}) 
exceeds the scaled template significantly at any wavelength
$\lambda_{\rm rest}$\,$\gtrsim$\,1\,$\mu$m. This includes the
presumably AGN
powered dust emission at short infrared wavelengths ($\lambda_{\rm
rest}$\,$\lesssim$\,20\,$\mu$m) as well as the FIR emission possibly
powered by additional star formation.

The observed discrepancy between the average SEDs in Figure 
\ref{average_seds} is somewhat puzzling. Both groups of objects 
have similar UV/optical properties (on average), indicating that 
their black holes grow at comparable rates. Now, assuming that the 
NIR and MIR emission is powered by the AGN, the objects with 
individual {\it Herschel} detections (red symbols) convert a higher 
fraction of their accretion luminosity into re-processed dust
emission. This could, in principle, be caused by a different dust 
geometry (or dust content) in the inner parts of these objects. This 
in turn could also lead to increased contributions of AGN powered dust 
emission at FIR wavelengths. If AGN-powered dust emission 
extends further into the FIR than anticipated by the torus models we
utilize here, the values for the inferred 
star-formation luminosity (and star formation rates) in these objects 
(Tab.\,\ref{results}) would be overestimated. 

While a detailed 
discussion of this issue is beyond the scope of 
this paper, it is worth keeping in mind that the AGN may be contributing to 
the heating of the FIR dust depending on the dust distribution and geometry.

\section{Summary and conclusions}

New {\it Herschel} observations of eleven $z>5$ quasars 
with detections at 1.2\,mm are combined with data at other 
wavelengths to analyze their full SEDs covering the rest 
frame wavelength range of $\sim$0.1-400\,$\mu$m. Our results 
can be summarized as follows:

\begin{itemize}

\item Six out of the eleven objects are detected in at least two of the 
  five {\it Herschel} bands. Five of them have sufficient data 
  coverage and quality to allow full SED fits.

\item In all cases where such fits could be performed, AGN powered 
  emission from a dusty torus is not sufficient to explain the observed 
  FIR fluxes. An additional FIR component is required to model the SEDs. 
  Similar to other studies of luminous (but lower redshift) quasars, 
  we notice the need for a hot ($\sim$1300\,K) dust component to 
  account for the strong rest frame NIR emission.

\item The additional FIR component was modeled as a modified black body 
  and shows temperatures of T\,$\sim$\,40-60\,K. We interpret this 
  emission as being powered by star formation with luminosities of 
  L$_{8-1000\mu{\rm m}}$\,$\sim$\,10$^{13}$\,L$_{\odot}$ which translate 
  into star-formation rates of several thousand solar masses per year.
    
\item Our fits also allow us to estimate that the contributions of the 
  AGN powered dust to the infrared SED are small at wavelengths $\lambda_{\rm
    rest}$\,$\gtrsim$\,50\,$\mu$m.  For the redshifts of our objects this
  implies that the star-formation powered FIR component can be isolated and 
  characterized adequately by single component fits if only photometric measurements at 
  $\lambda_{\rm obs}$\,$\gtrsim$\,350\,$\mu$m are considered. This explains
  the good match for the temperature and luminosity of the 
  FIR emission with previous studies of such objects which relied on 
  (ground-based) observations at $\lambda_{\rm obs}$\,$\gtrsim$\,350\,$\mu$m. 

 \item By stacking the {\it Herschel} data of individually undetected
  sources we recover a signal in PACS but not in SPIRE. We use this stacking 
  approach to construct average SEDs for objects with and without individual 
  {\it Herschel} detections. We find that the high-redshift objects which are 
  individually undetected with {\it Herschel} show an SED similar to to quasar 
  templates constructed from lower redshift and lower luminosity objects. 

\item The average SED of the {\it Herschel} detected objects, on 
  the other hand, shows a surplus of NIR and MIR emission relative to the 
  UV/optical when compared to the {\it Herschel} nondetections or to 
  quasar templates. This may suggest a 
  correlation between strong FIR emission (here modeled as star-formation 
  powered) and strong MIR emission (here modeled as AGN powered), and possibly 
  indicates significant AGN contributions to the FIR emission.

\end{itemize}

\acknowledgments

This publication makes use of data products from the Wide-field
Infrared Survey Explorer,  which is a joint project of the University
of California, Los Angeles, and the Jet Propulsion
Laboratory/California Institute of Technology, funded by the National
Aeronautics and Space Administration.  This work is based in part on
data obtained from the UKIRT Infrared Deep  Sky Survey (UKIDSS). 
CL acknowledges funding through DFG grant LE 3042/1-1. XF acknowledges 
supports from NSF grants AST 08-06861 and 11-07682. MH is supported by 
the Nordrhein-Westf\"alische Akademie der Wissenschaften und der
K\"unste. We thank the anonymous referee for helpful comments on 
the paper.

{\it Facilities:} \facility{Herschel}, \facility{Spitzer}.

\appendix

\section{Appendix material}
\subsection{Comments on individual objects}
Object names are given in the format J{\it hhmm+ddmm}. For 
full, NED compatible object designations see Table\,\ref{sample}. 

{\it J0203+0012 ($z=5.72$)}.\quad The combined optical and NIR 
spectroscopy by \citet{mort09} reveals broad absorption line features in this quasar which could be
one explanation for the abrupt change in SED slope observed at
$\lambda$$_{\rm rest}$\,$\lesssim$\,0.3$\mu$m
(Fig.\,\ref{sed_fits_nondetected}). It is fairly radio bright at
1.4\,Ghz compared to most $z \sim 6$ quasar \citep{wan08a} and is
detected at 250\,GHz with a flux of $1.85 \pm 0.46$\,mJy
\citep{wan11}. These authors did not detect the quasar in CO(6-5) nor in 
the corresponding continuum. While clearly seen in all our {\it Spitzer}
observations, no detection was achieved with {\it Herschel}.  Several
nearby galaxies can be identified in SDSS and IRAC, some of which
are prominent in most infrared channels. The closest bright
object is   $\sim$30\arcsec~northeast of the quasar. From SDSS
spectroscopy of some  galaxies in this area it appears that they might
belong to a foreground  cluster at redshift of $z \sim 0.077$.

{\it J0338+0021 ($z=5.03$)}.\quad This source was detected at 250\,GHz
($3.7 \pm 0.3$\,mJy), but remained undetected  at 1.4\,GHz
\citep{car01}. The 850\,$\mu$m flux is measured to be  $11.9 \pm
2.0$\,mJy \citep{pri03}. \citet{mai07} report CO(5-4) emission from
the quasar, but no continuum at 95.6\,GHz is detected. Inspection of
the field at optical through FIR wavelengths reveals a nearby source
$\sim$15\arcsec~to the west of the quasar. This object  is clearly
visible in the SDSS $r$ and $i$ bands, as well as with  IRAC and
MIPS. In all these filters, the quasar is typically the brighter
source, a situation which reverses at 100 and 160\,$\mu$m where the
nearby  object is $\sim$\,1.4 and $\sim$\,1.3 times brighter,
respectively, than the  quasar. In all SPIRE bands,  however, we  only
detect a single source. Interestingly, the quasar and the nearby
object can both be seen in ground-based 350\,$\mu$m observations
\citep{wan08b}  and the flux of the quasar ($17.7 \pm 4.4$\,mJy/beam)
is very  comparable to the SPIRE 350\,$\mu$m flux determined for the
single detection  ($18.5 \pm 6.0$\,mJy). This suggests that the SED of
the source close to the QSO peaks somewhere around the PACS bands but
does not contribute significantly at wavelengths $\gtrsim 350\,\mu$m.

{\it J0756+4104 ($z=5.11$)}.\quad The 250\,GHz flux ($5.5 \pm
0.5$\,mJy)  has been presented by \citet{pet03} who also detect the
source at 1.4\,GHz  and constrain the size of the radio emission to
$<2.3$\arcsec. \citet{pri08}  report detections at 850\,$\mu$m ($11.2
\pm 1.0$\,mJy) and 450\,$\mu$m  ($16 \pm 5$\,mJy) where ``the source
clearly appears elongated'' in the  850\,$\mu$m map at a position
angle of $\sim$\,70 degrees. We detect the quasar in all our five {\it
Herschel} bands, but the significance is often marginal.  Our flux at
350\,$\mu$m is consistent with the ground-based measurements of
\citet{wan08b}.

{\it J0818+1722 ($z=6.00$)}.\quad Radio continuum emission at 1.4\,GHz
is clearly detected \citep{wan07} and the 250\,GHz continuum is
observed  at the 3$\sigma$ level ($1.19 \pm 0.38$\,mJy;
\citealt{wan08a}). The  close inspection of our multi-wavelength
images reveal a resolved foreground  galaxy
$\sim$\,6\arcsec~north-east of the quasar. Both objects, the galaxy
and the quasar,  are individually  detected by MIPS at 24\,$\mu$m and
in shorter wavelength bands. However, only a single  source is
detected in PACS and SPIRE (note that the spatial resolution of  {\it
Herschel}/PACS at 100\,$\mu$m is comparable to {\it Spitzer}/MIPS at
24 \,$\mu$m).  From the relative positions of other sources in the
field we can identify the detection  at 100, 160, and possibly at
250\,$\mu$m with the foreground galaxy. It is conceivable that the
faint (3.5$\sigma$) detection at 350\,$\mu$m is also due to this
source. No detection is achieved at 500\,$\mu$m. In the light of these
results,  higher resolution mm observation are  clearly needed to
determine the source of the 250\,GHz continuum emission.


{\it J0840+5624 ($z=5.84$)}.\quad We do not detect this source in our PACS
nor in the SPIRE data. The quasar has been detected at 250\,GHz ($3.20
\pm 0.64$\,mJy), but not at 1.4\,GHz \citep{wan07}. CO emission is
seen in this source   (but no continuum at either 85\,GHz or 101\,GHz)
and  the 350\,$\mu$m emission is "marginally detected" from the ground
\citep{wan10}, which is consistent with our SPIRE 350\,$\mu$m upper
limit. \citet{wan10} also report the presence of another source
visible at 350\,$\mu$m as well as at 1.4\,GHz located
$\sim$\,30\arcsec~north-west of the quasar. This source is also
detected in all our infrared bands. Inspection of the IRAC maps
reveals this detection to coincide with two close objects which  could
be two slightly resolved galaxies separated by $\sim$1.8\arcsec as
seen on an archival {\it HST}/WFC3 image in the F105W filter.

{\it J0927+2001 ($z=5.77$)}.\quad Previously detected at 250\,GHz
($4.98 \pm 0.75$\,mJy; \citealt{wan07}) as well as in CO and  in the
90\,GHz continuum \citep{car07}. The 350\,$\mu$m observations  by
\citet{wan08b} show the quasar ($17.7 \pm 5.7$\,mJy\,beam$^{-1}$) and
a secondary source of  equal brightness 15\arcsec~to the
southeast. While the quasar detection was confirmed in \citet{wan10}
with better sensitivity ($11.7 \pm 2.4$\,mJy\,beam$^{-1}$), the
secondary source was not. We detect the quasar with SPIRE, but  not
with PACS.

{\it J1044$-$0125 ($z=5.78$)}.\quad This well studied quasar shows a
broad CIV absorption  feature in its spectrum
\citep[e.g.][]{mai01,goo01}. The continuum emission is detected at
850\,$\mu$m ($5.6 \pm 1.0$\,mJy; \citealt{pri08}) and at 250\,GHz
($1.82 \pm 0.43$\,mJy; \citealt{wan08a}). \citet{wan10} report the detection 
of CO (6-5) but can only give an upper limit on the continuum at 102\,GHz. 
The quasar is not seen in ground-based observations at 350\,$\mu$m \citep{wan10} 
and 1.4\,GHz \citep{pet03}. We detect the quasar with PACS, but not with SPIRE.



{\it J1048+4637 ($z=6.23$)}.\quad
While clearly detected in the available {\it Spitzer} bands, this
quasar  remains undetected in our {\it Herschel}
photometry. \citet{wan08b} only  give an upper limit on the
350\,${\mu}$m flux, just like \citet{rob04}  at 450 and
850\,${\mu}$m. These authors, however note that based on the
published 1.2\,mm detection ($3.0 \pm 0.4$\,mJy; \citealt{ber03a}),
the SCUBA 850\,${\mu}$m  observations are deep enough to enable the
detection of the source  with 4$\sigma$ significance given a dust
temperature of 40\,K. At even longer  wavelengths, \citet{wan10}
detect CO(6-5) emission as well as the continuum  at
96\,GHz. Observations at 1.4\,GHz only provide an upper limit on the
source flux \citep{wan07}.

%

{\it J1148+5251 ($z=6.42$)}.\quad
This famous object was the highest redshift quasar known for half
a decade \citep{fan03,wil07} and as such has been studied at many
wavelengths, including the mm and sub-mm regime. We also have
observed this quasar previously  with {\it Herschel}/PACS and reported
detections at 100 and 160\,$\mu$m  \citep{lei10a}. Surprisingly, we
discovered a secondary object  $\sim$10\arcsec~north-west of the
quasar which is brighter at 160\,$\mu$m  but can still be identified
at 100\,$\mu$m. Ground-based data at 350\,$\mu$m  \citep{bee06} and
250\,GHz \citep{ber03a} revealed an intriguing elongation  of the
quasar detection in the direction of the second source. We argued in
our  previous paper that this could be an indication for the presence
of the  secondary source. A possible counterpart is also seen in the
24\,$\mu$m images.  In the IRAC band, three sources are detected
around the position of this secondary source,  two of which can also
be identified on deep {\it Hubble} Space Telescope ({\it HST})  images
with the ACS camera in the F850LP filter. Recently, new deep {\it HST}
images  from WFC3 in the NIR revealed also the third source seen in
IRAC. This object is clearly detected,  but faint in F125W and gets
significantly brighter in F160W. This is our best candidate for a
counterpart of the secondary source seen at 24, 100 and
160\,$\mu$m. Surprisingly, however, follow-up observations with the
Plateau de Bure Interferometer (PdBI) at 1.2\,mm at  $\sim$\,1\arcsec
resolution did not yield a detection and the 3$\sigma$ upper limit we
derive is 0.9\,mJy.

Since our initial photometry \citep{lei10a} we have re-observed
the quasar with {\it Herschel} and  obtained new images at 70, 160,
250, 350, and 500\,$\mu$m. While the quasar itself is faintly detected
at 70\,$\mu$m, there is no sign for a secondary source. The  new
160\,$\mu$m observations confirm our earlier findings that the flux
appears  to come from two sources. The source complex is also detected
in all SPIRE bands,  but the spatial resolution is too low to identify
a possible double source. Combining the multi-wavelength photometry 
of the secondary source, we find that the SED is consistent with a 
star-forming galaxy at $z\sim2$ with ULIRG-like luminosity 
(L$_{8-1000\mu{\rm m}}$\,$\sim$\,few times 10$^{12}$\,L$_{\odot}$).

{\it J1335+3533 ($z=5.90$)}.\quad
The optical spectrum of this source is quite unusual as it shows 
a typical quasar continuum but virtually no Ly$\alpha$ emission 
\citep{fan06}. At longer wavelengths, the quasar is seen in the 
{\it Spitzer} bands, but not in our {\it Herschel} data. \citep{wan10} 
report the detection of the CO(6-5) transition and give upper limits on 
the continuum at 350\,$\mu$m and 100\,GHz. The 250\,GHz ($2.34 \pm 0.50$\,mJy) 
and 1.4\,GHz continuum was detected by \citet{wan07}. 


{\it J2054$-$0005 ($z=6.04$)}.\quad
This is the only source in our sample for which we do not have {\it
Spitzer} observations.  The SDSS imaging featured a sufficient number
of sources that could  also be identified on the PACS maps to
determine the position of the quasar accurately. The QSO is detected
at 250\,GHz ($2.38 \pm 0.53$\,mJy), but not at 1.4\,GHz
\citep{wan08a}. CO observations  revealed the (6-5) transition but no
continuum at 98\,GHz \citep{wan10}.  We see the source at 160\,$\mu$m
with PACS, but can only give an upper  limit on the 100\,$\mu$m
flux. A faint 3-4$\sigma$ source is visible at  250 and
350\,$\mu$m. At 500\,$\mu$m we run into confusion issues with a source
located $\sim$30\arcsec~north of the quasar's nominal position, which
can also be  identified (separate from the quasars) in the other SPIRE
bands and at  160\,$\mu$m. Since our photometry indicates that the FIR
peak of the quasar  in F$_{\nu}$ falls close to the 250\,$\mu$m band,
we do not expect significant  flux in the 500\,$\mu$m channel. In a
NIR spectrum, \citet{rya09} see a  very strong MgII absorber at
$z_{\rm abs}=2.598$.

\end{document}